\documentclass[a4paper,11pt]{article}
\pdfoutput=1

\usepackage{jheppub}
\usepackage[T1]{fontenc}
\usepackage{amsmath}
\usepackage{cernunits}
\usepackage{xspace}
\usepackage{heppennames2}

\newcommand{\beq}{\begin{equation}}
\newcommand{\eeq}{\end{equation}}
\newcommand{\qcut}{Q_{\mathrm{cut}}}
\newcommand{\GeV}{\mathrm{GeV}}

\newcommand{\alphaS}{\alpha_s}

\def\refeq#1{\mbox{(\ref{#1})}}

\def\reffi#1{\mbox{Fig.~\ref{#1}}}
\def\reffis#1{\mbox{Figures~\ref{#1}}}
\def\refta#1{\mbox{Table~\ref{#1}}}
\def\reftas#1{\mbox{Tables~\ref{#1}}}
\def\refse#1{\mbox{Section~\ref{#1}}}
\def\refses#1{\mbox{Sections~\ref{#1}}}
\def\refapp#1{\mbox{Appendix~\ref{#1}}}
\def\citere#1{\mbox{Ref.~\cite{#1}}}
\def\citeres#1{\mbox{Refs.~\cite{#1}}}

\def\ie{i.e.\ }
\def\cf{cf.\ }

\newcommand{\ri}{\mathrm{i}}
\newcommand{\rF}{\mathrm{F}}
\newcommand{\rR}{\mathrm{R}}
\newcommand{\rT}{\mathrm{T}}
\newcommand{\rd}{\mathrm{d}}
\newcommand{\MW}{M_\mathrm{W}}
\newcommand{\MZ}{M_\mathrm{Z}}
\newcommand{\MH}{M_\mathrm{H}}
\renewcommand{\perp}{\rT}

\def\mathswitchr#1{\relax\ifmmode{\mathrm{#1}}\else$\mathrm{#1}$\fi}
\newcommand{\psel}{\mathswitchr P}
\newcommand{\csel}{\mathswitchr C}
\newcommand{\ssel}{\mathswitchr S}
\newcommand{\Pt}{\mathswitchr t}
\newcommand{\lnln}{\ell\nu \ell\nu}
\newcommand{\mnen}{\mu^+\nu_\mu \Pe^-\bar\nu_{\Pe}}
\newcommand{\delres}{\Delta_\mathrm{res}}
\newcommand{\delqcd}{\Delta_\mathrm{QCD}}
\newcommand{\deltot}{\Delta_\mathrm{tot}}
\newcommand{\sigsr}{\sigma_\ssel}
\newcommand{\sigcr}{\sigma_\csel}
\newcommand{\dsc}{\delta_{\ssel/\csel}}
\newcommand{\met}{E\!\!\!/_\rT}

\newcommand{\NLOacc}{{\scshape Nlo}\xspace}
\newcommand{\MCatNLO}{{\scshape Mc@Nlo}\xspace}
\newcommand{\MEPSatNLO}{{\scshape Meps@Nlo}\xspace}
\newcommand{\Sherpa}{{\scshape Sherpa}\xspace}
\newcommand{\Comix}{{\scshape Comix}\xspace}
\newcommand{\Amegic}{{\scshape Amegic++}\xspace}
\newcommand{\Rivet}{{\scshape Rivet}\xspace}
\newcommand{\OpenLoops}{{\scshape OpenLoops}\xspace}
\newcommand{\SherpaOpenLoops}{{\scshape Sherpa+OpenLoops}\xspace}
\newcommand{\Collier}{{\scshape Collier}\xspace}
\newcommand{\ATLAS}{{\scshape Atlas}\xspace}
\newcommand{\CMS}{{\scshape Cms}\xspace}

\newcommand{\LOOPSQ}{{\scshape Loop$^2$}\xspace}
\newcommand{\LOOPSQfl}{{\scshape Loop$^2$\;$4\ell$}\xspace}
\newcommand{\LOOPSQflj}{{\scshape Loop$^2$\;$4\ell+1j$}\xspace}
\newcommand{\LOOPSQfljs}{{\scshape Loop$^2$\;$4\ell\,(+1j)$}\xspace}
\newcommand{\LOOPSQPS}{{\scshape Loop$^2$+PS}\xspace}
\newcommand{\LOOPSQPSfl}{{\scshape Loop$^2$+PS\:$4\ell$}\xspace}
\newcommand{\LOOPSQPSflj}{{\scshape Loop$^2$+PS\:$4\ell+1j$}\xspace}
\newcommand{\NLOfl}{{\scshape Nlo\:$4\ell$}\xspace}
\newcommand{\NLOflj}{{\scshape Nlo\:$4\ell+1j$}\xspace}
\newcommand{\NLOfljs}{{\scshape Nlo\:$4\ell\, (+1j)$}\xspace}
\newcommand{\LOfljs}{{\scshape Lo\:$4\ell\, (+1j)$}\xspace}
\newcommand{\MCatNLOfl}{{\scshape Mc@Nlo\:$4\ell$}\xspace}
\newcommand{\MCatNLOflj}{{\scshape Mc@Nlo\:$4\ell+1j$}\xspace}
\newcommand{\MEPSatNLOflj}{{\scshape Meps@Nlo\:$4\ell+0,1j$}\xspace}
\newcommand{\MEPSTOT}{{\scshape Meps@Nlo+Loop$^2$}\xspace}
\newcommand{\MEPSatLOOPSQ}{{\scshape Meps@Loop$^2$}\xspace}
\newcommand{\MEPSatLOOPSQflj}{{\scshape Meps@Loop$^2$\:$4\ell+0,1j$}\xspace}

\newcommand{\ir}{\mathrm{IR}}
\newcommand{\uv}{\mathrm{UV}}
\newcommand{\irct}{\mathrm{mod}}
\newcommand{\dZAb}{\delta Z_A^{(\mathrm{b})}}
\newcommand{\calIb}{\mathcal{I}^{(\mathrm{b})}}
\newcommand{\mll}{m_{\ell\ell'}}
\newcommand{\hww}{\PH\to\PW\PW^*}
\newcommand{\ptmax}{p_\rT^{\mathrm{max}}}
\newcommand{\ptmin}{p_\rT^{\mathrm{min}}}

\def\hwwatlaszero{./figures/plots-0j/HWW_ATLAS}
\def\hwwatlasone{./figures/plots-1j/HWW_ATLAS}
\def\hwwcmszero{./figures/plots-0j/HWW_CMS}
\def\hwwcmsone{./figures/plots-1j/HWW_CMS}
\def\hwwatlasratio{./figures/plots-loop2-ratio/HWW_ATLAS}
\def\hwwcmsratio{./figures/plots-loop2-ratio/HWW_CMS}

\graphicspath{{./feynmandiagrams/}}

\title{Precise Higgs-background predictions:\\
       merging NLO QCD and squared quark-loop\\
       corrections to four-lepton + 0,1 jet production}

\author[a]{F.~Cascioli,}
\author[b]{S.~H\"oche,}
\author[c]{F.~Krauss,}
\author[a]{P.~Maierh\"ofer,}
\author[a]{S.~Pozzorini,}
\author[d]{F.~Siegert}

\emailAdd{
cascioli@physik.uzh.ch,
shoeche@slac.stanford.edu,
frank.krauss@durham.ac.uk,
philipp@physik.uzh.ch,
pozzorin@physik.uzh.ch,
frank.siegert@cern.ch}

\affiliation[a]{Institut f\"ur Theoretische Physik, Universit\"at Z\"urich,  8057 Z\"urich, Switzerland}
\affiliation[b]{SLAC National Accelerator Laboratory, Menlo Park, CA 94025, USA}
\affiliation[c]{Institute for Particle Physics Phenomenology, Durham University, Durham DH1 3LE, UK}
\affiliation[d]{Physikalisches Institut, Albert-Ludwigs-Universit{\"a}t Freiburg, D-79104 Freiburg, Germany}

\abstract{%
  We present precise predictions for four-lepton plus jets production at the
  LHC obtained within the fully automated \SherpaOpenLoops framework.
  Off-shell intermediate vector bosons and related interferences
  are consistently included using the complex-mass scheme.
  Four-lepton plus 0- and 1-jet final states are described at NLO accuracy,
  and the precision of the simulation is further increased by squared
  quark-loop NNLO contributions in the $\Pg\Pg\to 4\ell$, $\Pg\Pg\to
  4\ell+\Pg$, $\Pg q\to 4\ell+q$, and $q\bar q\to 4\ell+\Pg$ channels.
  These NLO and NNLO contributions are matched to the \Sherpa parton shower,
  and the 0- and 1-jet final states are consistently merged using the
  \MEPSatNLO technique.
  Thanks to Sudakov resummation, the parton shower provides improved
  predictions and uncertainty estimates for exclusive observables.  This is
  important when jet vetoes or jet bins are used to separate four-lepton final
  states arising from Higgs decays, diboson production, and top-pair
  production.
  Detailed predictions are presented for the \ATLAS and \CMS
  $\hww$ analyses at 8 TeV in the 0- and 1-jet bins.
  Assessing renormalisation-, factorisation- and resummation-scale
  uncertainties, which reflect also
unknown subleading Sudakov logarithms
 in jet bins, we find that
  residual perturbative uncertainties are as small as a few percent.
}

\begin{document}

\preprint{
  \begin{minipage}[t]{20em}
    \begin{flushright}
      IPPP/13/66\\
      DCPT/13/132\\
      MCNET-13-12\\
      SLAC-PUB-15714\\
      ZU-TH 15/13\\
      LPN13-056\\
      FR-PHENO-2013-007
    \end{flushright}
  \end{minipage}
}

\maketitle
\flushbottom

\section{Introduction}
\label{Sec:Introduction}

Final states involving four leptons played a key role in the
discovery of the Higgs boson~\cite{Aad:2012tfa,Chatrchyan:2012ufa}
and will continue to be crucial in the understanding of its properties and
coupling structure.  There are two classes of final states of interest,
namely those consistent with 
$\PH\to \PZ\PZ^*$ decays
yielding four charged
leptons and those related to $\hww$ resulting in two charged
leptons and two neutrinos.
They have quite different backgrounds, and for the latter, the dominant and
large top-pair production background necessitates the introduction of jet
vetoes to render the signal visible.  More precisely, four-lepton final
states consistent with $\hww$ decays are split into exclusive bins with
0, 1 and 2 jets.  The separate analysis of the different jet bins permits to
disentangle Higgs production via gluon fusion from the vector-boson fusion (VBF)
production mode.  In addition, data-driven determinations of the
$\hww$ background take advantage of the fact that its two leading
components---diboson and top-pair production---deliver final states of
different jet multiplicity.  While diboson production represents the leading
background in the 0-jet bin, the top-production component becomes slightly
more important in the 1-jet bin and clearly dominant in the 2-jet bin.

Due to the absence of a mass peak and the high background cross section, the
experimental analyses suffer from  signal-to-background ratios as low
as around 10 percent.  It is thus clear that the precision of the employed
background-determination techniques, and the related error estimates, play a
crucial role for any Higgs-boson measurement in this channel.  In fact, with
the statistics available at the end of the LHC run at 8~TeV, systematic
errors resulting from the background subtraction already dominate the total
uncertainty.

In the $\hww$ analyses by \ATLAS~\cite{ATLAS-CONF-2013-030} and
\CMS~\cite{CMS-PAS-HIG-13-003} a data-driven approach is used to reduce
uncertainties in the simulation of the two leading backgrounds.  The
top-production contribution is fitted to data in a top-enriched control
sample.  Using Monte-Carlo tools, the top background is extrapolated to the
signal region and to an independent diboson-enriched control region.
This latter region is used to normalise the diboson background
after subtraction of the top contamination. The diboson background is
then extrapolated to the signal region using Monte-Carlo predictions.  
While this approach reduces theoretical uncertainties associated with the
background normalisation, the extrapolations between the
various control and signal sub-samples rely on Monte-Carlo modelling of the
background
shapes.  

Given that the accuracy of present Higgs-boson measurements requires
extrapolation uncertainties at the percent level, it is clear that Monte-Carlo 
simulations should include all available correction effects and
appropriate error estimates.  
In this context, due to various nontrivial features of the $\hww$
analyses, the requirements in terms of theoretical precision go beyond the
mere inclusion of higher-order corrections to inclusive four-lepton
production. 
First, a reliable modelling of the various jets associated to the
four-lepton final state requires higher-order QCD corrections up to the
highest relevant jet multiplicity.  Second, in order to describe potentially
large Sudakov logarithms and related uncertainties, which arise from jet
vetoes and exclusive jet bins, fixed-order predictions should be matched to
parton showers or supplemented by appropriate resummations.  Third, vector
bosons are produced well below their mass shell in $\hww\to\lnln$
decays.  Theoretical predictions for background processes should thus
account for corresponding off-shell effects, including non-resonant
channels and related interferences.

In this paper we will concentrate on diboson production, which represents about 75
and 40 percent of the $\hww\to\lnln$ background in the 0- and 1-jet bins,
respectively.  While we are especially interested in the Higgs-boson analyses, 
diboson production plays an important role also for precision
tests of the Standard Model, vector-boson scattering, searches for anomalous
couplings, or as a background in numerous searches.

Higher-order QCD corrections to diboson production at hadron colliders have
been extensively studied in the literature.  Next-to-leading order (NLO)
corrections to inclusive $\PW$-pair final states
\cite{Ohnemus:1991kk,Frixione:1993yp,Ohnemus:1994ff,Dixon:1999di,Campbell:1999ah,Campbell:2011bn}
amount to roughly 50\% at the LHC and can be further enhanced in the tails
of distributions or reduced by jet vetoes.  Due to the gluon--(anti)quark channels, which
start contributing to $\Pp\Pp\to\PWp\PWm$ only at NLO, the size of the
corrections largely exceeds estimates based on leading-order (LO) scale variations.
The matching of NLO predictions for WW production to parton showers was
first studied in~\citere{Frixione:2002ik} using the MC@NLO
method~\cite{Frixione:2002ik}, while the
POWHEG matching~\cite{Frixione:2007vw} for WW, WZ and ZZ production,
including spin-correlated leptonic decays with non-resonant contributions,
was presented in~\citere{Melia:2011tj}.  Similar predictions for ZZ
production based on the MC@NLO method can be found in
\citere{Frederix:2011ss}.

The NLO corrections to $\Pp\Pp\to\PWp\PWm j$ were presented in
\citeres{Campbell:2007ev,Dittmaier:2007th,Dittmaier:2009un}, including
spin-correlated leptonic decays and off-shell effects associated with the
Breit--Wigner distributions of the resonant W-bosons.  At the 14 TeV LHC with
rather inclusive cuts the corrections are slightly above 30\%.  Also in this case,
due to the opening of the $\Pg\Pg\to\PWp\PWm q\bar q$ channel at NLO, the
corrections largely exceed LO scale variations.  This means that
uncertainty estimates based on scale variations start to be meaningful only
at NLO. The inclusion of QCD corrections is thus essential in order
to improve both, theoretical predictions and error estimates.
The matching of NLO $\Pp\Pp\to\PWp\PWm j$ calculations to parton
showers remains to be addressed in the literature.

Higher-order QCD effects have been studied in quite some detail also for
$\Pp\Pp\to\PW\PW jj$ in the VBF- and QCD-production modes.  In the VBF 
case, NLO corrections including resonant and non-resonant leptonic
decays~\cite{Jager:2006zc} have been matched to parton showers
\cite{Jager:2013mu}.  For QCD-induced $\PWp\PWm jj$ production, NLO
predictions have been presented by two independent groups, including
spin-correlated leptonic decays as Breit--Wigner resonances
\cite{Melia:2011dw} or in narrow-width approximation~\cite{Greiner:2012im}. 
Depending on the scale choice and the collision energy, 
NLO effects at the LHC can range from a few percent to tens of
percent~\cite{Melia:2011dw}.  Up to date, only NLO QCD corrections to same-sign 
$\PW\PW jj$ production~\cite{Melia:2010bm,Denner:2012dz} have been matched to parton
showers~\cite{Jager:2011ms}.  Recently, NLO predictions became available
also for $\Pp\Pp\to WZ jj$~\cite{Campanario:2013qba}.

While full NNLO corrections to diboson production are not yet available, the
finite and gauge-invariant contribution from squared quark-loop
$\Pg\Pg\to\PWp\PWm$ amplitudes  was studied in detail in
\citeres{Binoth:2005ua,Binoth:2006mf,Campbell:2011cu,Melia:2012zg}.  Due to
the large gluon flux, such NNLO terms increase the inclusive $\PWp\PWm$
cross section by 3--5\% at the LHC.
Their relative importance is known to increase in the $\hww$
analysis.  While in presence of tight cuts it can reach up to $30\%$ 
\cite{Binoth:2005ua,Binoth:2006mf}, with the cuts currently applied by
the LHC experiments it remains around $10\%$
\cite{Campbell:2011cu,Melia:2012zg}, which corresponds to about half of the
Higgs-boson signal.  In spite of the tiny Higgs-boson width, the interference of the
$\Pg\Pg\to 4\ell$ continuum with the signal 
can reach order 10\% of the $\Pg\Pg\to 4\ell$ signal-plus-background cross section~\cite{Binoth:2006mf,Campbell:2011cu}.
This interference contribution arises almost entirely above threshold, \ie at invariant masses $M_{\PW\PW}>2\MW$, 
and is strongly suppressed at small dilepton invariant mass as well as 
in the transverse-mass region $m_\perp\lesssim M_{\PH}$~\cite{Campbell:2011cu,Kauer:2012hd}.
In~\citere{Melia:2012zg}
it was shown that also $\Pp\Pp\to \PWp\PWm j$ receives a significant 
$\Pg\Pg\to\PWp\PWm \Pg$ contribution from squared quark-loop amplitudes,
which can reach 6--9\% when Higgs-search cuts are applied.

In this paper we present new precise predictions for four-lepton plus 0- and
1-jet production,\footnote{First partial
results of this study were anticipated
in~\cite{Heinemeyer:2013tqa}.} obtained within the fully automated \SherpaOpenLoops
framework~\cite{Gleisberg:2008ta,Cascioli:2011va}.  
The \OpenLoops~\cite{Cascioli:2011va} algorithm is an 
automated generator of virtual QCD corrections to Standard-Model processes,
which uses the \Collier library~\cite{collier} for the numerically stable evaluation of
tensor integrals~\cite{Denner:2002ii,Denner:2005nn} and scalar
integrals~\cite{Denner:2010tr}.  Thanks to a fully flexible interface of
\Sherpa with \OpenLoops, the entire generation chain---from process
definition to collider observables---is fully automated and can be steered
through \Sherpa run cards.

The simulation presented in this paper is the first phenomenological 
application of  \SherpaOpenLoops.
It comprises all previously known QCD contributions to
$\Pp\Pp\to 4\ell$ and $\Pp\Pp\to 4\ell+1j$,
and extends them in various respects.  For both
processes, NLO corrections are matched to the
\Sherpa parton shower~\cite{Gleisberg:2008ta} using the fully colour-correct
formulation~\cite{Hoeche:2011fd,Hoeche:2012ft} of the \MCatNLO
method~\cite{Frixione:2002ik}.\footnote{In the following, \MCatNLO refers to
the algorithm of~\citeres{Hoeche:2011fd,Hoeche:2012ft}, which is an extension of the
original \MCatNLO method by Frixione and Webber~\cite{Frixione:2002ik}.
In particular, we never refer to the \MCatNLO event generator.}
Using the recently developed multi-jet merging at
NLO~\cite{Gehrmann:2012yg,Hoeche:2012yf}, the two \MCatNLO samples are
consistently merged in a single simulation, which preserves the logarithmic accuracy of the 
shower and simultaneously guarantees NLO accuracy in
the 0- and 1-jet bins.  Also squared quark-loop contributions
to $\Pp\Pp\to 4\ell+0,1$ jets are included.  In addition to the pure gluonic
channels, $\Pg\Pg\to 4\ell$ and $\Pg\Pg\to 4\ell+\Pg$, also the quark-induced $qg\to 4\ell+q$, $\bar q g\to
4\ell+\bar q$, and $q\bar q \to 4\ell+\Pg$ channels are taken into account.  Moreover,
the various squared quark-loop contributions are matched to the parton shower and
merged in a single sample.  To guarantee an exact treatment of spin
correlations and off-shell vector bosons, the complex-mass scheme~\cite{Denner:2005fg}
is used, and all resonant and non-resonant four-lepton plus jets topologies are
taken into account.

Detailed predictions are presented for the case of W-pair plus jets
production as a signal, as well as for the irreducible background to the \ATLAS
and \CMS $\hww$ analyses in the 0- and 1-jet bins.
To illustrate the relative importance of the various contributions, merged
NLO predictions are contrasted with an inclusive \MCatNLO simulation of
$\Pp\Pp\to 4\ell$, with separate NLO results for four-lepton plus 0- and 1-jet
production, and with squared quark-loop contributions.
Residual perturbative uncertainties are assessed by means of scale
variations.  In addition to the usual renormalisation- and factorisation-scale
variations, also the resummation scale of the \Sherpa parton shower is
varied.  This reflects subleading Sudakov logarithms beyond the shower
approximation, which renders error estimates more realistic in presence of
jet vetoes.

The presented simulation involves various interesting improvements for the
$\hww$ analyses.  The NLO matching and merging of $\Pp\Pp\to 4\ell+0,1$ jets
provides NLO accurate predictions and Sudakov resummation in the first two
exclusive jet bins.  The inclusion of $\Pp\Pp\to 4\ell+1j$ at NLO, which
contributes, as a result of merging, both to the 0- and 1-jet bins,
guarantees that all $q\bar q$, $q g$, $\bar q g$ and $\Pg\Pg$ channels are
open.  In this situation scale variations can be regarded as more realistic
estimates of theoretical uncertainties.  Matching and merging render squared
quark-loop $\Pg\Pg\to 4\ell$ contributions to exclusive jet bins more
reliable.  In fact, if not supplemented by shower emissions, the
parton-level $\Pg\Pg\to 4\ell$ channel completely misses the Sudakov
suppression induced by the jet veto.  Matching $\Pg\Pg\to 4\ell$ to the
parton shower automatically implies fermion-loop processes with
initial-state quarks, like $q\Pg\to 4\ell+q$, which result from $q\to q \Pg$
shower splittings.  The corresponding quark-induced matrix elements, which
are included for the first time in this study, provide an improved
description of hard jet emission.

Finally we point out that, while the presented simulation deals only with
$\mnen+$jets final states, the employed tools allow for a fully automated
generation of any other combination of charged leptons and neutrinos.

The paper is organised as follows.  In \refse{Sec:Framework} we discuss the
calculation of one-loop amplitudes with \OpenLoops and \Collier as well as NLO matching
and merging in \Sherpa.  Details of the Monte-Carlo simulations can be found
in \refse{se:mcsample}.  In \refse{Sec:WWinc} we present results for
inclusive WW-signal cuts, with emphasis on squared quark-loop contributions,
merging aspects and jet-veto effects.  \refse{Sec:HWW} is devoted to a
detailed discussion of the $\hww$ analyses at the LHC.  Our
conclusions are presented in \refse{Sec:Conclusions}. 
Appendix~\ref{app:b-treatment} describes the treatment of bottom- and top-quark contributions,
and the $\hww$ selection cuts are documented in Appendix~\ref{se:hww_setup}.

\section{NLO, matching and merging with \Sherpa and \OpenLoops}
\label{Sec:Framework}

This section is devoted to the automation of NLO calculations in
\SherpaOpenLoops and to methodological aspects of matching and merging of
NLO and squared quark-loop corrections.

\subsection{Loop amplitudes with \OpenLoops and \Collier}

For the calculation of virtual corrections we employ
\OpenLoops~\cite{Cascioli:2011va}, a fully automated generator of Standard-Model 
scattering amplitudes at one loop.  The \OpenLoops method has been
designed in order to break various bottlenecks in multi-particle one-loop
calculations.
The algorithm is formulated in terms of Feynman diagrams and tensor
integrals, 
which allows 
for very high CPU efficiency to be achieved.
While this was already known from $2\to 4$ NLO calculations based on algebraic 
methods~\cite{Bredenstein:2009aj,Bredenstein:2010rs,Denner:2010jp,Denner:2012yc},
the idea behind \OpenLoops is to replace algebraic manipulations of Feynman
diagrams by a numerical recursion, which results in order-of-magnitude
reductions both in the size of the numerical code and in the time needed to
generate it.  Thanks to these improvements, which are accompanied by a
further speedup of loop amplitudes at runtime, \OpenLoops is able to
address large-scale problems, such as NLO simulations for classes of
processes involving a large number of multi-leg partonic channels.

The \OpenLoops recursion is based on the well known idea that one-loop
Feynman diagrams can be cut-opened in such a way that the resulting
tree-like objects can be generated with automated tree algorithms.  However,
rather than relying on conventional tree algorithms, the recursion is
formulated in terms of loop-momentum polynomials called ``open loops''.  An
analogous idea was proposed in \citere{vanHameren:2009vq} in the framework
of Dyson-Schwinger off-shell recursions.  Diagrams involving $N$ loop
propagators are built by reusing components from related diagrams with $N-1$
loop propagators in a systematic way.  Together with other techniques to speed up
colour and helicity summations~\cite{Cascioli:2011va}, this allows to handle
multi-particle processes with up to $\mathcal{O}{(10^4-10^5)}$ one-loop
diagrams.

The algorithm is completely general, since the kernel of the
reduction depends only on the Feynman rules of the model at hand, and once
implemented it is applicable to any process.  Similarly, the so-called $R_2$
rational terms~\cite{Draggiotis:2009yb} are generated as counterterm-like
diagrams from corresponding Feynman rules.

For the numerical evaluation of one-loop tensor integrals, \OpenLoops is
interfaced to the \Collier library~\cite{collier}, which implements the
Denner--Dittmaier reduction methods~\cite{Denner:2002ii,Denner:2005nn} and
the scalar integrals of~\citere{Denner:2010tr}.  Thanks to a variety of
expansions in Gram determinants and other kinematic
quantities~\cite{Denner:2005nn}, the \Collier library systematically avoids
spurious singularities in exceptional phase-space regions.  This allows for
a fast and numerically stable evaluation of tensor integrals in double
precision.
Alternatively, OPP reduction~\cite{Ossola:2006us} can be used instead of tensor integrals.

The present implementation of \OpenLoops can handle one-loop QCD corrections
to any Standard-Model process with up to six particles attached to the
loops.\footnote{Final-state lepton pairs couple to QCD loops only via
electroweak vector bosons and should thus be counted as a single particle.}
Virtual QCD corrections are computed exactly, and the full set of Feynman
diagrams contributing to a given process is taken into account by default. 
For final states involving four leptons, the complex-mass
scheme~\cite{Denner:2005fg} is used for a consistent description of resonant
and non-resonant vector-boson propagators and their interferences. 
\OpenLoops can also be used to compute squared one-loop
matrix elements,
such as the various squared quark-loop amplitudes considered in
this paper. The correctness of one-loop amplitudes generated with
\OpenLoops has been tested systematically against an independent in-house
generator for more than one hundred different parton-level processes,
and agreement at the level of 12-14 digits on average was found.
The first public version of the code will be released in the course of 2013.

\subsection{Matching to parton shower and merging in \Sherpa}
\label{sec:match_n_merge}

The combination of fixed-order calculations and resummation is essential for
the analysis of exclusive cross sections.  Parton showers implement
resummation in a simple, yet effective way.  While formally only correct to
leading-logarithmic accuracy, they include a number of features that are
important for a realistic prediction of exclusive jet spectra.  Firstly, the
strong coupling factors associated to quark and gluon emissions are
evaluated at scales set by the transverse momenta in the parton branchings. 
This choice sums higher-logarithmic corrections, originating in the enhanced
probability for soft and collinear radiation.  Secondly, modern parton
showers naturally implement local four-momentum conservation in each
individual parton emission, which leads to a realistic description of the
kinematics in multi-particle final states.  Thirdly, most parton showers
include higher-logarithmic corrections in an effective approximation known
as angular ordering.  This method yields the correct jet rates in $\Pep\Pem$
annihilation to hadrons~\cite{Catani:1991hj}, as well as the production of
Drell--Yan lepton pairs in hadronic collisions~\cite{Catani:1990rr}.

Cross sections in jet bins 
as analysed here 
are strongly sensitive
to real radiative corrections, or their suppression.  Such corrections are
dominated by 
Sudakov double 
logarithms of
the jet-veto scale, which can have a large impact both on exclusive cross
sections and related uncertainty estimates.  A priori it is not clear if
renormalisation- and factorisation-scale variations provide a meaningful
estimate of NLO cross sections in jet bins.  In fact conventional scale
variations can turn out to be artificially small as a consequence of
accidental cancellations between Sudakov-enhanced logarithms and
contributions that do not depend on the jet veto~\cite{Stewart:2011cf}.  In
this respect, fixed-order calculations matched to a parton shower allow for
more reliable predictions and error estimates.  In particular,
factorisation- and renormalisation-scale uncertainties can be supplemented
by independent variations of the resummation scale, \ie the scale that
enters Sudakov logarithms and corresponds to the starting point of the
parton-shower evolution.  Resummation-scale variations reflect the
uncertainties associated with subleading Sudakov logarithms beyond the
shower approximation, and independent variations of the factorisation,
renormalisation and resummation scales provide a more reliable assessment of
theoretical errors in presence of jet bins.

The parton shower used for our calculation is based on Catani--Seymour dipole
subtraction~\cite{Catani:1996vz}.  It was described in detail
in~\citeres{Schumann:2007mg,Hoeche:2009xc}.  Splitting kernels are given by
the spin-averaged dipole-insertion operators, taken in the large-$N_\mathrm{c}$
limit.  
The momentum mapping in branching processes
is defined by
inversion of the kinematics in the dipole-subtraction scheme.  The
parameters of the parton shower are given by its infrared cutoff, by the
resummation scale, and by the precise scale at which the strong coupling is
evaluated.  This latter scale must be proportional to the transverse
momentum $k_\perp$ in the splitting process, but it may be varied using a
prefactor, $b$, of order one.  In practice, the explicit form of $k_\perp$
is dictated by the dipole kinematics, and different prefactors are used for
final-state and initial-state evolution.  The resummation scale can be
chosen freely in principle, but at leading order it must be equal to the
factorization scale.

The matching of NLO calculations and parton showers in the \MCatNLO
method~\cite{Frixione:2002ik} is based on the idea that
$\mathcal{O}(\alphaS)$ expansions of the parton shower can provide local
subtraction terms (called MC counterterms), which 
cancel all
infrared singularities in real-emission matrix elements.  The subtracted
result is a finite remainder.  When combined with the parton shower it gives
the correct $\mathcal{O}(\alphaS)$ distribution of emissions in the
radiative phase space.  The total cross section is obtained to NLO accuracy
by adding virtual corrections and integrated MC counterterms to the Born
cross section and combining them into a common seed for the parton shower. 
The matching procedure effectively restricts the role of the parton shower
to QCD emissions beyond NLO.

This method needs to be modified in processes with more than three coloured
particles at Born level, because of non-factorisable soft-gluon insertions
at real-emission level.  Spin correlations further complicate the picture. 
This problem is solved by using a variant of the original \MCatNLO
technique~\cite{Hoeche:2011fd,Hoeche:2012ft}.  Like \Sherpa's parton shower
itself, this method is based on the dipole-subtraction formalism by Catani
and Seymour~\cite{Catani:1996vz}, and it is implemented in \Sherpa in a
fully automated way.  It supplements the parton shower with spin and colour
correlations for the first emission and therefore extends it systematically
beyond the large-$N_{\mathrm{c}}$ approximation.

We combine \MCatNLO calculations of varying jet multiplicity into inclusive
event samples using the \MEPSatNLO
method~\cite{Gehrmann:2012yg,Hoeche:2012yf}.  This technique is based on
partitioning the phase space associated to QCD emissions into a soft and a
hard regime.  The soft region is filled by the parton shower alone, while
the hard region is described in terms of fixed-order calculations, to which
the parton shower has been matched.  In case of the \MCatNLO simulation with
the highest jet multiplicity, $N_{\mathrm{max}}$, the parton shower is
allowed to fill the entire phase space.  The phase-space separation is
achieved in terms of a kinematical variable analogous to the jet criterion
in longitudinally-invariant $k_\perp$-clustering
algorithms~\cite{Hoeche:2009rj}.  We will denote the separation cut by
$Q_{\rm cut}$.  It should be chosen smaller than the minimum jet transverse
momentum.  In this manner, the prediction for inter-jet correlations
involving up to $N_{\mathrm{max}}$ jets is always NLO accurate, and
augmented by resummation as implemented in the parton shower.

The choice of the renormalisation scale in the \MEPSatNLO approach is based
on the CKKW technique, 
a  
multi-jet merging
algorithm for tree-level matrix elements~\cite{Catani:2001cc}.  
Each 
shower emission is associated with a factor
$\alphaS(b\,k_\perp^2)$, where the scale is dictated by the resummation.
The smooth 
transition between parton-shower and matrix-element regimes at the
merging scale $\qcut$ requires a similar scale choice also in matrix
elements.  To this end, 
multi-jet events are clustered 
into a 
$2\to 2$ core process.
The clustering 
algorithm is defined as an exact inversion of the
parton shower, 
such that clusterings are determined according to the
parton-shower
 branching probabilities~\cite{Hoeche:2009rj}.  The coupling
factors resulting from the various QCD emissions are then evaluated at
scales $\mu^2=b\,k^2_\perp$, where $k_\perp$ is the nodal scale of the
corresponding branching, while the $\alphaS^K(\mu^2)$ term associated with
the core process is taken at the usual scale $\mu=\mu_\rR$.  This latter can
be chosen freely as in fixed-order calculations.

In practice, in the \MEPSatNLO algorithm all $\alphaS$ terms are first
evaluated at the scale $\mu_\rR$, and the CKKW prescription is implemented
via weight-correction factors,
\begin{equation}
\label{eq:as_mepsatnlo}
\frac{\alpha_s(b\, k_\perp^2)}{\alphaS(\mu_\rR^2)}\approx 1-
\frac{\alphaS(\mu_\rR^2)}{2\pi}b_0
\ln
\left(\frac{b\,k_\perp^2}{\mu_\rR^2}\right),
\end{equation}
for each branching. More precisely, in LO and NLO matrix elements the left-
and right-hand sides of \refeq{eq:as_mepsatnlo} are used, respectively.  For
the hard remainder function in the \MCatNLO calculations contributing to the
\MEPSatNLO result the renormalisation scale is always evaluated according to
the most likely underlying Born configuration, classified according to the
branching probability in the parton shower.

The fact that the CKKW scale choice adapts to the jet kinematics can improve
the description of jet emission also at high transverse momentum.  In this
region, where jet emission is typically associated to CKKW coupling factors
$\alphaS(p^2_\rT)$, fixed-order calculations based on a global
renormalisation scale $\mu_\rR$ involve a relative factor
$\alphaS(\mu^2_\rR)/\alphaS(p^2_\rT)$, which can significantly overestimate
the jet rate if $\mu_\rR$ does not adapt to the jet transverse momentum and
$p_\rT\gg \mu_\rR$.  This factor tends to be compensated by NLO corrections,
but 
in \MCatNLO simulations with fixed jet multiplicity $N$ it 
remains uncompensated
for the $(N+1)$-th jet, whose description relies on real-emission LO
matrix elements.  Within \MEPSatNLO, if $N<N_{\mathrm{max}}$ such
real-emission matrix elements are confined at transverse momenta below the
merging scale and replaced by an \MCatNLO simulation with $N+1$ jets above
$\qcut$.  In this way NLO accuracy is ensured for the first
$N_{\mathrm{max}}$ jets, and the problem remains present only for the
subsequent jet.  A simple solution consists of including
$(N_{\mathrm{max}}+1)$-jet LO matrix elements in the merging procedure.  In
this way, also the $(N_{\mathrm{max}}+1)$-th jet receives a CKKW coupling
factor $\alphaS(p^2_\rT)$ above the merging scale.  As discussed in
\refse{se:matchedsamples}, for the \MEPSatNLO simulation of $\Pp\Pp\to
4\ell+0,1j$ we adopt a dynamical scale $\mu_\rR$ that depends only on the
W-boson transverse energy and does not adapt to extra jet emissions.  The
above discussion is thus relevant for the high-$p_\rT$ tail of the second
jet, where it's likely that $\mu_\rR \ll p_\rT$, since the two jets
typically recoil against each other and the transverse energy of the W bosons
tends to remain of the order of $\MW$.

In order to guarantee a complete treatment of scale uncertainties,
renormalisation-scale variations in the \MEPSatNLO merging approach are
performed simultaneously in the fixed-order calculation and in the parton
shower.  The same rescaling factors are applied to the CKKW scales and to
the scale $\mu_\rR$ used in the $\alphaS$ terms associated with the core
process.

\subsection{Merging of squared quark-loop contributions to four-lepton production}
\label{sec:merge_loop_induced}

We present here, for the first time, a combination of the squared quark-loop
contributions to $\Pp\Pp\to 4\ell+0,1j$ using the ME+PS merging technique
of~\citere{Hoeche:2009rj}.  At matrix-element level we consider all squared
one-loop amplitudes that involve a closed quark loop.  While squared
quark-loop corrections to $4\ell$ final states involve only $\Pg\Pg$ initial
states, $4\ell+1j$ production involves, in addition to $\Pg\Pg\to
4\ell+\Pg$, also $q\Pg\to 4\ell+q$, $\bar q\Pg\to 4\ell+\bar q$ and
$q\bar{q}\to 4\ell+\Pg$ contributions.  For these quark-initiated channels
we require that all final-state leptons are connected to the quark loop via
vector-boson exchange, \ie we exclude topologies where vector bosons couple
to the external quark line.  The inclusion of these quark-initiated channels
is mandatory for a consistent merging of the $4\ell+0,1j$ samples.  This is
due to the fact that gluon- and quark-initiated channels are intimately
connected via $q\to q \Pg$ and $\Pg\to q\bar q$ parton-shower splittings. 
Including the $q\Pg$ and $\bar q\Pg$ channels ensures that all splitting
functions used in the shower are replaced by matrix elements in the hard-jet
region.  The finite contribution from the $q\bar{q}\to 4\ell+\Pg$ channel is
added for consistency.  While the $\Pg\Pg$-induced channels have already
been discussed in the
literature~\cite{Binoth:2005ua,Binoth:2006mf,Campbell:2011cu,Melia:2012zg,Agrawal:2012df},
the squared quark-loop contributions to the $q\Pg$-, $\bar q\Pg$- and $q\bar
q$-channels are investigated for the first time in this paper.

To merge the $4\ell+0,1j$ final states we can use the tree-level techniques
of~\citere{Hoeche:2009rj} since all involved matrix elements are infrared
and ultraviolet finite.  In particular, the merging scale $\qcut$ acts as an
infrared cutoff that avoids soft and collinear divergences of $4\ell+1j$
matrix elements, and the phase-space region below $\qcut$ is filled by
$\Pg\Pg\to 4\ell$ matrix elements plus shower emissions.  As discussed in
\refse{se:matchedsamples}, while squared quark-loop corrections represent
NNLO contributions to inclusive $4\ell+0,1j$ production, their intrinsic
accuracy 
is only leading order.  Consequently, as we
will see in \refses{Sec:WWinc}--\ref{Sec:HWW}, squared quark-loop terms are
more sensitive to renormalisation- and resummation-scale variations as
compared to \MEPSatNLO predictions.

\section{Monte-Carlo simulations}
\label{se:mcsample}

In the following we 
discuss
 input parameters and theoretical ingredients 
of the Monte-Carlo simulations presented in \refses{Sec:WWinc} and \ref{Sec:HWW}.

\subsection{Input parameters and process definition}

The presented results refer to $\Pp\Pp\to\mnen+X$ at a centre-of-mass energy of
$8\UTeV$. Predictions at NLO and squared quark-loop corrections are 
evaluated using the five-flavour CT10 NLO parton distributions~\cite{Lai:2010vv} 
with the respective running strong coupling $\alphaS$.
At LO we employ the CT09MCS PDF set.
For the vector-boson masses we use
\beq
\MW = 80.399 \UGeV,\qquad
\MZ = 91.1876\UGeV,
\eeq
and in order to guarantee NLO accurate $\PW\to\ell\nu$ branching fractions
we use NLO input widths
\beq
\Gamma_{\PW} = 2.0997\UGeV,\qquad
\Gamma_{\PZ} = 2.5097\UGeV.
\eeq
The electroweak mixing angle is obtained 
from 
the ratio of the complex W- and
Z-boson masses
as~\cite{Denner:2005fg} 
\beq
\cos^2{\theta_{\mathrm{w}}}=\frac{\MW^2-\ri\Gamma_{\PW}\MW}{\MZ^2-\ri\Gamma_{\PZ}\MZ},
\eeq
and the electromagnetic fine-structure constant
is derived from the Fermi constant $G_\mu= 1.16637\cdot 10^{-5}\UGeV^{-2}$ in
the so-called $G_\mu$-scheme, which results in
\beq
\alpha^{-1} = \frac{\pi}{\sqrt{2}\, G_\mu \MW^2}\left(1-\frac{\MW^2}{\MZ^2}\right)^{-1}=
132.348905\,.
\eeq
Since quark-mixing effects cancel almost completely~\cite{Dittmaier:2009un}, 
we set the CKM matrix equal to one.

Partonic channels with initial- and final-state b quarks are not included in
order to avoid any overlap with $\Pt\bar \Pt$ and $\Pt\PW$ production.  At
NLO this separation is nontrivial since $\PWp\PWm+1j$ production receives
$\Pp\Pp\to\PWp\PWm\Pb\bar\Pb$ real-emission contributions that involve
top-quark resonances.  At the same time, $\PWp\PWm\Pb\bar\Pb$ final states
are intimately connected to the virtual corrections to $q\bar q\to
\PWp\PWm\Pg$ via cancellations of collinear singularities that arise from
$\Pg\to\Pb\bar\Pb$ splittings~\cite{Dittmaier:2009un}.  This is discussed in
detail in Appendix~\ref{app:b-treatment}, where we introduce a prescription
to separate $\PWp\PWm+$jets from single-top and top-pair production processes
in such a way that each contribution is infrared finite and free from large
logarithms associated to $\Pg\to\Pb\bar\Pb$ splittings.  This prescription
is not unique, and we estimate the related ambiguity to be of order $1\%$. 
It can be eliminated by a consistent matching of $\PWp\PWm+$jets and
$\PWp\PWm\Pb\bar\Pb$ production as explained in
Appendix~\ref{app:b-treatment}.

\subsection{Fixed-order ingredients of the calculation}
\label{se:foing}

\begin{figure}
  \begin{center}
    \includegraphics{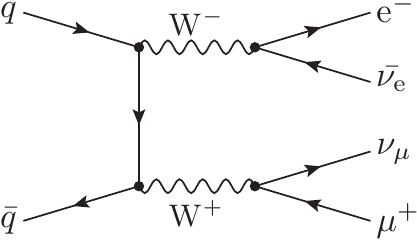}\qquad\includegraphics{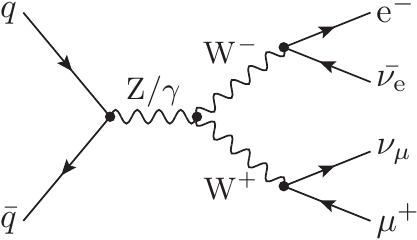}\qquad\includegraphics{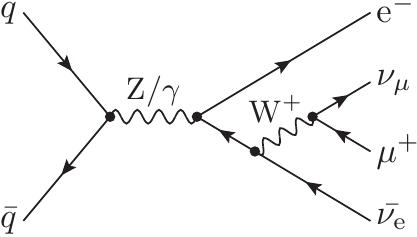}\\[5ex]
    \includegraphics{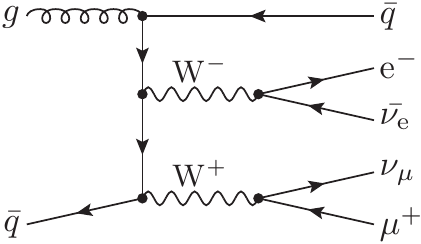}\qquad\includegraphics{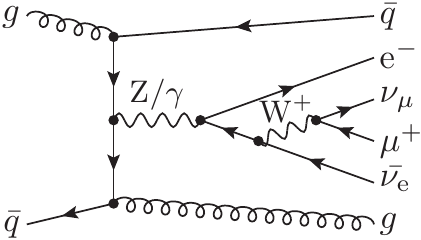}\qquad\includegraphics{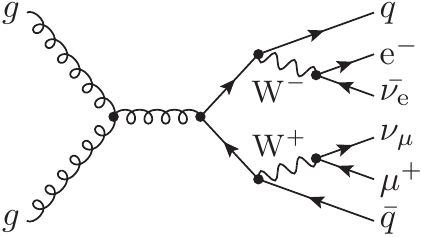}\\[5ex]
    \includegraphics{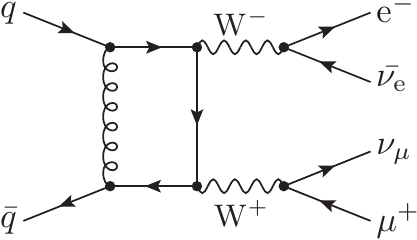}\qquad\includegraphics{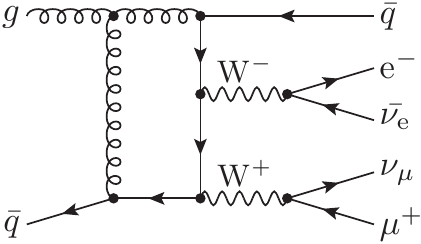}\qquad\includegraphics{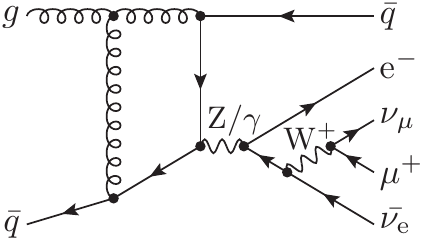}
  \end{center}
  \caption{Sample Feynman diagrams contributing to 
$\Pp\Pp\to \mnen$ and $\Pp\Pp\to \mnen+1j$ at NLO.
}
\label{fig:nlodiags}
\end{figure}

\begin{figure}
  \begin{center}
\includegraphics{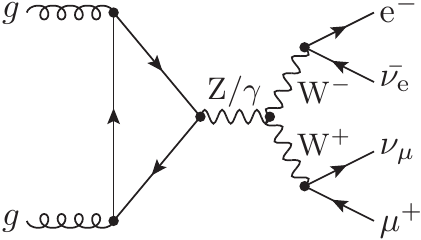}
\qquad\includegraphics{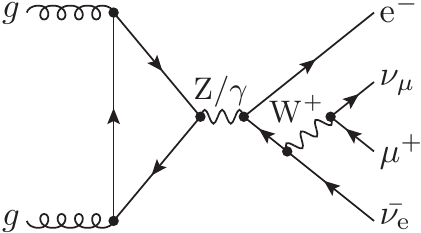}
\qquad\includegraphics{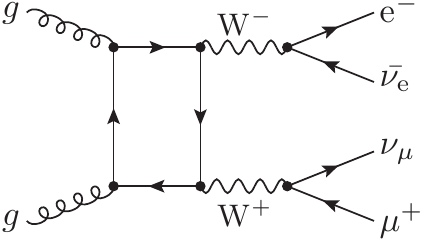}
\\[5ex]
\includegraphics{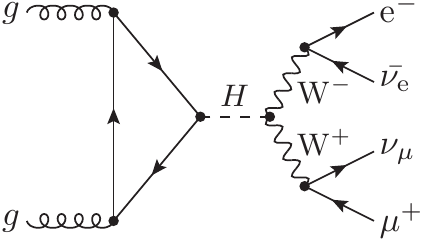}
\qquad\includegraphics{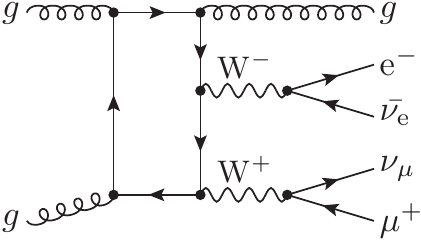}
\qquad\includegraphics{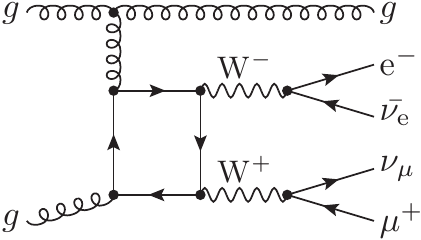}
\\[5ex]
\includegraphics{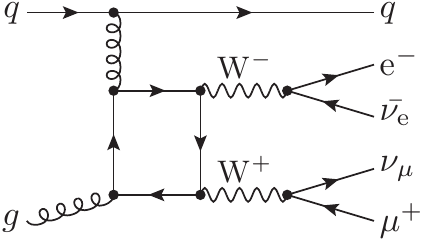}
\qquad\includegraphics{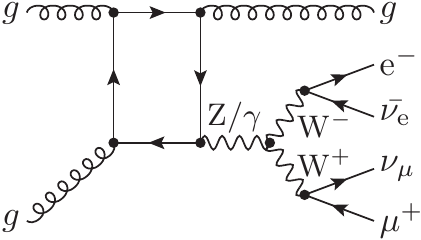}
\qquad\includegraphics{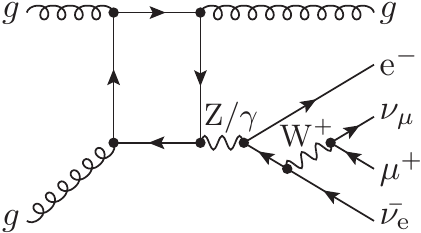}
\end{center}
\caption{Sample Feynman diagrams involved in the
squared quark-loop NNLO contributions to $\Pp\Pp\to \mnen$ and $\Pp\Pp\to\mnen+1j$.
}
\label{fig:loop2diags}
\end{figure}

Sample Feynman diagrams contributing to the fixed-order building blocks of
the calculation are shown in \reffis{fig:nlodiags} and \ref{fig:loop2diags}. 
For brevity $\mnen$ configurations are often denoted as 
$\lnln$ or $4\ell$ final states in the following.
The first figure illustrates NLO QCD corrections to $\Pp\Pp\to 4\ell$ and
$\Pp\Pp\to 4\ell+1j$, which involve various $\bar q q$, $q\Pg$, $\bar q\Pg$ and
$\Pg\Pg$ partonic channels. The complete set of Feynman diagrams and related interferences 
is taken into account, including single-resonant $\PZ/\gamma^*\to \Pem \bar
\nu_{\Pe}\PWp(\to\mu^+\nu_\mu)$ sub-topologies.  Pentagons represent the most
involved one-loop topologies.

In addition to NLO corrections also squared quark-loop contributions to the
partonic channels $\Pg\Pg\to 4\ell$, $\Pg\Pg\to 4\ell+\Pg$, $\Pg q\to 4\ell+q$,
$\Pg\bar q\to 4\ell+\bar q$, and $q\bar q\to 4\ell+\Pg$ are computed. 
Corresponding Feynman diagrams are shown in~\reffi{fig:loop2diags}.  
The most involved diagrams are again pentagons.  As discussed in
\refse{sec:merge_loop_induced}, the inclusion of the quark-induced channels
is mandatory for a correct description of the full spectrum of jet emission
based on the merging of $4\ell$ and $4\ell+j$ simulations.
Contributions where the leptons are coupled to 
quark triangles via $Z/\gamma^*$ exchange, like
in the first two diagrams of \reffi{fig:loop2diags},
vanish due to electroweak Ward identities~\cite{Campbell:2011bn}.
In contrast, related topologies with an extra gluon in the final state,
like the last two diagrams in \reffi{fig:loop2diags},
yield non-vanishing contributions.
The various NLO and
squared quark-loop amplitudes generated for the present study comprise
all relevant Higgs-boson  contributions, including the interference of the 
Higgs signal with the four-lepton continuum. 
However, for the background
predictions presented in \refses{Sec:WWinc}--\ref{Sec:HWW} all Higgs-boson contributions
have been decoupled by setting $\MH\to\infty$.

A series of checks has been performed to validate all ingredients of the 
QCD corrections.  To check the correctness of the $q\bar q\to 4\ell+0,1 \Pg$
\OpenLoops matrix elements we used an independent computer-algebra
generator, originally developed for the calculations
of~\citeres{Bredenstein:2009aj,Denner:2010jp}.  The squared quark-loop
$\Pg\Pg\to4\ell+0,1\Pg$ amplitudes have been checked against
MCFM~\cite{Campbell:2010ff} and \citere{Melia:2012zg}.  The NLO and
squared quark-loop integrated cross sections for $\Pp\Pp\to4\ell+0,1j$ and
$\Pg\Pg\to4\ell+0,1\Pg$ have been found to agree with various results in the
literature~\cite{Binoth:2006mf,Melia:2011tj,Melia:2012zg}.  Finally, the NLO
cross sections for hadronic $4\ell+0,1j$ production have been reproduced with
sub-permil statistical precision using an independent Monte-Carlo
generator, which was developed by S.~Kallweit in the framework of the
$\Pp\Pp\to\PWp\PWm\Pb\bar\Pb$ calculation of~\citere{Denner:2010jp}.

The calculation of tree-level matrix elements is performed either 
by the \Amegic~\cite{Krauss:2001iv} or the \Comix~\cite{Gleisberg:2008fv} 
matrix-element generator, where \Comix is used only for
$\Pp\Pp\to 4\ell+2j$ subprocesses. Integrated and real 
subtraction terms are computed with the method of Catani and 
Seymour~\cite{Catani:1996vz}, using the automated implementation 
in \Amegic~\cite{Gleisberg:2007md}.

\subsection[Matching to the parton shower, multi-jet NLO merging, and scale variations]{Matching to the parton shower, multi-jet NLO merging,\\and scale variations}
\label{se:matchedsamples}

The perturbative content of the various fixed-order, matched and merged
simulations that are presented in \refses{Sec:WWinc} and \ref{Sec:HWW} is
illustrated in~\refta{Tab:accuracy}.  Parton-level NLO predictions for
$\Pp\Pp\to4\ell+0,1j$ are denoted as \NLOfl and \NLOflj.  Their NLO
predictive power is restricted to the 0- and 1-jet bins,
respectively.\footnote{In this discussion of the perturbative accuracy we
refer to jet bins in the inclusive sense.  The 0-, 1- and 2-jet bins should
namely be understood as final states with $\ge 0$, $\ge 1$ and $\ge 2$ jets,
or equivalently as observables that explicitly or implicitly involve a
corresponding number of jets.} In bins with one extra jet 
with respect to 
the simulated process, the precision
decreases to LO, and higher-multiplicity bins are not populated at all.

\begin{table}
  \begin{center}
    \begin{tabular}{|c||c|c|c|}
\hline
  \textbf{\NLOacc simulations}
& \textbf{0-jet}
& \textbf{1-jet}
& \textbf{2-jet}
\\[1mm]\hline\hline
   \NLOfl
& NLO 
& LO
& -
\\[1mm]\hline
  \NLOflj
& - 
& NLO
& LO
\\[1mm]\hline
  \MCatNLOfl
& NLO+PS
& LO+PS
& PS
\\[1mm]\hline
  \MCatNLOflj     
& -
& NLO+PS
& LO+PS
\\[1mm]\hline
  \MEPSatNLOflj
& NLO+PS
& NLO+PS
& LO+PS
\\[1mm]\hline\hline
  \textbf{\LOOPSQ simulations} 
& \textbf{0-jet}
& \textbf{1-jet}
& \textbf{2-jet}
\\[1mm]\hline\hline
\LOOPSQfl
& LO
& -
& -
\\[1mm]\hline
\LOOPSQflj
& -
& LO
& -
\\[1mm]\hline
\LOOPSQPSfl
& LO+PS
& PS
& PS
\\[1mm]\hline
\LOOPSQPSflj
& -
& LO+PS
& PS
\\[1mm]\hline
\MEPSatLOOPSQflj
& LO+PS                           
& LO+PS     
& PS   
\\[1mm]\hline
\end{tabular}
    \caption{
Perturbative accuracy of various fixed-order, matched and merged simulations
for final states with 0, 1 and 2 jets.}
\label{Tab:accuracy}
\end{center}
\end{table}

This is overcome by matching \NLOfl or \NLOflj matrix
elements to the parton shower.  Corresponding predictions are denoted as
\MCatNLOfl and \MCatNLOflj.  The shower radiates an arbitrary
number of extra jets, which effectively resums large Sudakov logarithms that
arise when QCD radiation is constrained by tight cuts, such as in presence
of jet vetoes.  Similarly as the underlying NLO matrix elements, \MCatNLO
predictions provide NLO precision only for one particular jet multiplicity. 
In the following sections we will consider only \MCatNLOfl
predictions.  This corresponds to the usual inclusive NLO+PS samples used in
experimental studies, where observables involving one jet are only LO
accurate, and the emission of additional jets is entirely based on the
parton-shower approximation.

Our best 
NLO
predictions are denoted as \MEPSatNLOflj and result from
merging \MCatNLOfl and \MCatNLOflj samples. This provides
shower-improved NLO precision in the first two jet bins. 
To ensure that the formal NLO accuracy is preserved in the 0- and 1-jet
bins, the merging scale $\qcut$ should not exceed the $p_\rT$-threshold used for jet binning.
On the other hand, in the limit of small $\qcut$ the fact that 
higher-logarithmic terms in the fixed-order \NLOflj calculation are not resummed
in the Sudakov form factor gives rise to a logarithmic sensitivity to the 
merging scale.
Such logarithms are beyond the shower accuracy but can be numerically 
non-negligible~\cite{Lonnblad:2012ng,Lonnblad:2012ix}. 
Thus the merging scale should not be set too far below the jet-$p_\rT$ threshold. 
Following this reasoning the value
$\qcut=20\UGeV$ has been used as merging scale, and the stability
of the results with respect to this technical parameter has been tested using variations in the range 
$15\UGeV\le\qcut\le 35\UGeV$. The corresponding uncertainties are discussed 
in \refse{Sec:HWW} for the case of the $\hww$ analysis, where they 
turn out to be at the percent level.
The \MEPSatNLO
$4\ell+0,1j$ sample is further improved by including LO matrix elements with
two jets in the merging procedure.  As explained in
\refse{sec:match_n_merge}, this guarantees a better (CKKW-type) scale choice 
for the $\alphaS$ factor associated with the second jet emission.

In order to gain insights into the importance of parton-shower and merging
effects, we will present systematic comparisons of NLO, \MCatNLO and
\MEPSatNLO predictions.  While Sudakov resummation effects due to the parton
shower show up in the difference between \NLOfl and \MCatNLOfl, comparing
\MCatNLOfl to \MEPSatNLOflj allows one to assess NLO corrections to the first
emission.

As already mentioned, squared quark-loop terms included in our simulation represent
NNLO contributions to $\Pp\Pp\to 4\ell+(0)1j$. On the other
hand, since NNLO is the first order at which the $\Pg\Pg\to 4\ell+0(1)\Pg$ 
channels start contributing to $4\ell+(0)1j$ production, these corrections can also be
regarded as LO contributions. 
As 
indicated in \refta{Tab:accuracy}, squared quark-loop terms
behave as LO predictions 
also
for what concerns the number of external QCD partons.  
In fact, fixed-order squared quark-loop predictions, which we denote as \LOOPSQfl
and \LOOPSQflj, populate only a single jet bin.  In
particular, \LOOPSQfl predictions completely
miss exclusive jet emission and suppression effects resulting from jet vetoes.
A first realistic estimate of jet-veto effects is obtained by showering
squared quark-loop contributions.  The corresponding predictions are labelled
as \LOOPSQPSfl and \LOOPSQPSflj, depending on the jet
multiplicity of the underlying matrix elements. 
Merging the \LOOPSQPS simulations with 0 and 1 jets results in
a single \MEPSatLOOPSQflj sample, which provides a reliable description
of the full spectrum of jet emission, from soft to hard regions.
This merged squared quark-loop simulation comprises also partonic channels 
with initial-state quarks.
To assess their relative importance, in \refse{Sec:WWinc}, 
full \MEPSatLOOPSQflj predictions are compared to corresponding predictions
involving only initial-state gluons.

As a default renormalisation ($\mu_\rR$), factorisation ($\mu_\rF$) and
resummation ($\mu_Q$) scale we adopt the average W-boson 
transverse energy
\beq
\mu_0 = \frac{1}{2}\left(E_{\rT,\PWp}+E_{\rT,\PWm}\right),
\label{eq:scale}
\eeq
where $E^2_{\rT,\PW}=\MW^2+(\vec{p}_{\rT,\ell}+\vec{p}_{\rT,\nu})^2$.
As discussed in \refse{Sec:WWinc}, motivated by the comparison of hard-jet
emission from parton shower and matrix elements, in the case of squared quark-loop
contributions we decided to reduce the resummation scale by a factor two,
\ie we set $\mu_Q=\mu_0/2$.

Renormalisation- and factorisation-scale uncertainties are assessed by
applying independent variations $\mu_\rR=\xi_\rR\mu_0$ and
$\mu_\rF=\xi_\rF\mu_0$, with factor-two rescalings $(\xi_\rR,\xi_\rF)$ = $(2,2)$,
$(2,1)$, $(1,2)$, $(1,1)$, $(1,0.5)$, $(0.5,1)$, $(0.5,0.5)$.  The
renormalisation scale is varied in all $\alphaS$ terms that arise in matrix
elements or from the shower.  
In \NLOacc and \MCatNLO predictions all $\alphaS$ terms arising from matrix elements
are evaluated at
$\mu_\rR=\xi_\rR\mu_0$, while in \MEPSatNLO the scale $\mu_0$ is used only in
tree and loop contributions to the $\Pp\Pp\to 4\ell$ core process, which
results from $4\ell+$jets configurations via clustering of all hard jets. 
For the $\alphaS$ factors associated with jet emissions a CKKW scale choice
is applied, as discussed in \refse{sec:match_n_merge}.  
As a consequence, \MEPSatNLO predictions are less sensitive to the choice of the
central scale $\mu_0$.
Also in \MEPSatLOOPSQ merging the scale of $\alphaS$ factors associated to QCD emissions
is dictated by the CKKW prescription. In this case the core process involves 
a term $\alphaS^2(\mu_\rR)$, which renders squared quark-loop corrections 
more sensitive to the choice of the central scale $\mu_0$.

In addition to usual QCD-scale studies, the \Sherpa framework allows also
for automated variations of the resummation scale $\mu_Q$, which corresponds
to the starting scale of the parton shower.  This scale is varied by factors
$\mu_Q/\mu_0=1/\sqrt{2},1,\sqrt{2}$, while keeping $\mu_{\rR}$ and
$\mu_{\rF}$ fixed.  As discussed in \refse{sec:match_n_merge}, this reflects 
uncertainties related to subleading
logarithms beyond the shower approximation and yields more realistic error
estimates for exclusive observables such as jet-vetoed cross sections.
In order to quantify the total scale uncertainty we will regard
$(\mu_\rR,\mu_\rF)$ and $\mu_Q$ variations as uncorrelated and add them in 
quadrature.\footnote{ 
Another natural way of combining these two sources of uncertainty is to
consider simultaneous variations of $(\mu_\rR,\mu_\rF,\mu_Q)$, excluding
rescalings in opposite directions as usual.  The variations resulting from
this alternative approach are likely to be even smaller than those obtained
by adding QCD- and resummation-scale uncertainties in quadrature.}
Uncertainties related to the
PDFs, $\alphaS(\MZ)$, hadronisation, and underlying event are not
considered in this study.

The presented results were obtained with a \Sherpa2.0 pre-release
version\footnote{This pre-release version corresponds to SVN revision 21825
and the main difference with respect to the final \Sherpa2.0
release version is the tuning of parton shower, hadronisation and multiple
parton interactions to experimental data.}. 
First partial results of this simulation have been presented
in~\citere{Heinemeyer:2013tqa}.  In addition to the squared quark-loop
contributions, which were not included in~\citere{Heinemeyer:2013tqa}, in
this paper we investigate various new observables.  Due to the difference
between \refeq{eq:scale} and the scale choice $\mu_0=M_{\lnln}$
in~\citere{Heinemeyer:2013tqa}, results presented here 
should not be directly compared to those of~\citere{Heinemeyer:2013tqa}.

\section{Analysis of inclusive \texorpdfstring{$\lnln+0,1$}{lvlv+0,1} jets production}
\label{Sec:WWinc}

As a first application of our simulation we study $\mnen$ and $\mnen+1$ jet
production without any Higgs-analysis specific cuts.  To this end we adopt
the cuts of the \verb!MC_WWJETS!  truth analysis provided with the Rivet
Monte-Carlo validation framework~\cite{Buckley:2010ar}.  Specifically, we
require charged leptons with $p_{\perp,\ell}>25\UGeV$ and $|\eta_\ell|<3.5$. 
Missing transverse energy is identified with the vector sum of the neutrino
transverse momenta and required to fulfil $E\!\!\!/_T>25\UGeV$.  Jets are
defined using the anti-$k_\perp$ algorithm~\cite{Cacciari:2008gp} with a
distance parameter of $R=0.4$.  No jet-rapidity cuts are applied.

To illustrate the importance of the various corrections and the respective
scale uncertainties, we present cross sections and distributions at the
different levels of simulation introduced in \refse{se:matchedsamples}.  In
\refse{se:NLOWW} we compare fixed-order predictions to 
matched and merged NLO simulations. Squared quark-loop corrections are
discussed in \refse{se:sqlWW}.

\subsection{Fixed-order, matched and merged NLO simulations}
\label{se:NLOWW}

Rates for the inclusive analysis and when requiring (at least) one jet with
$p_\perp>30\UGeV$ are shown in \refta{Tab:InclusiveXS1}. Fixed-order LO and
NLO predictions for $\Pp\Pp\to 4\ell$ or $4\ell+1j$, depending on the
jet bin, are compared to the inclusive \MCatNLOfl simulation and to the NLO
merged simulation of $4\ell+0,1j$.  
For 0- and 1-jet production we observe
positive NLO corrections of 50\% and 38\%, respectively,
consistent with the typical size of $K$-factors in the literature.
At NLO, scale uncertainties range from 3 to 5 percent, which is twice as
large as compared to our previous Higgs-background predictions
in exclusive jet bins~\cite{Heinemeyer:2013tqa}.  This can be attributed 
to the new scale choice \refeq{eq:scale} and to the
fact that results in \refta{Tab:InclusiveXS1} correspond
to inclusive jet bins. In fact, as shown in~\citere{Dittmaier:2009un}, the choice
of the central scale and a jet veto can have a strong impact on scale
uncertainties in $4\ell+1j$ production~\cite{Dittmaier:2009un}.  In this
respect, we note that the central scale used
in~\citere{Heinemeyer:2013tqa}, \ie the total four-lepton invariant mass, is
more than a factor two higher than the transverse-energy scale
\refeq{eq:scale} adopted for the present study.

\begin{table}
  \begin{center}
    \begin{tabular}{|@{\,}c@{\,}||@{\,}c@{\,}|@{\,}c@{\,}|@{\,}c@{\,}|@{\,}c@{\,}|@{\,}c@{\,}|}
\hline
  Analysis
& \textbf{\LOfljs}
& \textbf{\NLOfljs}
& \textbf{\MCatNLOfl}
& \textbf{\MEPSatNLOflj}
\\[1mm]\hline
   $\geq 0$ jets
& $217.99(2)\;^{+1.9\%}_{-2.8\%}$
& $328.08\;^{+3.1\%}_{-2.4\%}$
& $326.70(29)\;^{+4.5\%}_{-2.8\%}$$\;^{+0.0\%}_{-0.2\%}$
& $356.01(58)\;^{+1.3\%}_{-0.8\%}$$\;^{+1.8\%}_{-0.0\%}$
\\[1mm]\hline
   $\geq 1$ jets
& $73.61(1)\;^{+14.5\%}_{-11.6\%}$
& $101.70\;^{+5.2\%}_{-4.9\%}$
& $83.23(15)\;^{+9.9\%}_{-9.0\%}$$\;^{+2.4\%}_{-4.6\%}$
& $103.45(28)\;^{+2.8\%}_{-3.7\%}$$\;^{+3.3\%}_{-0.5\%}$
\\[1mm]\hline
\end{tabular}
\caption{Cross-section predictions in femtobarns for the 
$\mnen$ analyses requiring $\ge 0$ and $\ge 1$ jets.
Fixed-order LO and NLO results 
for the $\ge 0$-jet and $\ge 1$-jet analyses 
correspond to $4\ell$ and $4\ell+1j$ production, respectively.
They are compared to an inclusive
\MCatNLOfl simulation and to \MEPSatNLOflj predictions.
Uncertainties associated to variations of the QCD scales ($\mu_\rR,
\mu_\rF$) and the resummation scale ($\mu_Q$) are shown separately as
$\sigma\pm\delta_\mathrm{QCD}\pm\delta_\mathrm{res}$.  Statistical errors
are given in parenthesis.}
  \label{Tab:InclusiveXS1}
  \end{center}
\end{table}

Comparing the \MCatNLO and \NLOacc simulations we observe one-percent
level agreement and rather similar uncertainties in the inclusive analysis.  
This agreement, as well as the tiny resummation-scale uncertainties of
\MCatNLO, reflect the unitarity of the parton shower for inclusive
observables.  In contrast, in the 1-jet bin \MCatNLO predictions exhibit a
deficit of about 20\% and much larger uncertainties as compared to \NLOacc. 
This is due to the fact that the inclusive matched calculation is only LO
accurate in the 1-jet bin.

The inclusive \MEPSatNLO cross section is found to be roughly $30\Ufb$
larger as compared to the \NLOacc calculation, which can be interpreted as a
result of NLO corrections to the first emission in the merged sample. 
In fact, the shift of $30\Ufb$ is comparable to the difference between the
\NLOacc and \MCatNLO cross sections with $\ge 1$ jets, which corresponds to
NLO effects in the 1-jet bin.  Finally, variations of the QCD and
resummation scales in \MEPSatNLO amount to only 1--3\% in both jet bins.  As
already mentioned, the fact that 
fixed-order NLO cross sections feature significantly larger scale variations 
is related to the choice of the central scale $\mu_0$.  This scale
plays only a marginal role in \MEPSatNLO, since the $\Pp\Pp\to4\ell$ core
process does not depend on the strong coupling, and $\alphaS$ terms
resulting from jet emissions are controlled by the CKKW prescription.

\begin{figure}
\begin{center}
\includegraphics[width=0.48\textwidth]{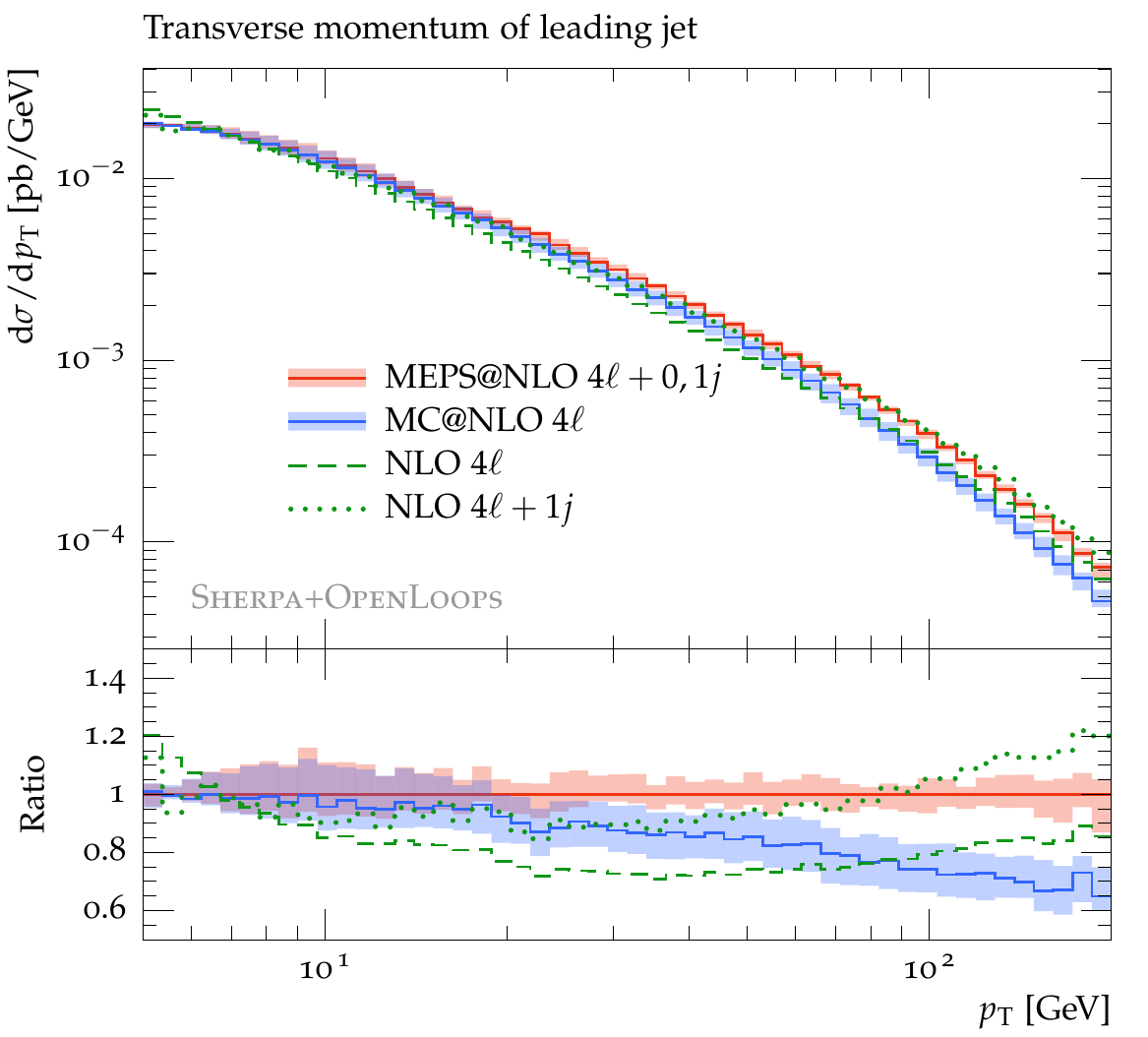}\nolinebreak
\includegraphics[width=0.48\textwidth]{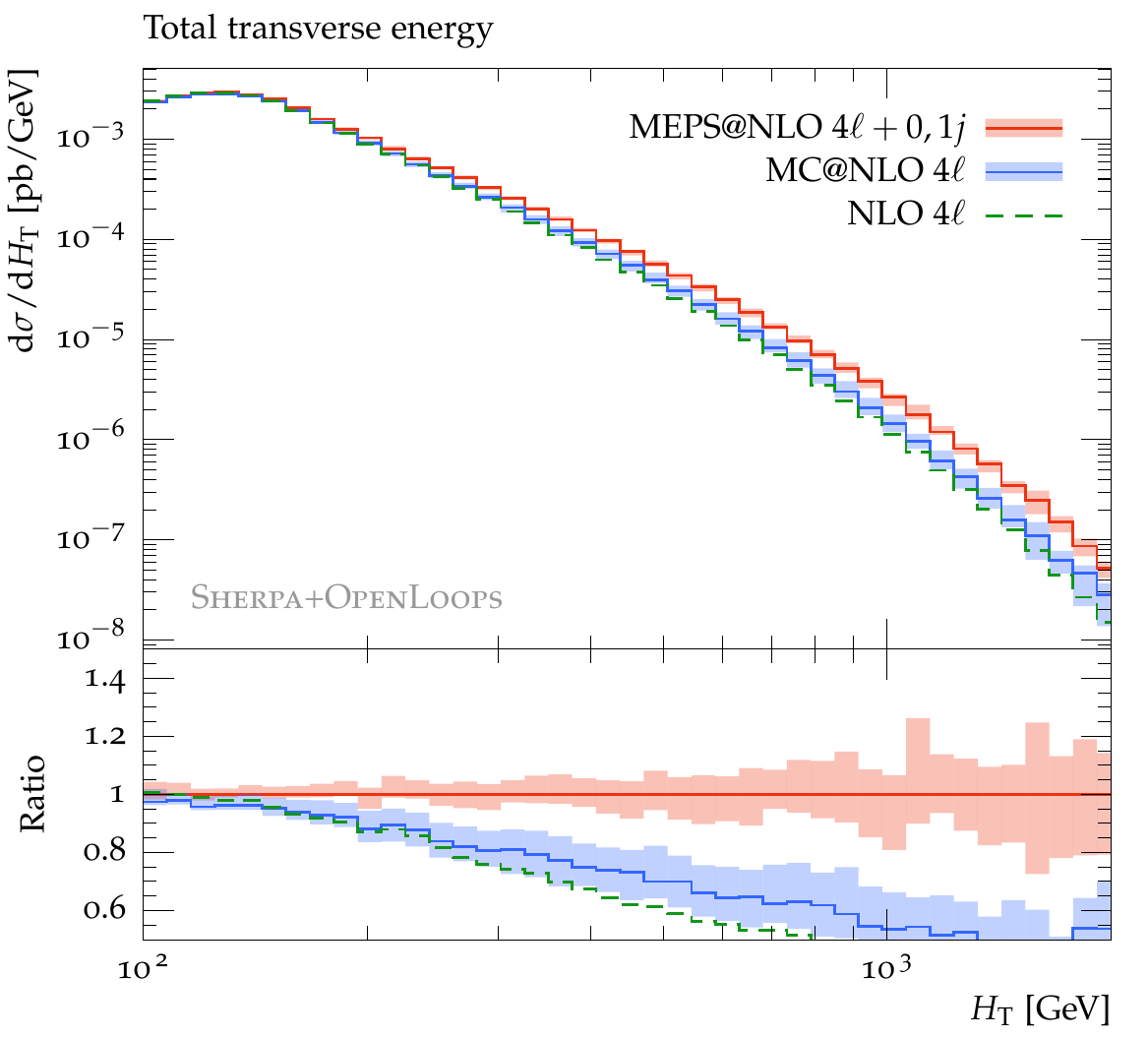}
\end{center}
\vspace*{-0.3cm}
\caption{Leading-jet transverse momentum (left) and total transverse energy (right):
\NLOfl (green dashed) and \NLOflj (green dotted) results are compared to
an inclusive \MCatNLOfl simulation (blue) and to
\MEPSatNLOflj predictions (red).
Uncertainty bands describe combined QCD- and resummation-scale
uncertainties (added in quadrature). 
}
\label{fig:inclusive_merging} 
\end{figure}

Distributions in the hardest-jet transverse momentum and in the
total transverse energy $H_\rT$---defined as the scalar sum of the
transverse momenta of leptons, missing $E_\rT$, and all reconstructed
jets---are displayed in \reffi{fig:inclusive_merging}.
The bands are obtained by adding QCD- and resummation-scale variations in quadrature.
The \MCatNLO and \MEPSatNLO $p_\rT$-distributions agree fairly well in the
soft region, but \MCatNLO develops an increasingly large deficit at higher
$p_\rT$, which reaches 30\% in the tail.   Similarly as \MCatNLO, also \NLOacc predictions 
for inclusive four-lepton production are only LO accurate in the first-jet emission and
tend to underestimate the tail.  The shapes of the
\MCatNLO and \NLOacc tails are however somewhat different.
This is due to the fact that, in the MC@NLO method, the 
weights of the first shower emission and of its MC-subtraction counterpart 
differ by an $\mathcal{O}(\alphaS)$ relative factor, which involves the
$\alphaS(p_\rT)/\alphaS(\mu_\rR)$ ratio as well as unresolved NLO
corrections.
This difference disappears above the 
resummation scale, i.e.~where the parton shower stops emitting.
This is however not visible in the plot, since due to the 
dynamical nature of the resummation-scale choice \refeq{eq:scale},
this transition takes place only far above the scale $\MW$.
In the $p_\rT\to 0$ limit, the
\NLOfl calculation involves an infrared singularity of the form
$\rd\sigma/\rd p_\rT \sim \alphaS \ln(p_\rT)/p_\rT$, which manifests itself
as a linear rise if the distribution is plotted against $\ln(p_\rT)$ as in
\reffi{fig:inclusive_merging}.a.  This feature is qualitatively clearly visible
but quantitatively very mild, and the corresponding 
enhancement does not exceed 20\% down to $p_\rT=5\UGeV$.
This signifies that the effect of resumming Sudakov
logarithms is important but not dramatic in the considered $p_\rT$-range. 
Higher Sudakov logarithms are partially included in the \NLOacc calculation
of $4\ell+1j$ production, which remains infrared divergent at $p_\rT\to 0$,
but turns out to be in better agreement with \MCatNLO and \MEPSatNLO
predictions for $p_\rT>5\UGeV$.  The \NLOflj distribution has a higher tail
with respect to inclusive \NLOacc and \MCatNLO predictions, as expected, but
for $p_\rT\gtrsim \MW$ it starts to be above the \MEPSatNLO curve as well.
This can be explained by the fact that, in contrast to the \MEPSatNLO
approach, in fixed-order predictions the scale of $\alphaS$ couplings
associated with jet emission is not adapted to the jet-$p_\rT$
(\cf discussion in \refse{sec:merge_loop_induced}).

The total transverse energy, plotted in \reffi{fig:inclusive_merging}.b, is
dominated by hard multi-jet emissions that cannot be properly described
neither by the inclusive \NLOacc calculation nor by the \MCatNLO approach and
its parton-shower emissions. This starts to be visible at $H_\rT\sim 200\GeV$ and
the deficit with respect to \MEPSatNLO approaches 50\% at $1\UTeV$.
%

\begin{figure}
\begin{center}
\includegraphics[width=0.48\textwidth]{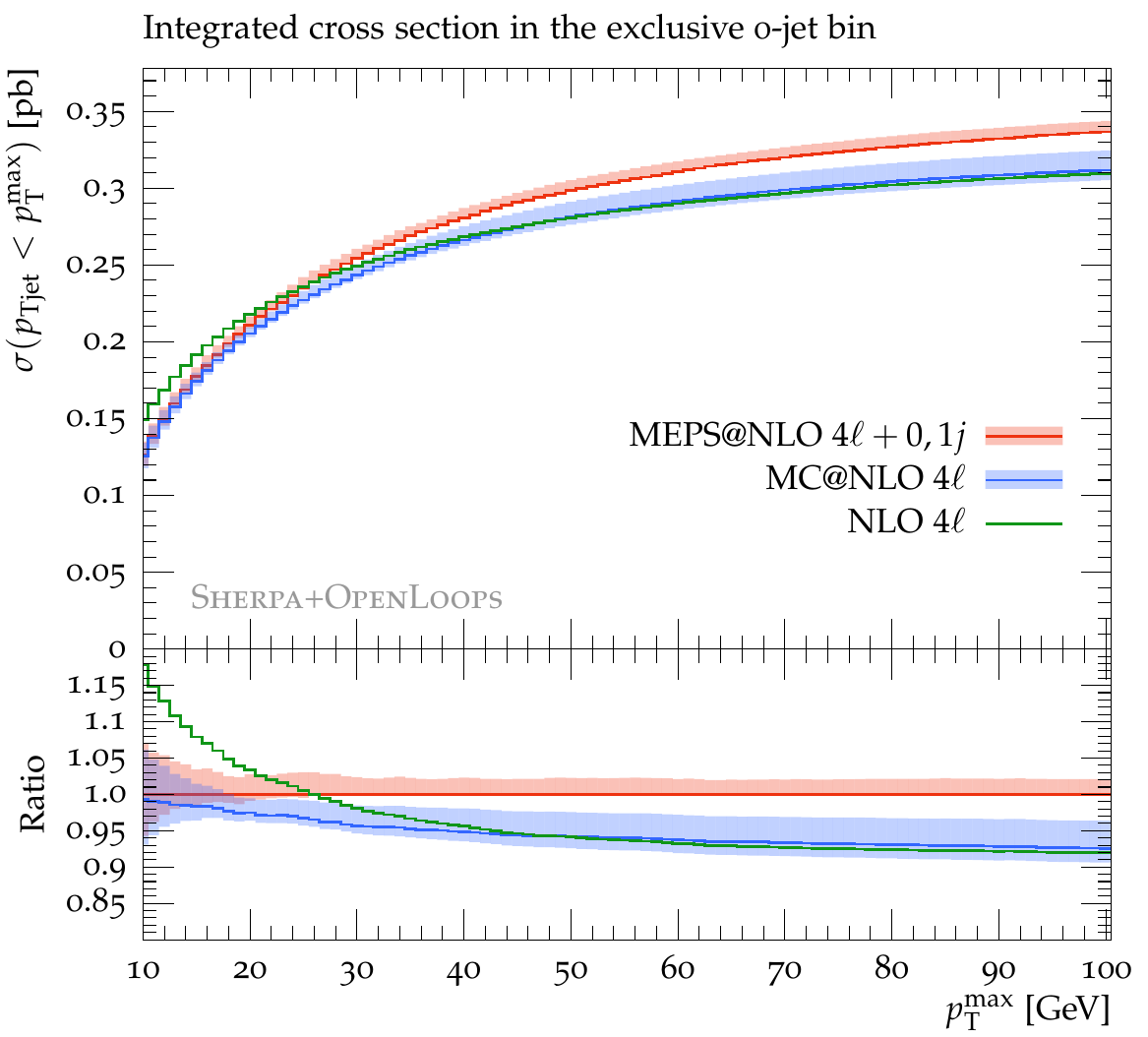}
\includegraphics[width=0.48\textwidth]{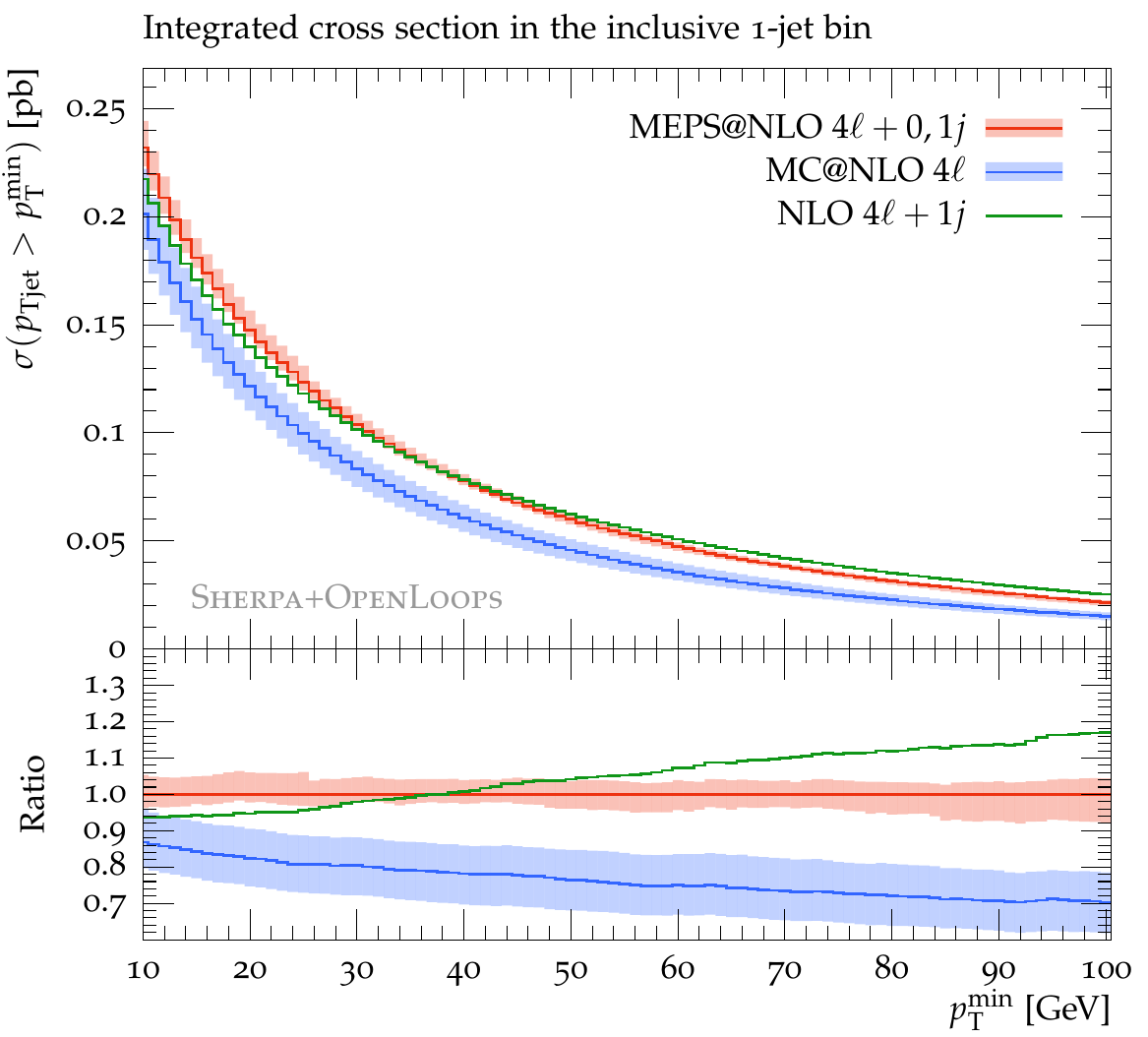}
\end{center} 
 \vspace*{-0.3cm} 
 \caption{
Integrated cross sections in the exclusive 0-jet bin (left)
and in the inclusive 1-jet bin (right) as a function of the respective 
transverse-momentum bounds, $\ptmax$ and $\ptmin$.
\NLOacc results with appropriate jet multiplicity (green) are
compared to \MCatNLOfl (blue) and \MEPSatNLOflj (red)
simulations.  Uncertainty bands correspond to  
QCD-scale variations combined with the resummation-scale
variations in quadrature.
}
\label{fig:inclusive_jetveto} 
\end{figure}

Matching and merging effects in presence of a jet veto and jet binning are
illustrated in \reffi{fig:inclusive_jetveto}, where the integrated cross
sections in the exclusive 0-jet bin ($p_\rT<\ptmax$) and in the inclusive
1-jet bin ($p_\rT>\ptmin$) are plotted as a function of the corresponding
upper and lower transverse-momentum bounds.  
In the 0-jet bin, \MCatNLO and \MEPSatNLO predictions agree well at small
jet-veto scales and differ by less than 10\% at large $\ptmax$.  The respective
uncertainties are as small as a few percent and nearly independent of
$\ptmax$.  For sufficiently inclusive jet-veto values, the \NLOacc
$\Pp\Pp\to 4\ell$ calculation is in excellent agreement with \MCatNLO.
In the $\ptmax\to 0$ limit, \NLOacc predictions develop a 
double-logarithmic singularity of the form $-\alphaS \ln^2(\ptmax/Q)$,
while \MCatNLO and \MEPSatNLO vetoed cross sections 
consistently
tend to  zero as a result of the exponentiation of Sudakov logarithms.
In this infrared regime, the exponentiation of double logarithms should manifest itself as a 
positive correction beyond NLO, while for $\ptmax\gtrsim 10\UGeV$
we observe that matched/merged predictions are still below the \NLOacc jet-vetoed cross section.
This is due to the fact that Sudakov logarithms are relatively mild in this region 
(\cf \reffi{fig:inclusive_merging}.a), and parton-shower effects are dominated
by subleading logarithms associated with the running of $\alphaS$ in the
$\alphaS(p_\rT) \ln(p_\rT)/p_\rT$ 
terms.  Double logarithms become dominant at
much smaller transverse momenta, and we checked that they drive the 
\NLOacc cross sections into the negative range only at $\ptmax\sim 2\UGeV$.
For $\ptmax\simeq 25$--$30\UGeV$, which corresponds to the jet-veto values
in the $\hww$ analyses at the LHC, fixed-order and matched/merged results
deviate by less than 5\%.  This 
represents the net effect of Sudakov logarithms
beyond NLO, and its smallness is due to the moderate size of the logarithmic
terms but also to cancellations between leading and subleading logarithms.
The uncertainty due to subleading Sudakov logarithms that are
not included in the \MCatNLO and \MEPSatNLO approximations are 
quantified via resummation-scale variations, which are reflected in the respective scale-variation bands, 
and turn out to be at the percent level.

As shown in \reffi{fig:inclusive_jetveto}.b, in the inclusive 1-jet bin the
discrepancies between the various approximations become more sizable.  The
inclusive \MCatNLO simulation underestimates the 1-jet cross
section by 20--30\% for $30\UGeV<\ptmin<100\UGeV$. For transverse-momentum 
thresholds up to $50\UGeV$,  the fixed-order $4\ell+1j$ cross section is in quite good
agreement with the \MEPSatNLO prediction as expected.
However, as already observed in \reffi{fig:inclusive_merging}.a, the \NLOacc cross section 
develops a significant excess in the tail.
The uncertainties of the \MEPSatNLO and \MCatNLO cross sections in the 1-jet bin 
are rather independent of the $p_\rT$-threshold and amount to 
about 5\% and 10\%, respectively.

\subsection{Squared quark-loop contributions}
\label{se:sqlWW}

\begin{table}
  \begin{center}
    \begin{tabular}{|@{\;}c@{\;}||@{\;}c@{\;}|@{\;}c@{\;}|@{\;}p{30mm}@{\;}|@{\;}p{30mm}@{\;}|}
\hline
  Analysis
&  \textbf{\LOOPSQfljs}
&  \textbf{\LOOPSQPSfl}
&  \textbf{\MEPSatLOOPSQ \newline $\Pp\Pp \to 4\ell+0,1j$}
&  \textbf{\MEPSatLOOPSQ \newline $\Pg\Pg \to 4\ell+0,1\Pg$}
\\[1mm]\hline
   $\geq 0$ jets
& $8.71(3)\;^{+28\%}_{-20\%}$
& $8.76(3)\;^{+28\%}_{-21\%}$$\;^{+0.2\%}_{-0.1\%}$
& $9.24(4)\;^{+31\%}_{-20\%}$$\;^{+20\%}_{-14\%}$
& $9.10(3)\;^{+28\%}_{-21\%}$$\;^{+15\%}_{-12\%}$
\\[1mm]\hline
   $\geq 1$ jets
& $3.98(7)\;^{+48\%}_{-30\%}$
& $1.75(1)\;^{+32\%}_{-25\%}$$\;^{+55\%}_{-51\%}$
& $2.75(3)\;^{+40\%}_{-24\%}$$\;^{+5\%}_{-6\%}$
& $2.01(2)\;^{+35\%}_{-25\%}$$\;^{+1.4\%}_{-3.2\%}$
\\[1mm]\hline
\end{tabular}
\caption{Squared quark-loop predictions in femtobarns for the $\mnen$
analyses requiring $\ge 0$ jets and $\ge 1$ jets.  Fixed-order results
(\LOOPSQ) with a number of jets corresponding to the actual analysis
are compared to an inclusive parton-shower simulation (\LOOPSQPSfl)
and to predictions
from the merged \MEPSatLOOPSQflj simulation with and without the
inclusion of quarks in the initial state.  Scale variations and statistical
errors are presented as in \refta{Tab:InclusiveXS1}. }
\label{Tab:InclusiveXS2} 
\end{center} 
\end{table}

Detailed results for the squared quark-loop cross sections in the inclusive
analysis and requiring one or more jets with $p_\perp>30\UGeV$ are presented in
\refta{Tab:InclusiveXS2}.  Fixed-order calculations for $4\ell$ or
$4\ell+1j$ production, depending on the jet bin, are compared to an
inclusive simulation obtained by showering four-lepton matrix elements
(\LOOPSQPSfl) and to merged predictions (\MEPSatLOOPSQflj).  Additionally,
to assess the importance of quark-induced channels, we show merged squared
quark-loop results that involve only gluon--gluon partonic channels and, for
consistency, only $\Pg\to\Pg\Pg$ splittings in the parton shower.

As compared to the \MEPSatNLO cross sections in \refta{Tab:InclusiveXS1},
squared quark loops represent a correction of about 3\%, 
both in the inclusive analysis and in the 
1-jet bin.
In the inclusive case, fixed-order and shower-improved predictions
are in excellent agreement, as expected from the unitarity of the shower.
In contrast, the \LOOPSQPS simulation---which corresponds to the approach typically
adopted in present experimental studies, where jet emission is entirely
based on the shower approximation---underestimates the squared quark-loop
cross section in the inclusive 1-jet bin by around 50\%.  Due to their
LO $\alphaS^2$ and $\alphaS^3$ dependence, squared quark-loop corrections 
feature a QCD-scale dependence of 30--40\%.  The resummation-scale uncertainty of the
\LOOPSQPS simulation is close to zero in the inclusive case (due to unitarity), while
in the 1-jet bin it is as large as 50\%, due to the fact that the 1-jet bin
is entirely filled by shower emissions.  

Comparing \LOOPSQPS predictions to the merged sample we observe that the
matrix-element description of jet emission significantly increases the cross
section, especially in the 1-jet bin.  The QCD-scale uncertainty remains at
30--40\% level, but resummation-scale variations change substantially: the
1-jet bin cross section becomes almost independent of the resummation scale,
since, as a result of merging, 1-jet events are described in terms of matrix
elements, and shower emissions induce only minor bin migrations.  In
contrast, in the inclusive analysis the merged simulation features a
significantly higher resummation-scale dependence of 
approximately
15\%, which can
be attributed to unitarity violations induced by the merging procedure: the
resummation-scale dependence that arises from the region below the merging
cut, where 0-jet matrix elements are combined with the Sudakov suppression
factor, is not compensated by an opposite dependence from above $\qcut$,
since the parton shower is superseded by 1-jet matrix elements in that
region.  We note that this kind of resummation-scale sensitivity is due to
the LO nature of squared quark-loop merging and is strongly reduced in the
case of NLO merging (\cf last column in \refta{Tab:InclusiveXS1}).  The
fact that the \MEPSatLOOPSQ cross section in the 1-jet bin is $30\%$
below the fixed-order result can be attributed to the CKKW scale choice in
the merging approach and is consistent with the size of
renormalisation-scale variations.  Finally, comparing the last two columns
in \refta{Tab:InclusiveXS2}, we observe that quark-induced channels account
for roughly 1.5\% and 30\% of the squared quark-loop corrections in the 0- and
1-jet bins, respectively.  This corresponds to about 0.5 permil and 1 percent of the
total cross section in the respective jet bins.  We note that the individual
impact of quark channels at matrix-element or parton-shower level is
significantly larger, \ie a naive merging approach based on pure-gluon
matrix elements plus a standard parton shower would lead to bigger
deviations with respect to the \MEPSatLOOPSQ results in
\refta{Tab:InclusiveXS2}.

\begin{figure}
\begin{center}
\includegraphics[width=0.48\textwidth]{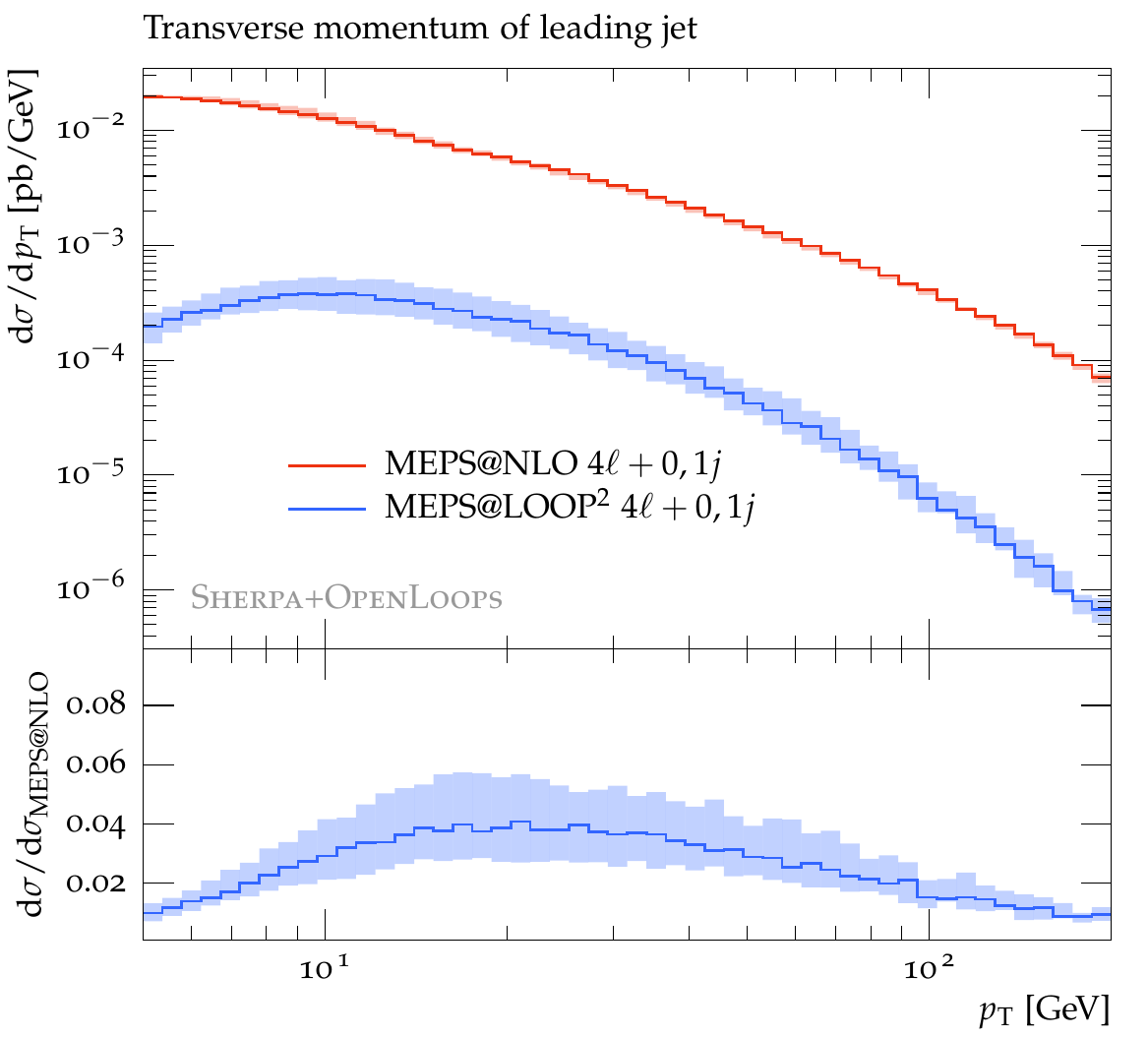}
\includegraphics[width=0.48\textwidth]{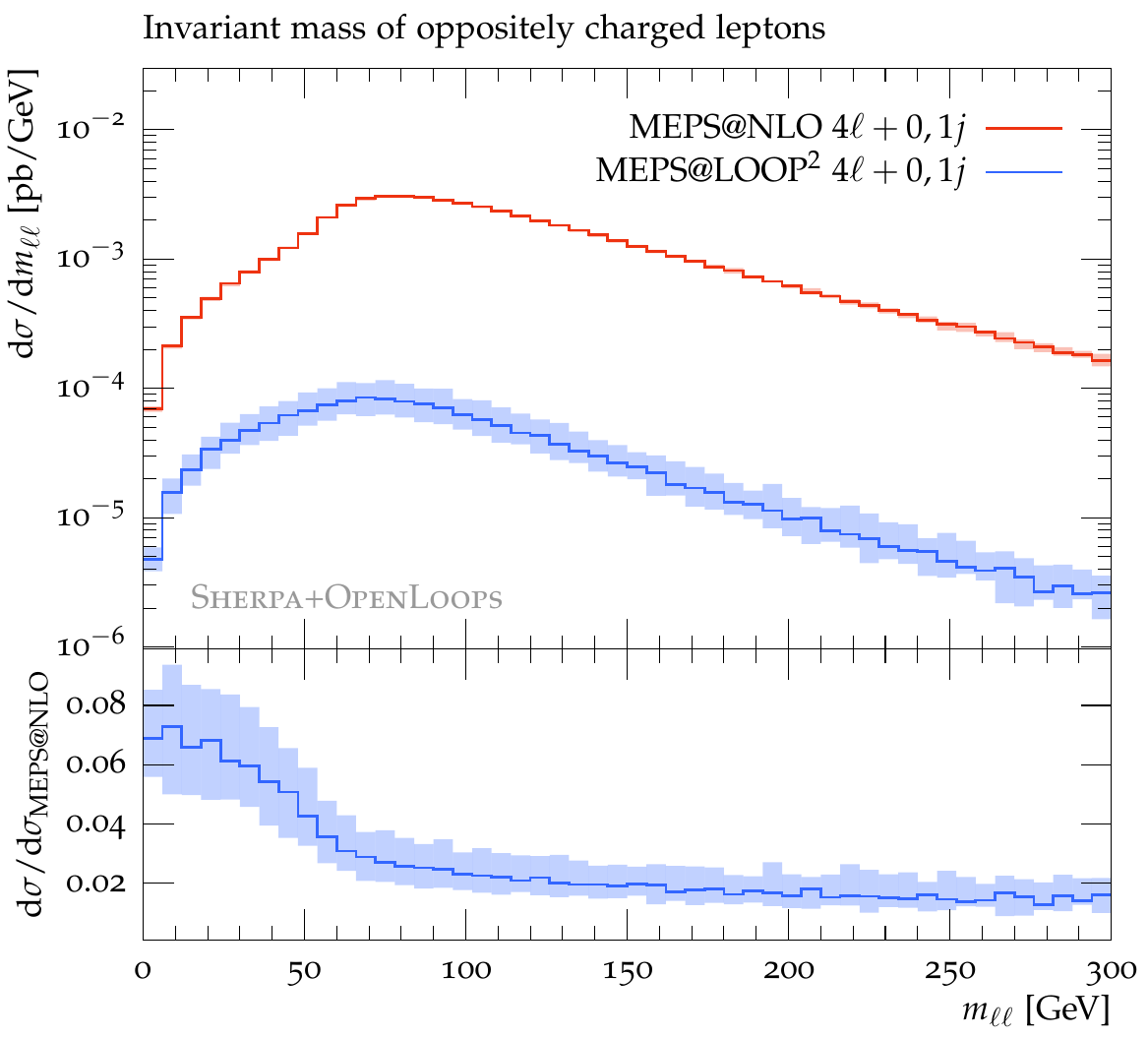}
\end{center}
\vspace*{-0.3cm}
\caption{Comparison of merged squared quark-loop (blue) and NLO (red)
predictions for $4\ell+0,1j$ production: transverse momentum of the
  leading jet (left) and invariant mass of the two charged leptons (right).} 
\label{fig:inclusive_loop2_mepsnlo}
\end{figure}

Squared quark-loop corrections to differential observables are 
compared to NLO merged predictions in \reffi{fig:inclusive_loop2_mepsnlo}.
As already found in \reftas{Tab:InclusiveXS1} and
\ref{Tab:InclusiveXS2}, their impact typically amounts to a
few percent.  Both for the leading-jet transverse momentum and for the
dilepton invariant mass they feature a rather different kinematic dependence
as compared to \MEPSatNLO results.  In the considered range their relative
importance varies from one to seven percent, and the maximum lies 
in the region of small dilepton mass, which corresponds to the signal region
of the $\hww$ analysis.

\begin{figure}
\begin{center}
\includegraphics[width=0.48\textwidth]{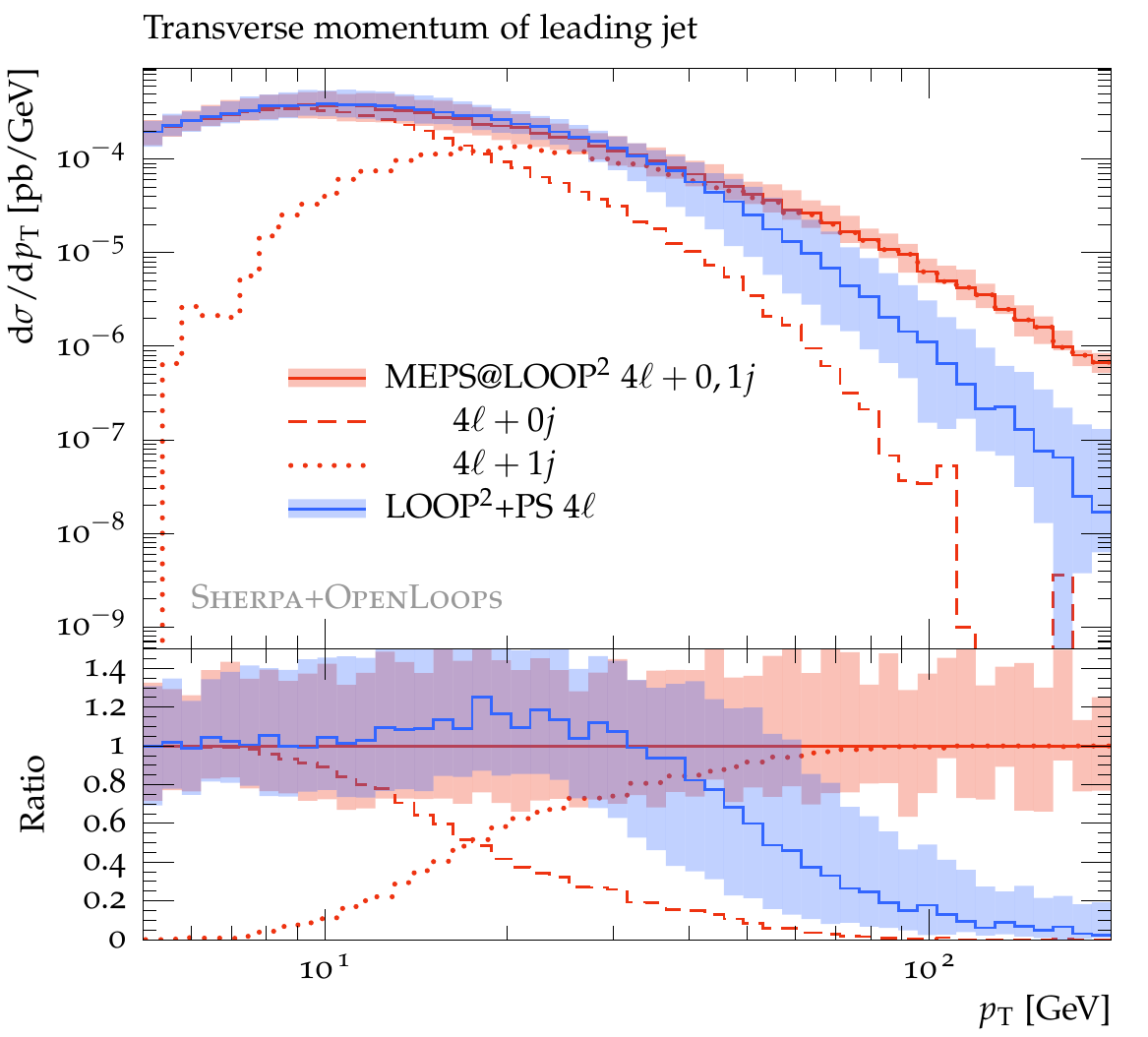}
\includegraphics[width=0.48\textwidth]{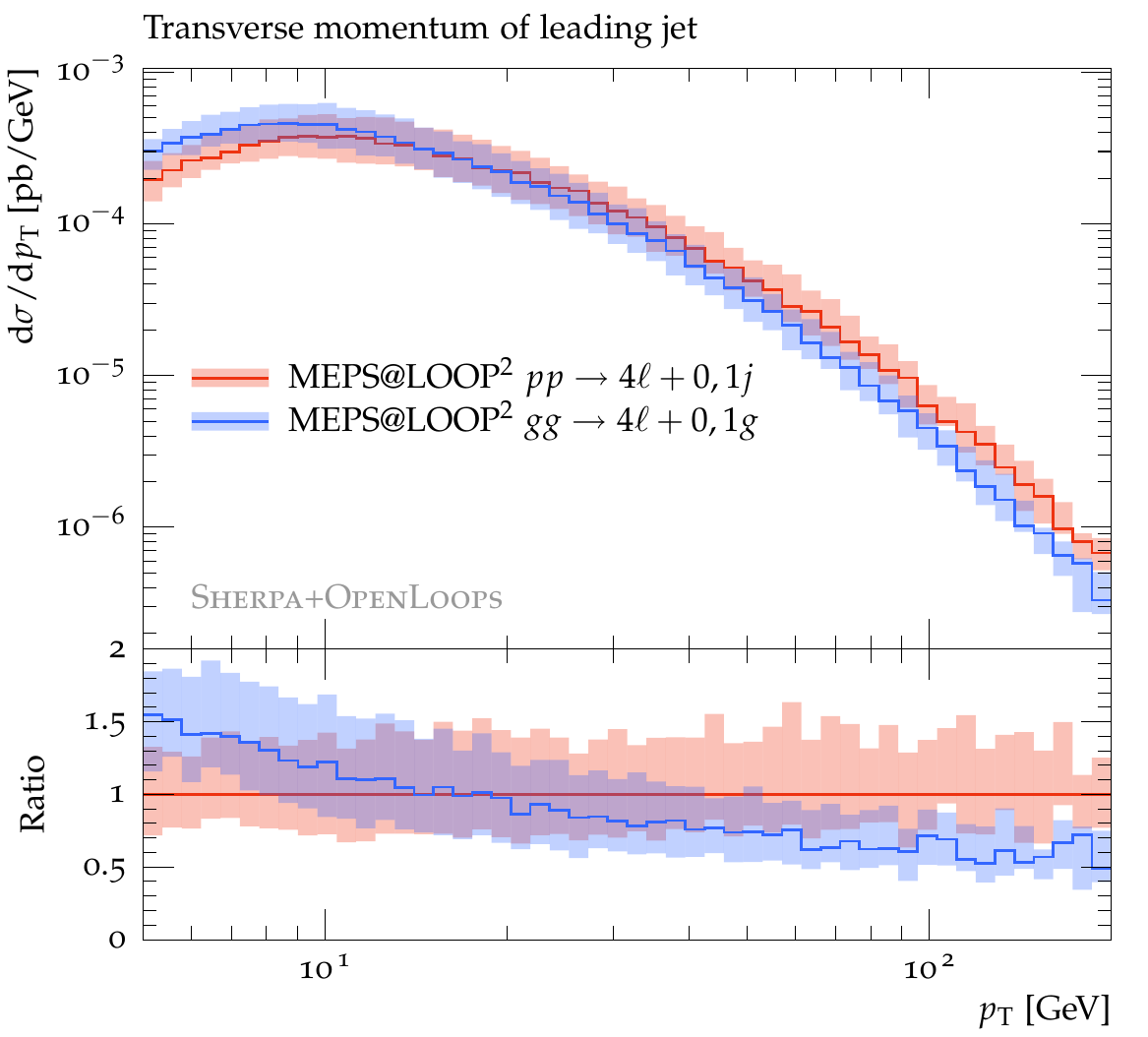}
\end{center}
\vspace*{-0.3cm}
\caption{Squared quark-loop corrections to the leading-jet
$p_\rT$-distribution: (a) a simulation based on $4\ell$ matrix elements plus parton shower (blue)
is compared to complete merged predictions (red solid). The latter are split into the contributions 
from $4\ell+0j$ (red dashed) and $4\ell+1j$ (red dotted) matrix elements;
(b) full merged predictions (red) are compared to a corresponding simulation involving only 
gluon contributions (blue).
Uncertainty bands correspond to the combination (in quadrature) of QCD- and resummation-scale 
variations.}
\label{fig:inclusive_loop2_details}
\end{figure}

Merging effects are illustrated in the left plot of
\reffi{fig:inclusive_loop2_details}, where predictions from the inclusive
squared quark-loop $\Pg\Pg\to 4\ell$ matrix element supplemented with a
regular parton shower (\LOOPSQPS) are compared to the merged $\Pp\Pp\to
4\ell+0,1j$ simulation (\MEPSatLOOPSQ).  The latter is decomposed into
contributions from $4\ell+0j$ and $4\ell+1j$ matrix elements.  In the region
well below the merging cut, $\qcut=20\UGeV$, merged predictions are
dominated by 0-jet matrix elements and agree almost perfectly with the \LOOPSQPS
curve.  The agreement remains better than 10\% up to $p_\rT\sim\qcut$, where
the \MEPSatLOOPSQ sample is characterised by the transition from the 0-jet
to the 1-jet matrix-element regime.  This supports the use of the 0-jet plus shower
approximation up to the merging scale.  Starting  from $p_\rT\gtrsim 40\UGeV$,
where 1-jet matrix elements dominate and render \MEPSatLOOPSQ predictions more reliable,
the parton-shower results feature a sizable deficit 
and are also strongly sensitive to the resummation scale.

Setting the resummation scale equal to the default scale \refeq{eq:scale},
we found that the slight excess of the parton shower at $p_\rT\sim\qcut$
propagates to higher transverse momenta reaching up to 40\% at $p_\rT\gtrsim
100 \UGeV$.  In order to avoid such an unnatural parton-shower excess at
high $p_\rT$, and a corresponding excess in the Sudakov suppression at low
$p_\rT$, as anticipated in \refse{se:NLOWW} we decided to evaluate squared
quark-loop contributions using a smaller resummation scale, $\mu_Q=\mu_0/2$. 
Of course the small value of $\mu_Q$ amplifies the natural deficit of the
shower at large $p_\rT$ and yields a quite small \LOOPSQPS cross section in
the 1-jet bin (\cf \refta{Tab:InclusiveXS2}).  However this side-effect is
compensated by 1-jet matrix elements in the \MEPSatLOOPSQ simulation.  The
bands describe the total scale uncertainty, obtained by adding QCD- and
resummation-scale variations in quadrature.  Apart from the suppressed
high-$p_\rT$ tail of the \LOOPSQPS distribution, we find a rather constant
uncertainty of about 30\%.

The right plot in \reffi{fig:inclusive_loop2_details} illustrates the impact
of quark-channel contributions on the leading-jet $p_\rT$-distribution. 
Plotted are full \MEPSatLOOPSQ results and corresponding predictions
involving only $\Pg\Pg$-induced matrix elements and $\Pg\to\Pg\Pg$ shower
splittings.  As is clearly visible from the ratio plot, the quark channels
enhance hard-jet emissions and induce a related Sudakov suppression at low
$p_\rT$.  The resulting distortion in the jet-$p_\rT$ distribution amounts
to $\pm 50\%$. When looking at \refta{Tab:InclusiveXS2}, such opposite behaviour
in the hard and soft regions explains why the quark-channel contribution 
reaches 30\% in the 1-jet bin but goes down to 1.5\% in the inclusive case.

\begin{figure}
\begin{center}
\includegraphics[width=0.48\textwidth]{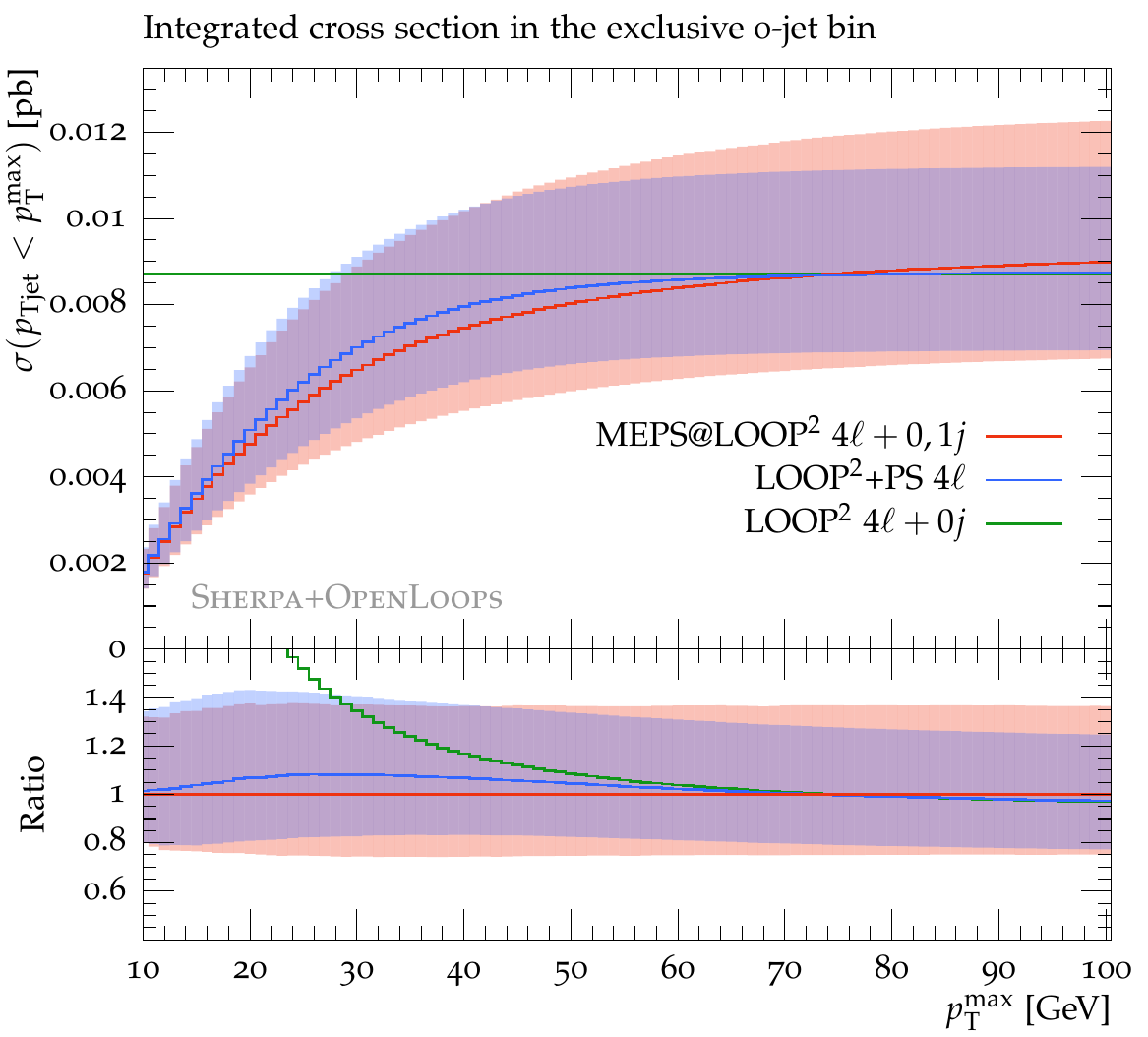}
\includegraphics[width=0.48\textwidth]{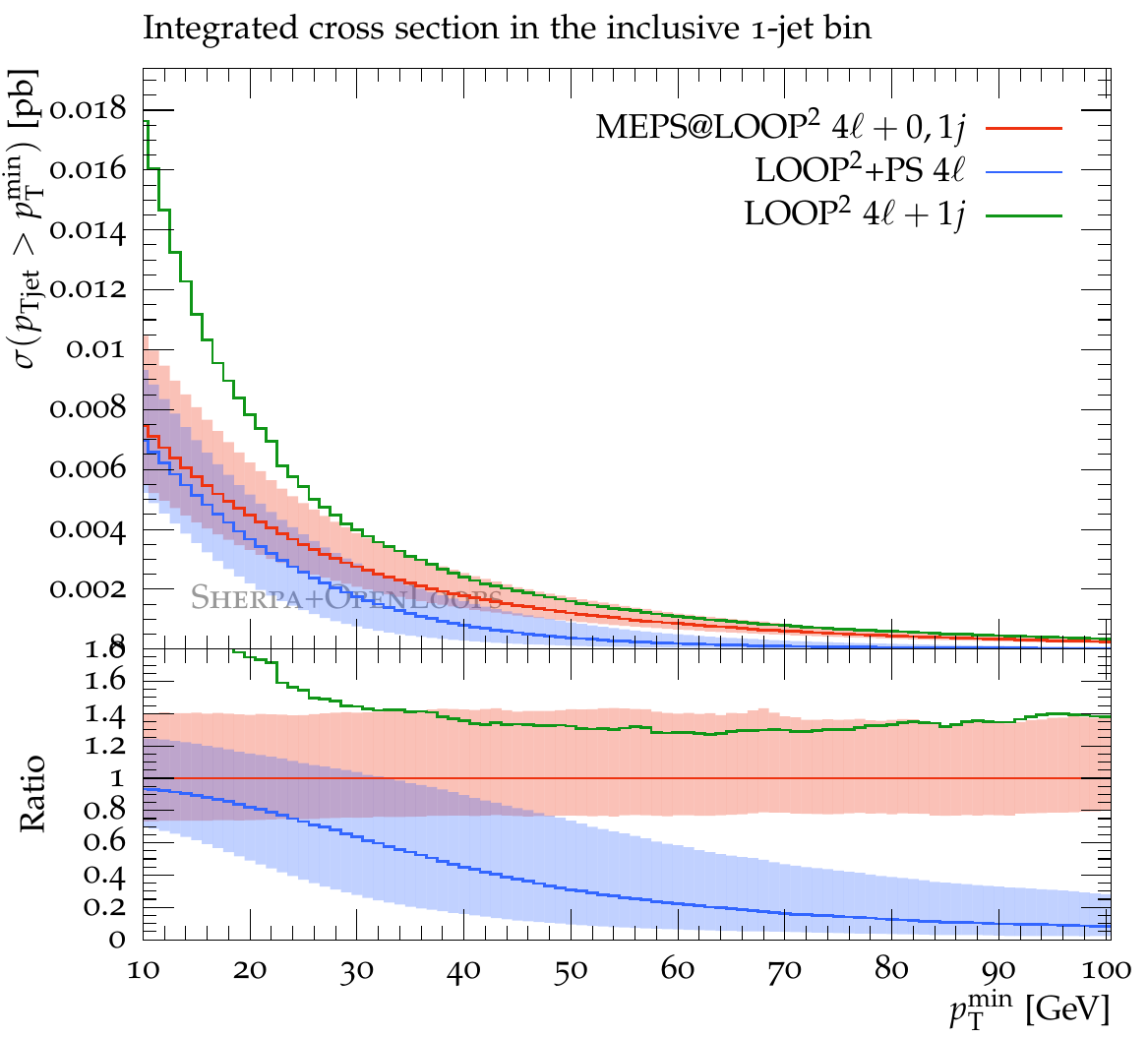}
\end{center}
\vspace*{-0.3cm}
\caption{
Integrated squared quark-loop cross sections in the exclusive 0-jet bin (left)
and in the inclusive 1-jet bin (right) as a function of the respective 
transverse-momentum bounds, $\ptmax$ and $\ptmin$.
Fixed-order \LOOPSQfljs results (green) are
compared to \LOOPSQPSfl (blue) and \MEPSatLOOPSQflj (red)
simulations.  Uncertainty bands correspond to
QCD-scale variations combined with resummation-scale
variations in quadrature.}
\label{fig:inclusive_loop2_resummation}
\end{figure}

Jet-veto and jet-binning effects on squared quark-loop contributions are
shown in \reffi{fig:inclusive_loop2_resummation}, where 
the integrated cross sections in the exclusive 0-jet bin ($p_\rT<\ptmax$) and in the 
inclusive 1-jet bin ($p_\rT>\ptmin$) are plotted as a function of $\ptmax$ and $\ptmin$.
In the 0-jet bin, apart from the minor excess around $30\UGeV$, \LOOPSQPS predictions
agree quite well with \MEPSatLOOPSQ ones for any jet-veto scale up to
$100\UGeV$.  The corresponding scale uncertainties are in the 20--40\% range.
As in \refta{Tab:InclusiveXS2}, \MEPSatLOOPSQ uncertainties tend to
be larger in the inclusive limit.
Fixed-order $\Pg\Pg\to 4\ell$ contributions are inherently inclusive and independent of $\ptmax$. 
Comparing them to 
the \MEPSatLOOPSQ and \LOOPSQPS curves we observe that 
jet-veto scales of 25--30$\UGeV$, as those used in the experimental $\hww$ analyses,
correspond to a moderate cross-section suppression of approximately 30\%.
In this regime the parton shower should provide a sufficiently 
reliable resummation of Sudakov logarithms.

The right plot of \reffi{fig:inclusive_loop2_resummation} compares
fixed-order, shower-improved and merged predictions in the inclusive 1-jet
bin.  For a jet threshold of $30\UGeV$, the various approximations agree only
marginally within the respective errors, while higher and smaller 
values of $\ptmin$ lead to very large discrepancies.  As compared to \MEPSatLOOPSQ
predictions, at large $p_\rT$ we observe a dramatic deficit of the shower
approximation, while the fixed-order squared quark-loop calculation
yields a rather constant 40\% excess as in \refta{Tab:InclusiveXS2}.  
The resummation of Sudakov logarithms becomes relevant
only for transverse-momentum thresholds below $30\UGeV$, where 
the excess of the fixed-order prediction grows up to 
150\% at 10\UGeV.

\section{\ATLAS and \CMS \texorpdfstring{H$\to$WW$^*$}{H->WW*} analyses in the 0- and 1-jet bins}
\label{Sec:HWW}

In this section we study the irreducible four-lepton background to the
\ATLAS~\cite{ATLAS-CONF-2013-030} and \CMS~\cite{CMS-PAS-HIG-13-003} $H \to
\PW\PW^*\to \mnen$ analyses at 8 TeV.  We restrict ourselves to the exclusive 
0- and 1-jet bins, which contain the bulk of the 
four-lepton background
associated with diboson production, 
and focus on opposite-flavour $\mnen+$jets final
states, which provide the highest sensitivity to the Higgs-boson signal. 
Technically, within the automated \SherpaOpenLoops framework, the
simulation of $\lnln+$jets production with same lepton flavour is almost
equivalent to the opposite-flavour case.  Also for what concerns QCD
corrections and uncertainties we do not expect any important difference
between opposite- and same-flavour channels.

In the following we apply the cuts listed in \refapp{se:hww_setup}, which
correspond to the \ATLAS~\cite{ATLAS-CONF-2013-030} and
\CMS~\cite{CMS-PAS-HIG-13-003} analyses at 8 TeV.  Let us remind that the
two experiments employ different definitions
of the $\PW\PW$ transverse mass, reported in eq.~\refeq{eq:mt},
and different anti-$k_\perp$ jet radii. Note also that \ATLAS 
employs a lower transverse-momentum threshold for central jets.
After a pre-selection, which basically requires 
two hard leptons and large missing energy, 
two complementary selections based on 
$p_{\perp,\ell\ell'}$,
$\Delta\phi_{\ell\ell'}$,
$m_{\ell\ell'}$ and $m_T$,
are used to define a
signal  and a control region.
The latter is exploited to normalise WW-background simulations to data.
Separate analyses are performed in the 0-, 1-, and 2-jet bins
in order to improve the sensitivity to the Higgs-boson signal and the data-driven
normalisation of the various background components.

In \refse{se:hWW_presel} we investigate
kinematic distributions that are relevant for the experimental selection
after pre-selection cuts. 
In \refse{se:hWW_dist} we consider the control and signal regions and
discuss the observables that are exploited in the final stage of the Higgs
analyses, namely the $\PW\PW$ transverse mass and the dilepton invariant
mass.  Finally, in \refse{se:hWW_XS} we present predictions 
for the 0- and 1-jet bin cross sections in the signal and
control regions, as well as uncertainties associated with variations of 
renormalisation, factorisation, resummation, and merging scales.

For each observable we present results for the \ATLAS and \CMS analyses in
the exclusive 0- and 1-jet bins and, to provide insights into the
convergence of the perturbative expansion and the size of Sudakov
logarithms in jet bins, we compare \NLOacc, \MCatNLO,
\MEPSatNLO and squared quark-loop predictions.  As discussed in \refse{se:mcsample}, 
in
\NLOacc predictions for the 0- and 1-jet bins 
we
always include the corresponding
number of jets at matrix-element level.  In contrast, \MCatNLO results refer
as usual
to a single simulation of inclusive $\mnen$ production, which is NLO
accurate in the 0-jet bin and only LO accurate in the 1-jet bin.  Only
\MEPSatNLO predictions are consistently matched to the parton shower and NLO
accurate in both jet bins.

\subsection{Kinematic distributions after pre-selection cuts}
\label{se:hWW_presel}

\begin{figure}
\begin{center}
\includegraphics[width=0.48\textwidth]{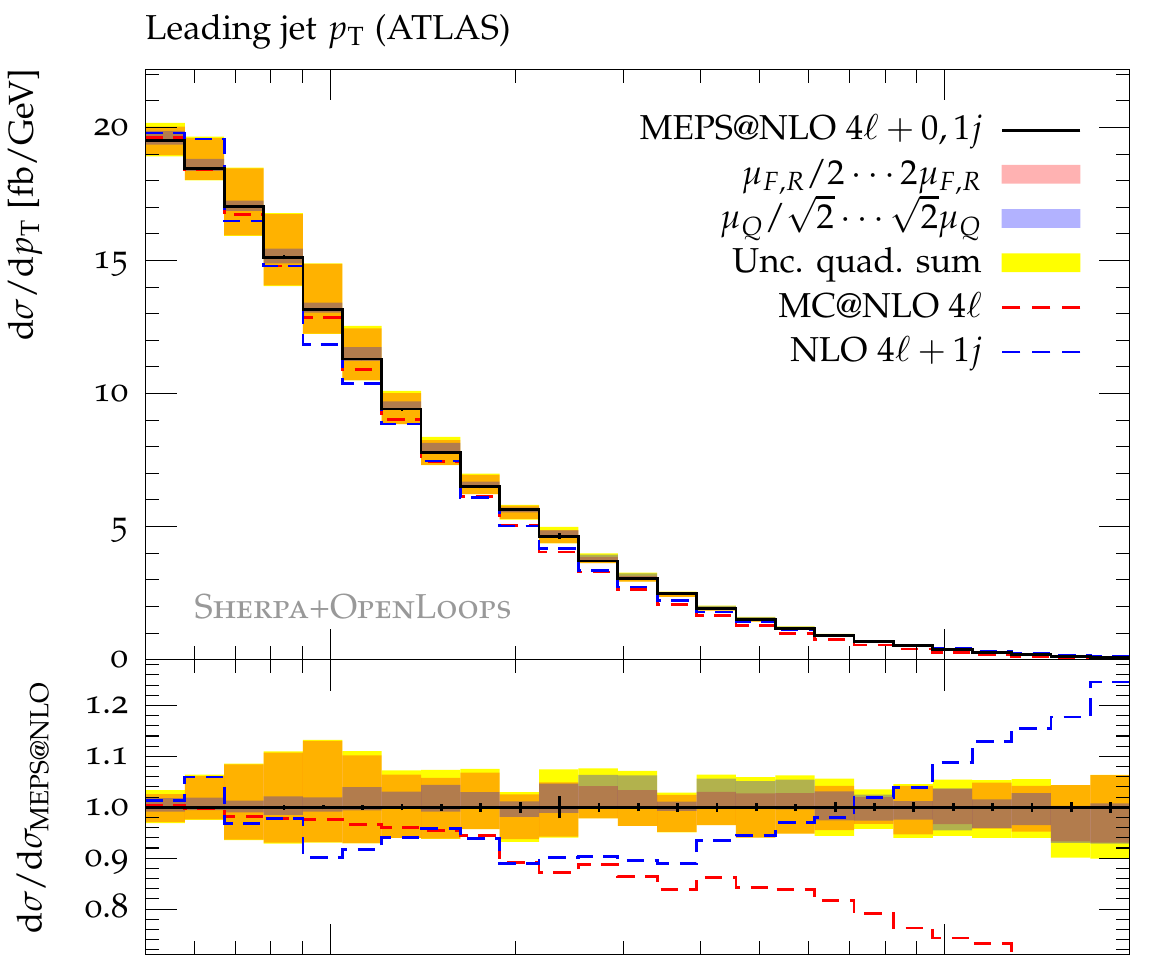}
\includegraphics[width=0.48\textwidth]{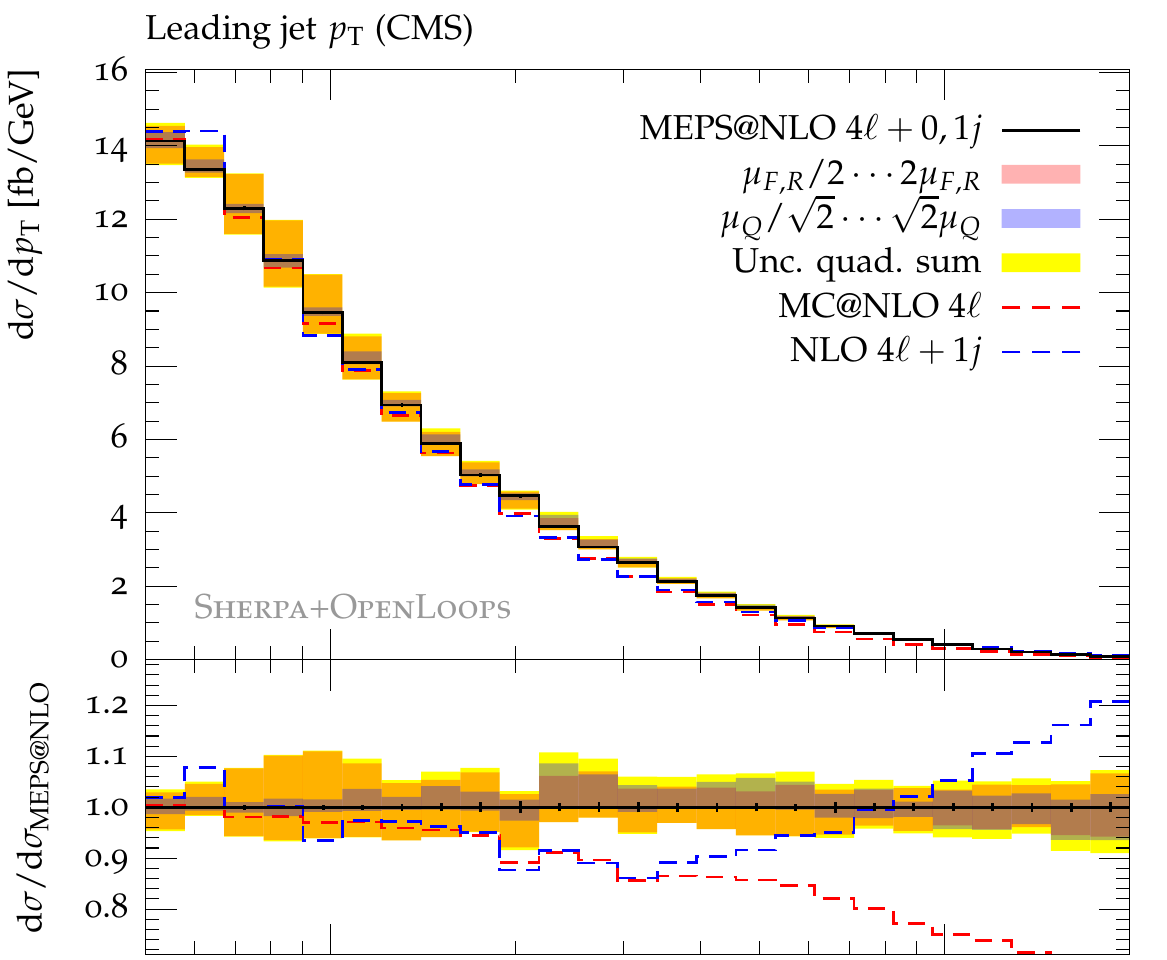}\\\vspace*{-3px}
\includegraphics[width=0.48\textwidth]{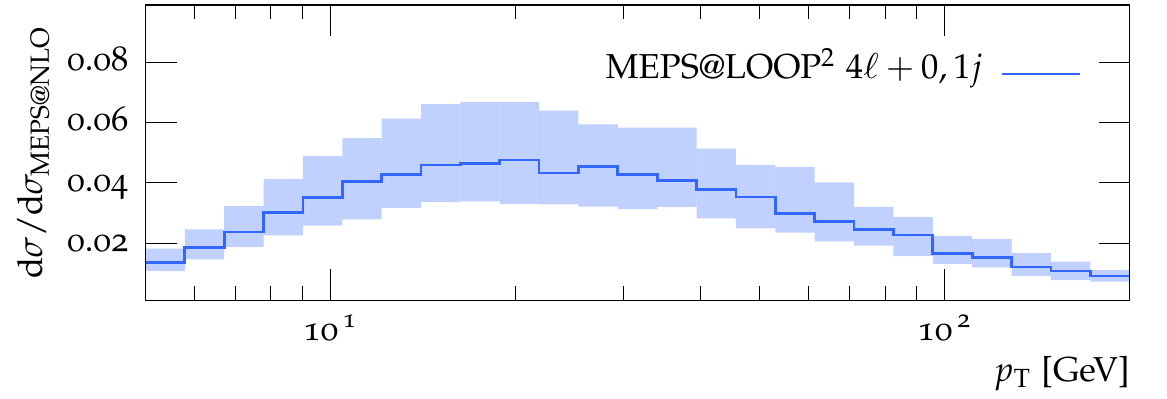}
\includegraphics[width=0.48\textwidth]{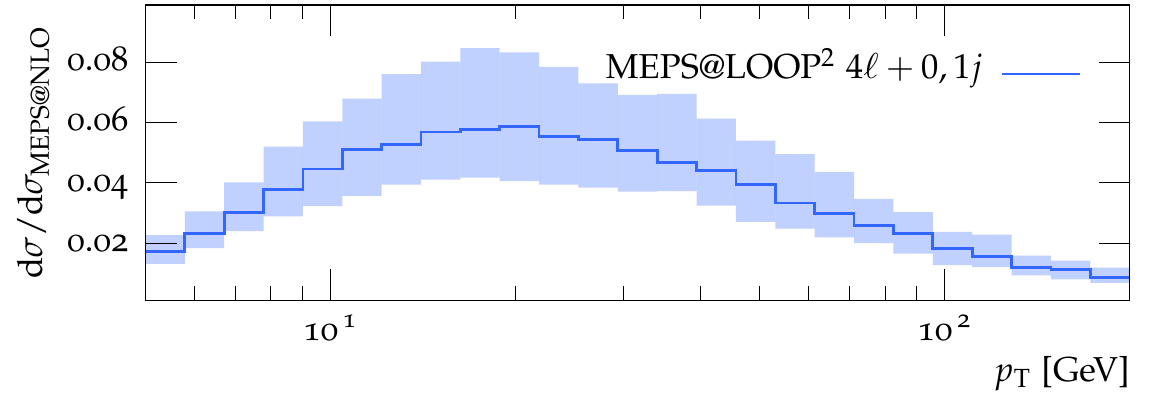}\\
\includegraphics[width=0.48\textwidth]{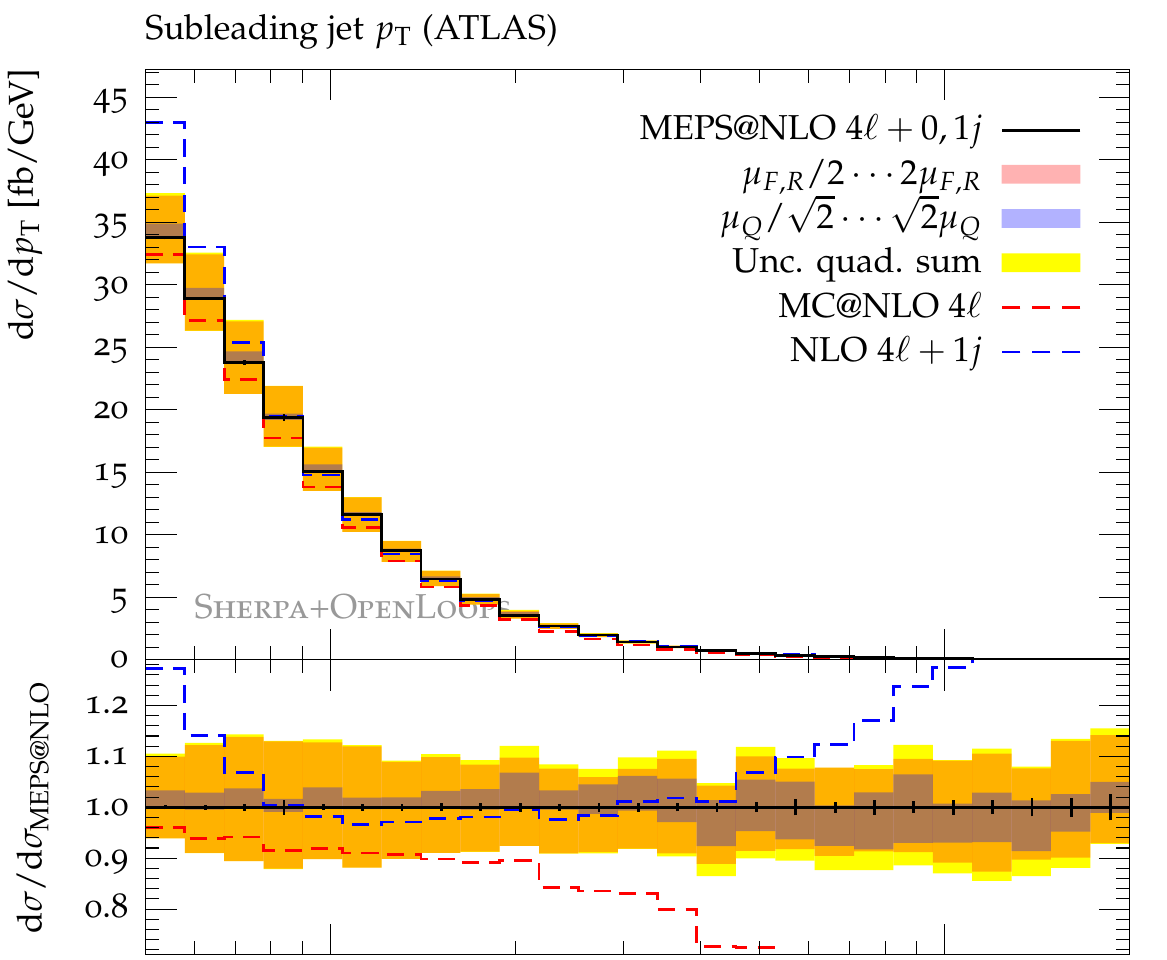}
\includegraphics[width=0.48\textwidth]{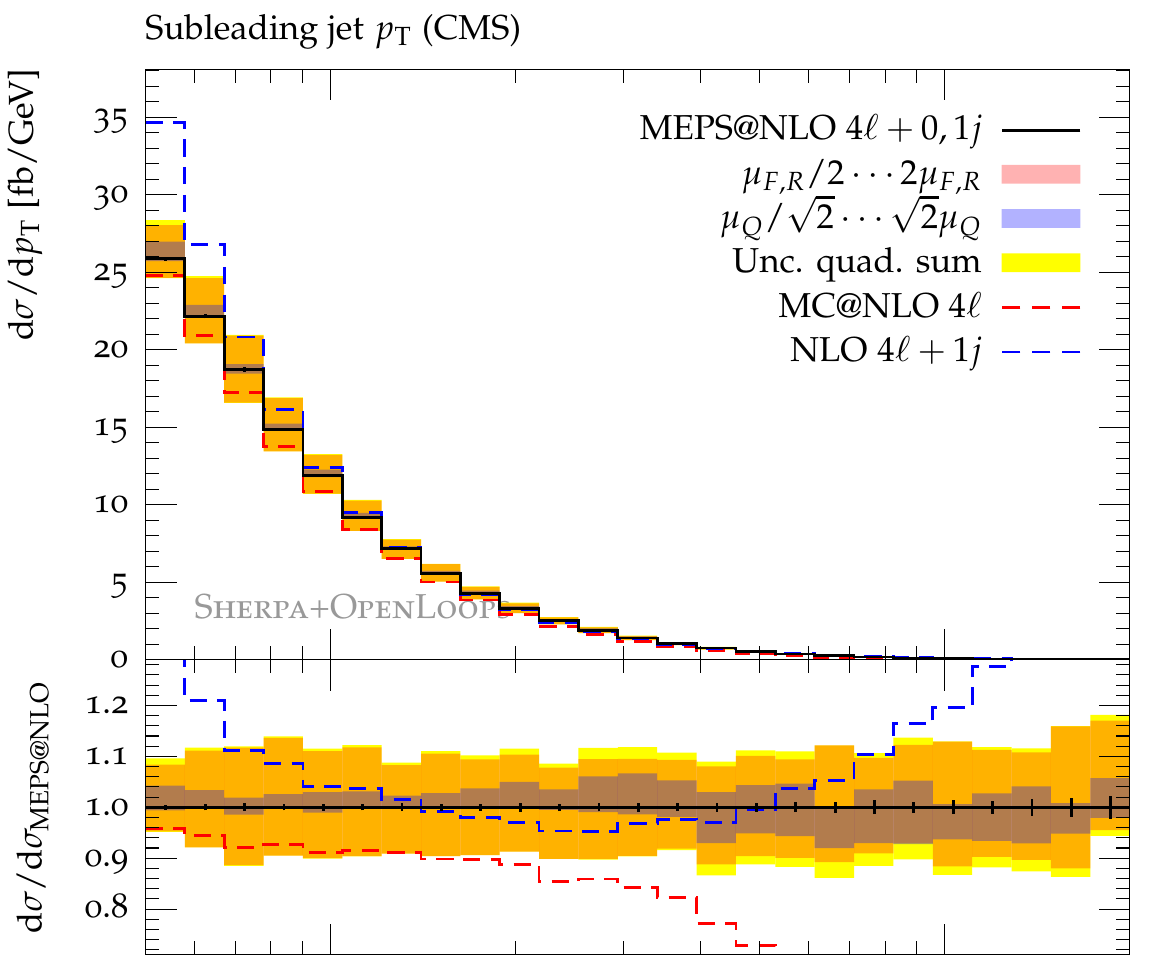}\\\vspace*{-3px}
\includegraphics[width=0.48\textwidth]{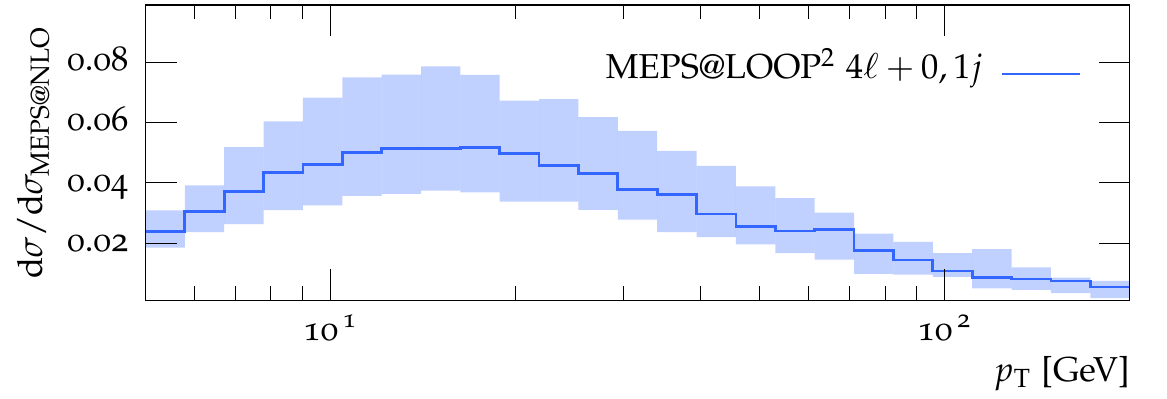}
\includegraphics[width=0.48\textwidth]{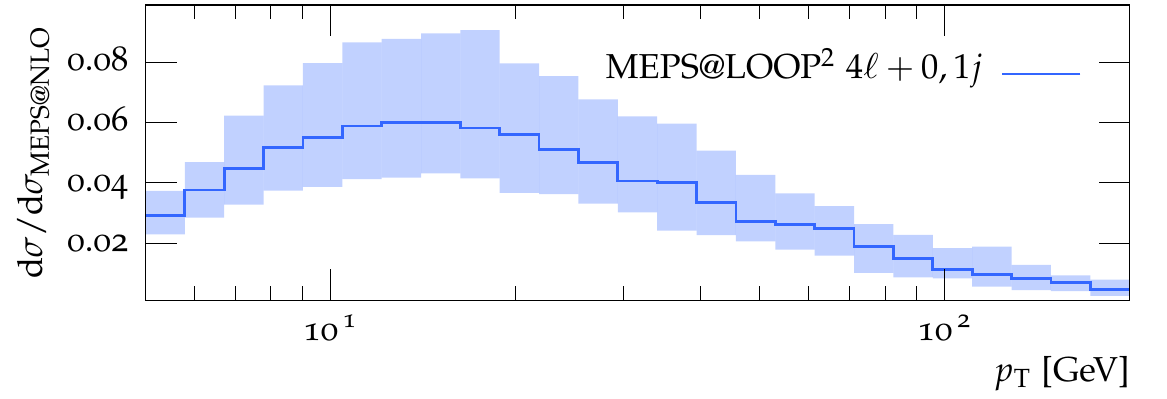}
\end{center}
\vspace*{-0.3cm}
\caption{\ATLAS (left)  and \CMS (right)
analysis at $8\UTeV$ after pre-selection cuts:
transverse-momentum distributions of the first (top) and second (bottom) jet.
\MEPSatNLO (black solid), inclusive \MCatNLO (red dashed), and
\NLOacc (blue dashed) predictions at the central scale.  The ratio plots in
the middle panels show relative uncertainties as well as \MCatNLO and \NLOacc deviations
with respect to \MEPSatNLO.
The lower panels display relative \MEPSatLOOPSQ corrections and uncertainties
normalised to \MEPSatNLO at the central scale.
The factor-two variations of $\mu_\rR$ and 
  $\mu_\rF$ (red band), and factor-$\sqrt{2}$ variations of $\mu_Q$ (blue band),
  are combined in quadrature (yellow band).
  Scale-variation bands are colour additive, \ie  yellow+blue=green, 
yellow+red=orange, and yellow+red+blue=brown.
} 
\label{fig:flj_hww_pre_ptjets}
\end{figure}

\begin{figure}
\begin{center}
\includegraphics[width=0.48\textwidth]{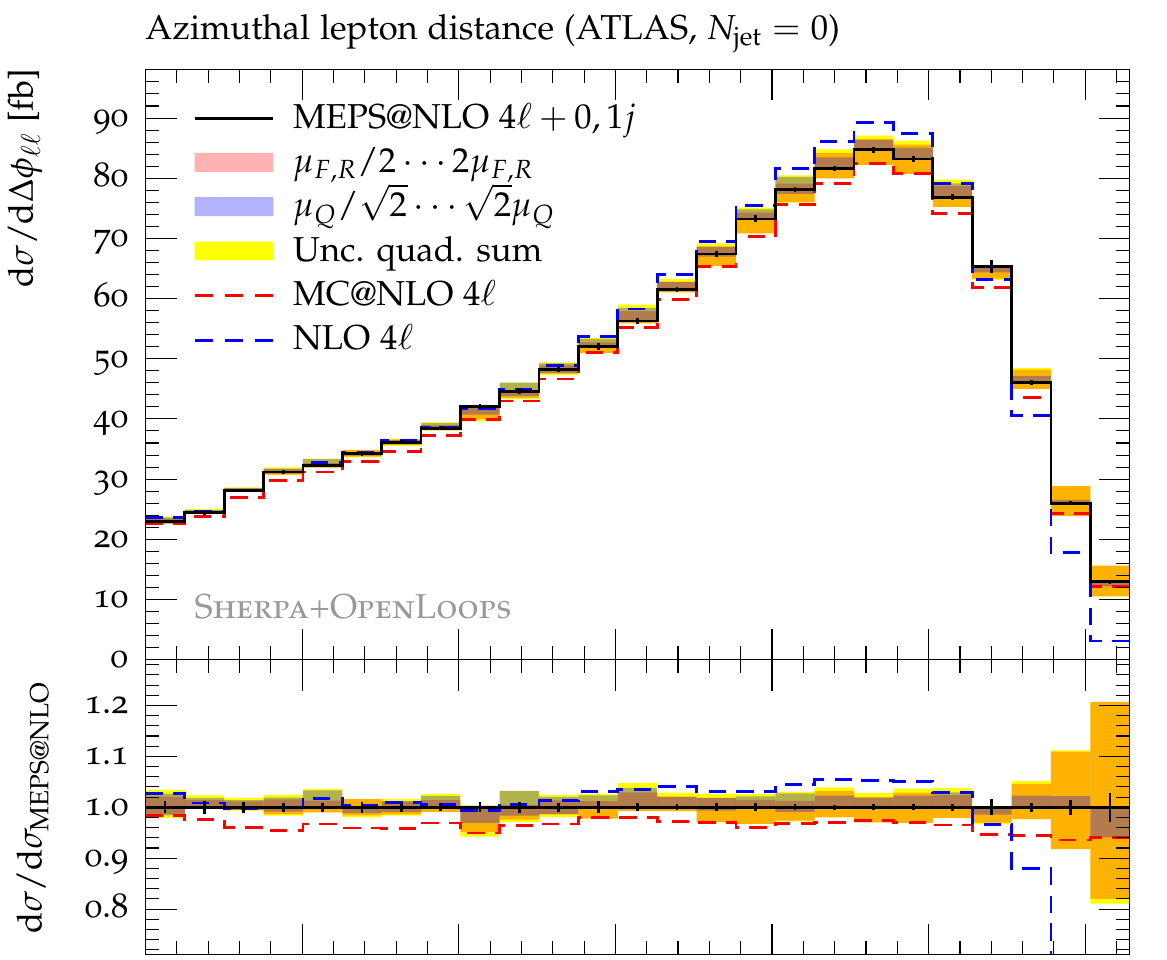}
\includegraphics[width=0.48\textwidth]{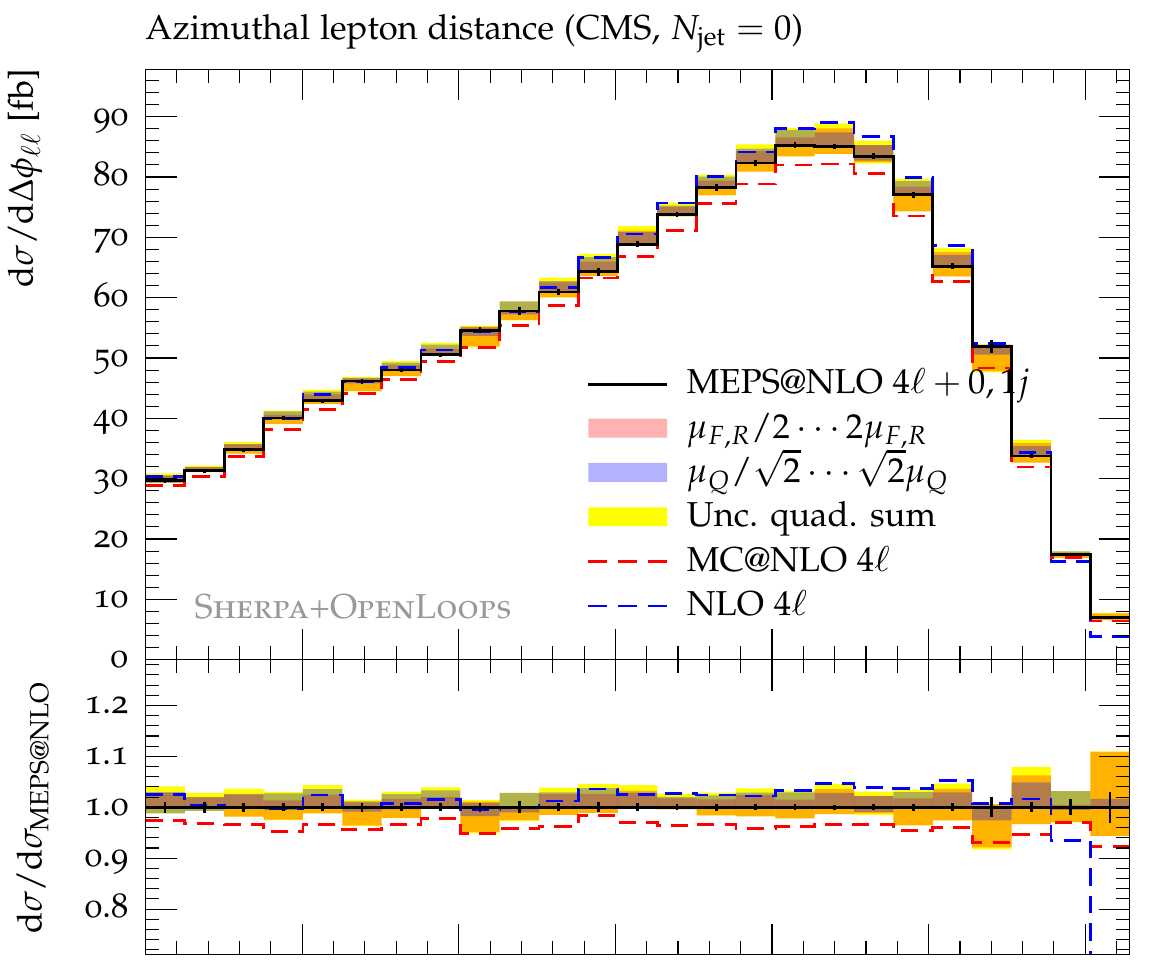}\\\vspace*{-3px}
\includegraphics[width=0.48\textwidth]{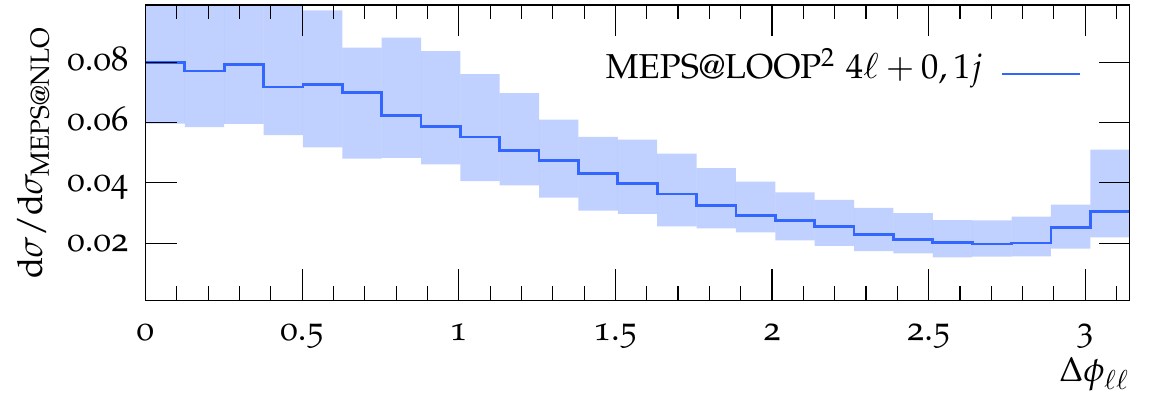}
\includegraphics[width=0.48\textwidth]{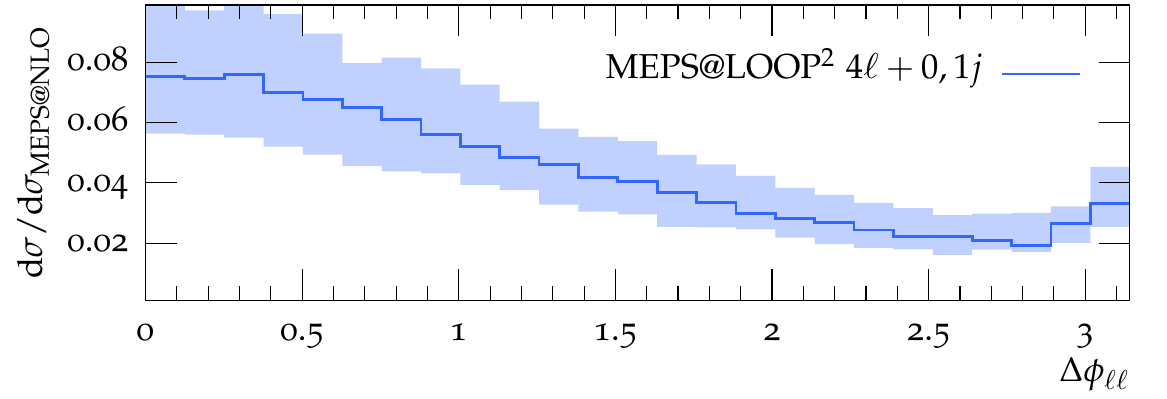}\\
\includegraphics[width=0.48\textwidth]{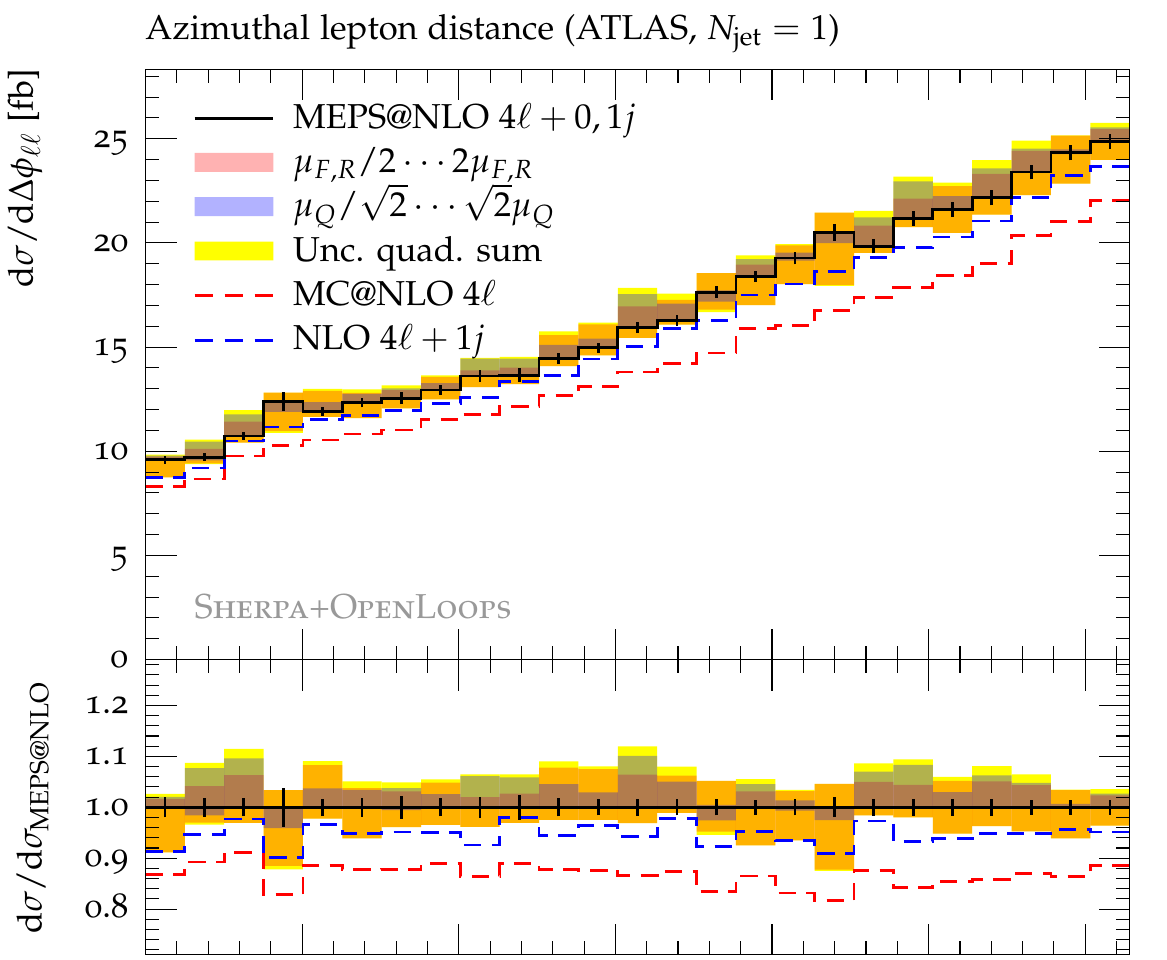}
\includegraphics[width=0.48\textwidth]{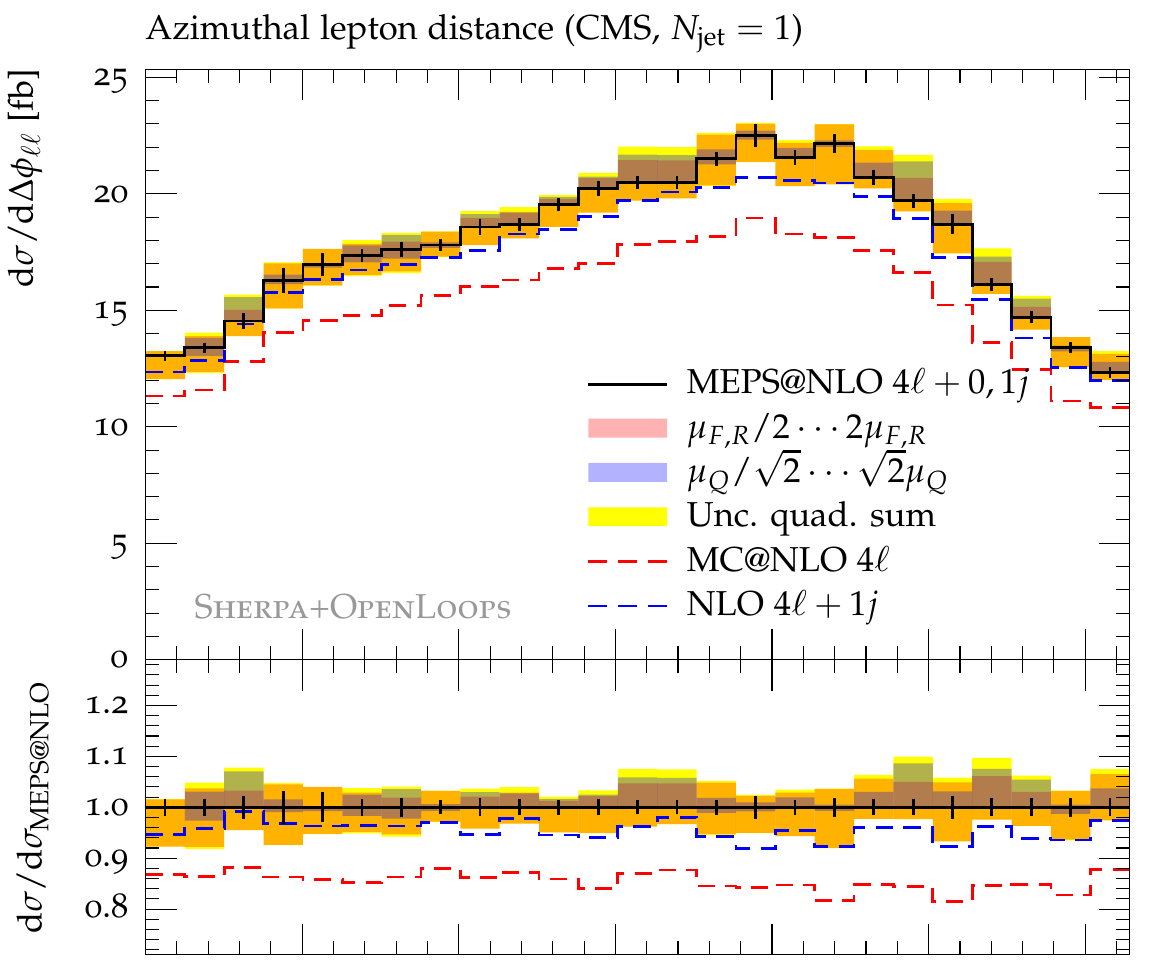}\\\vspace*{-3px}
\includegraphics[width=0.48\textwidth]{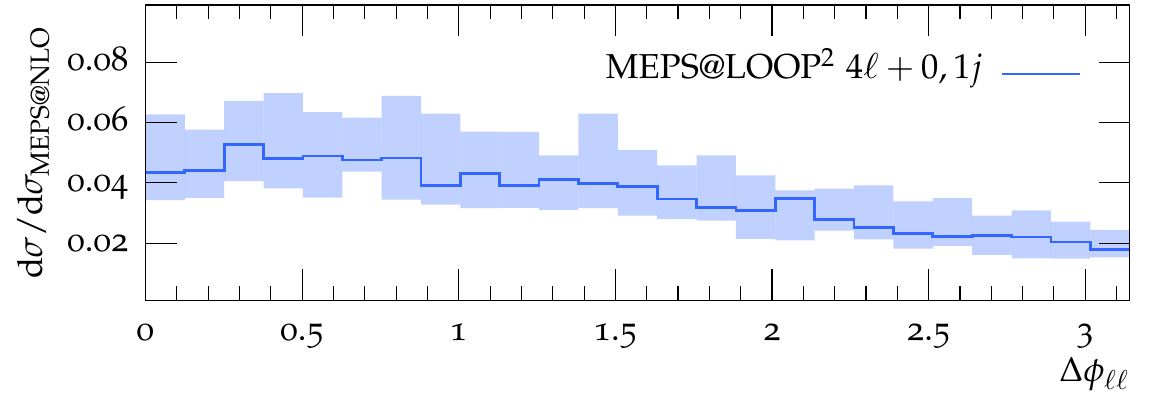}
\includegraphics[width=0.48\textwidth]{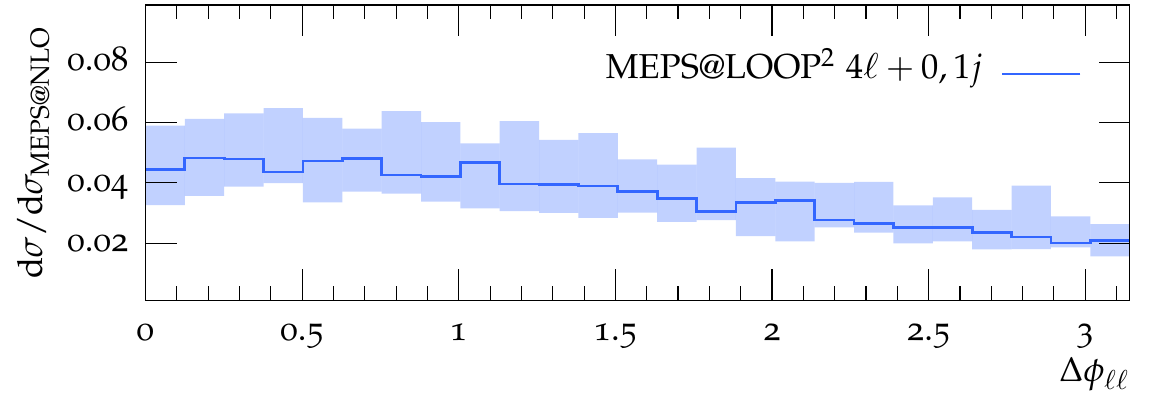}
\end{center}
\vspace*{-0.3cm}
\caption{\ATLAS (left)  and \CMS (right)
analysis at $8\UTeV$ after pre-selection cuts:
azimuthal separation of the charged leptons in the 0-jet (top) and 1-jet (bottom) bins.
Similar predictions and uncertainty bands as in \reffi{fig:flj_hww_pre_ptjets}.
}
\label{fig:flj_hww_pre_phill}
\end{figure}

\begin{figure}
\begin{center}
\includegraphics[width=0.48\textwidth]{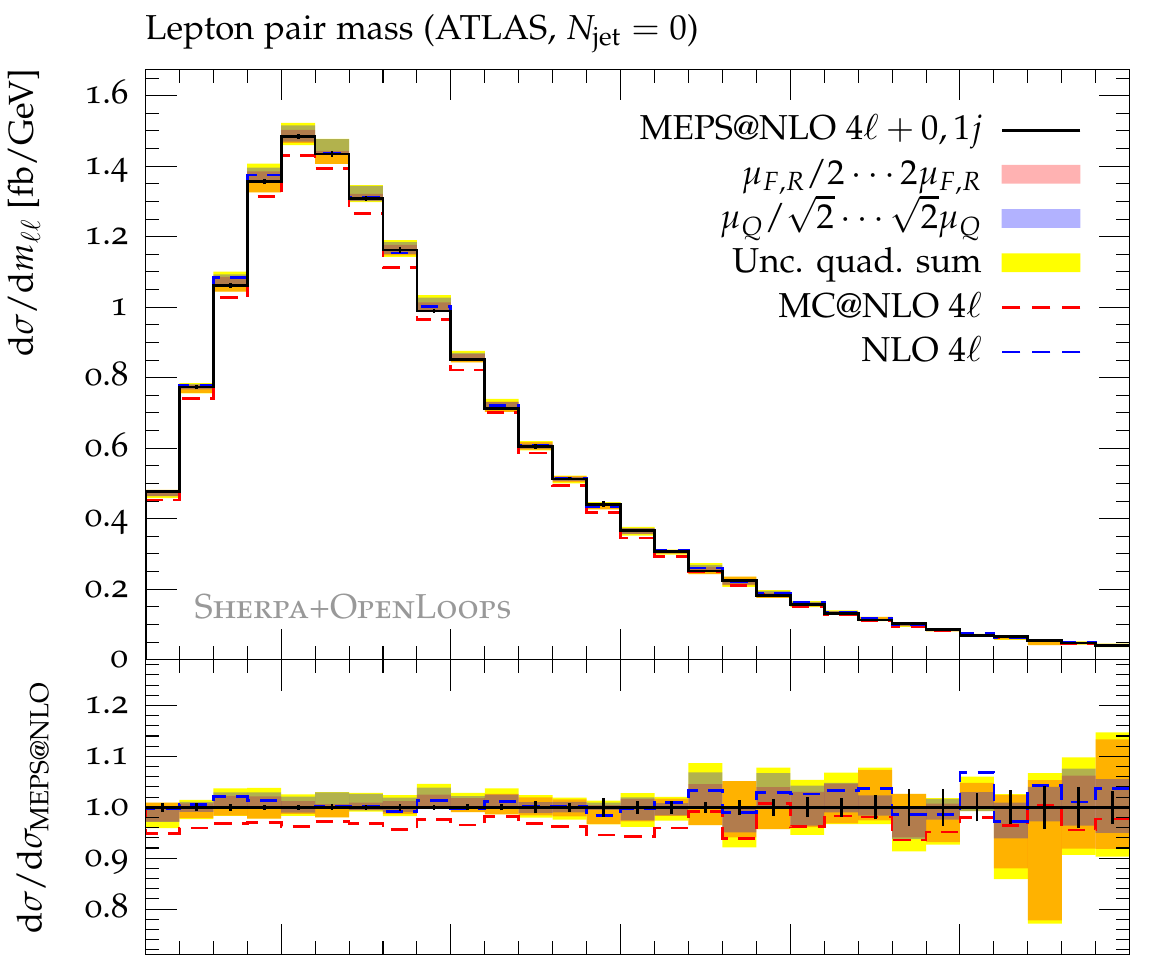}
\includegraphics[width=0.48\textwidth]{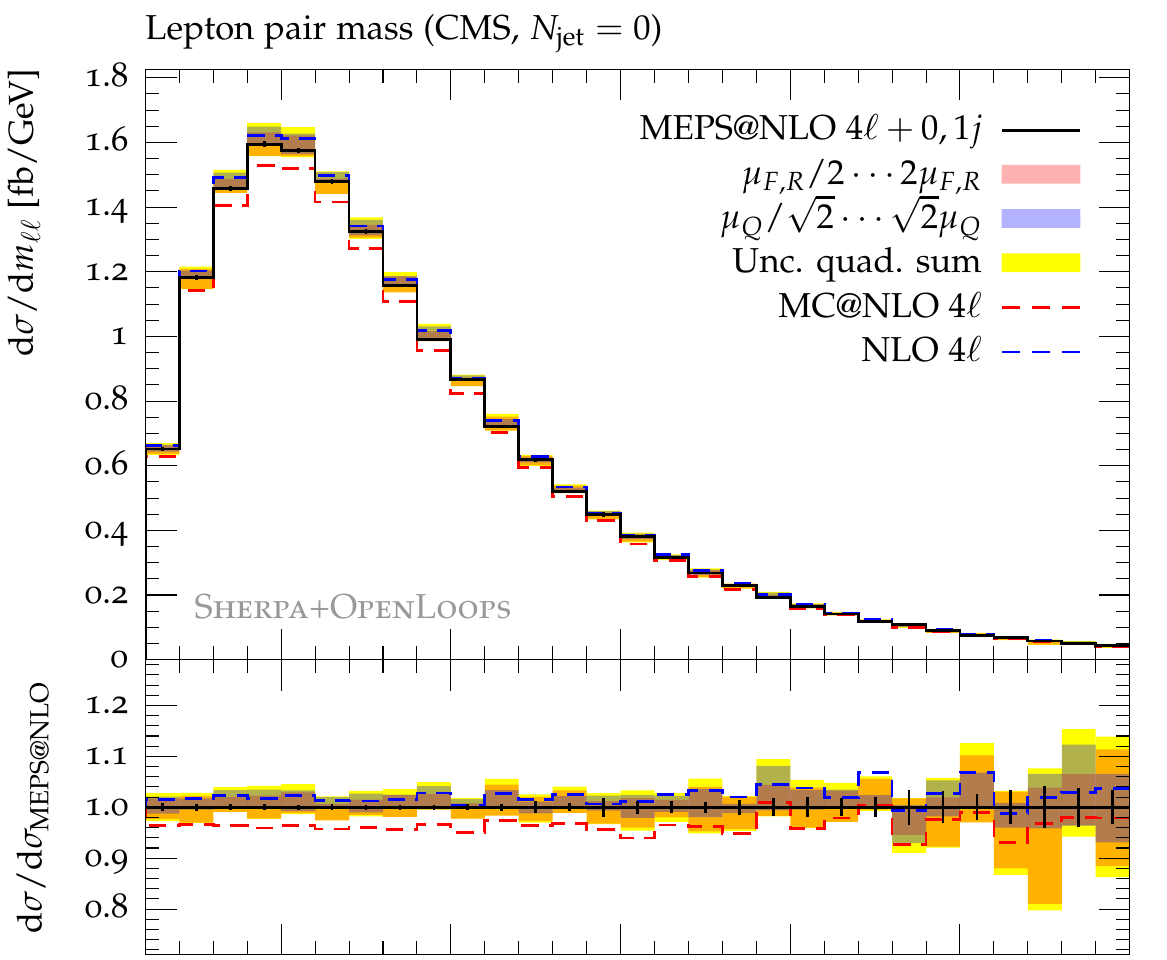}\\\vspace*{-3px}
\includegraphics[width=0.48\textwidth]{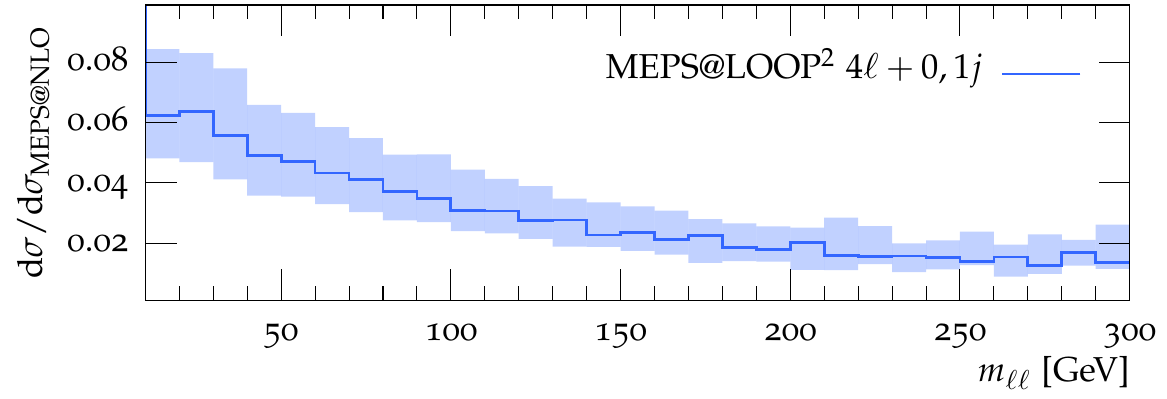}
\includegraphics[width=0.48\textwidth]{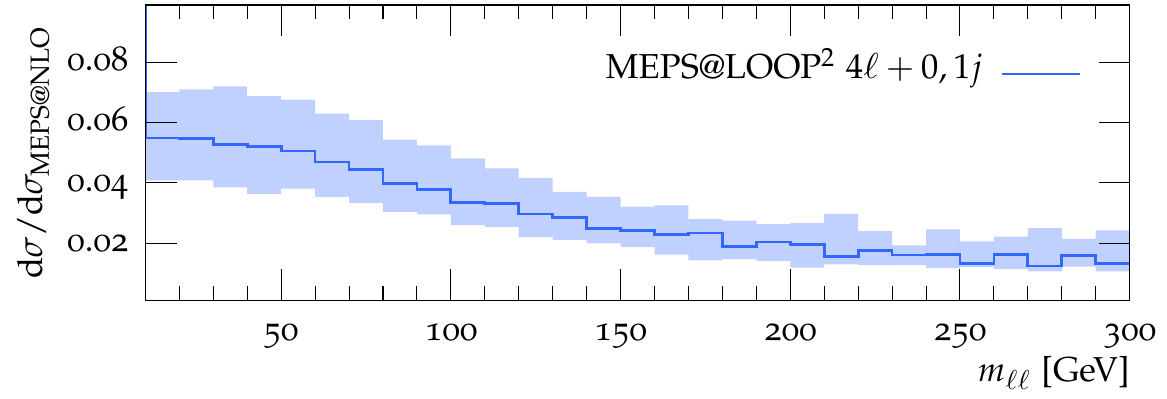}\\
\includegraphics[width=0.48\textwidth]{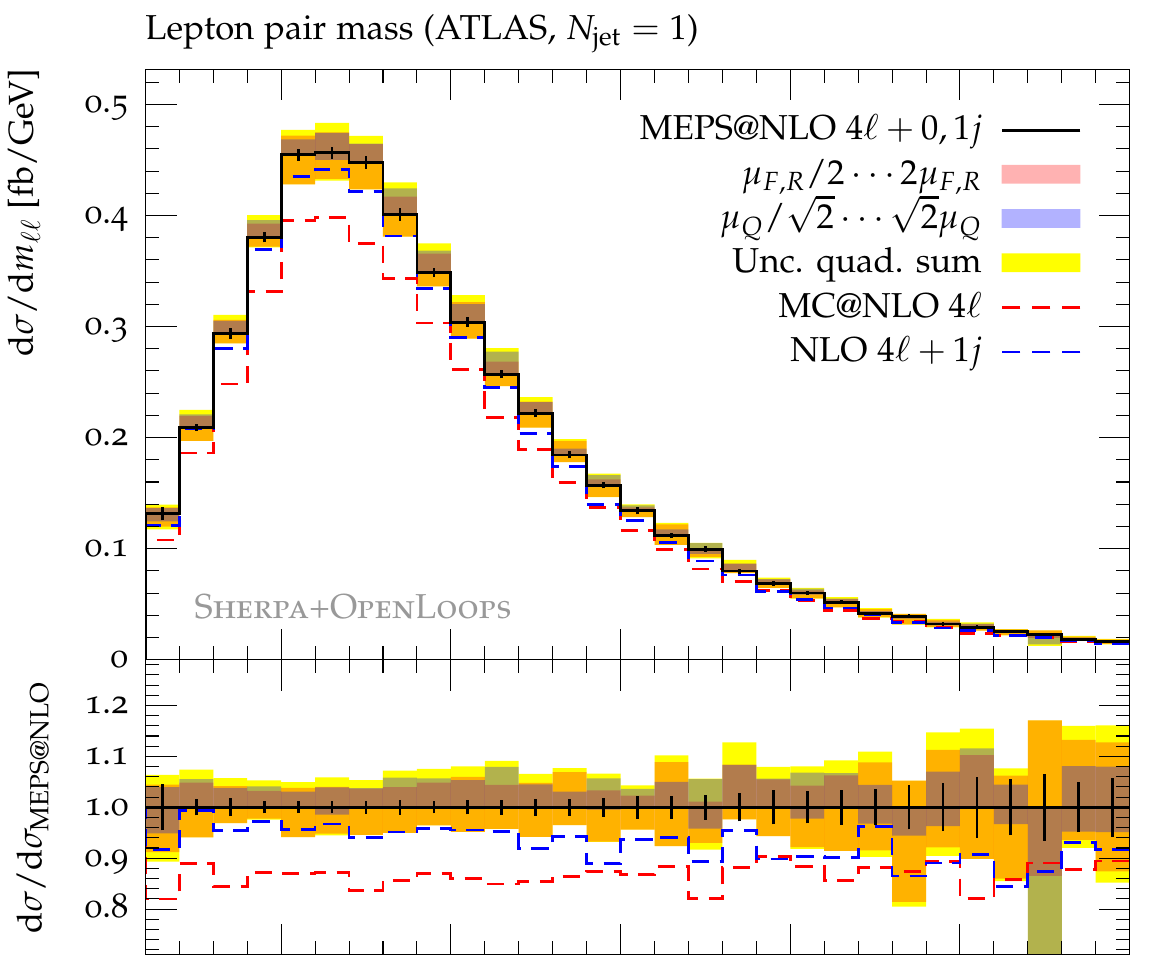}
\includegraphics[width=0.48\textwidth]{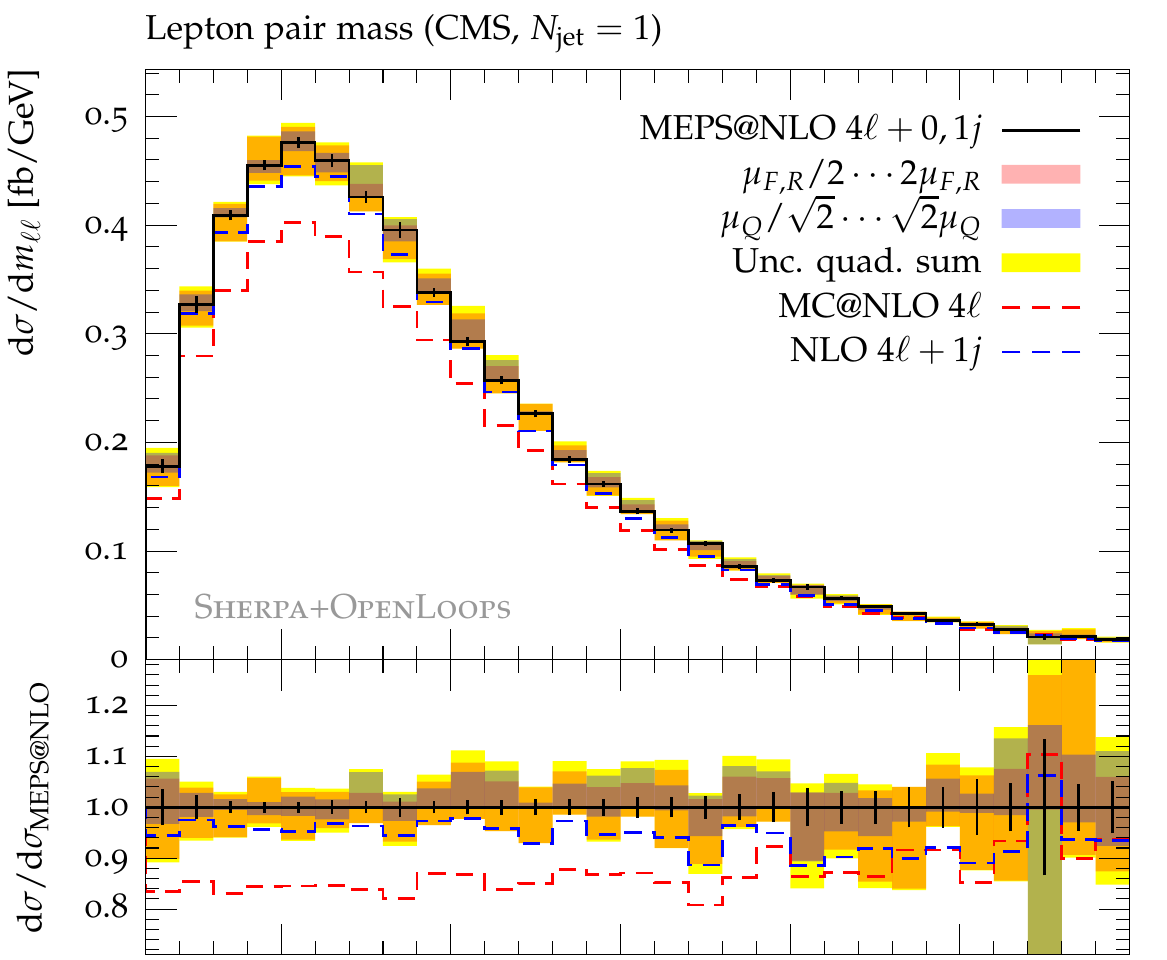}\\\vspace*{-3px}
\includegraphics[width=0.48\textwidth]{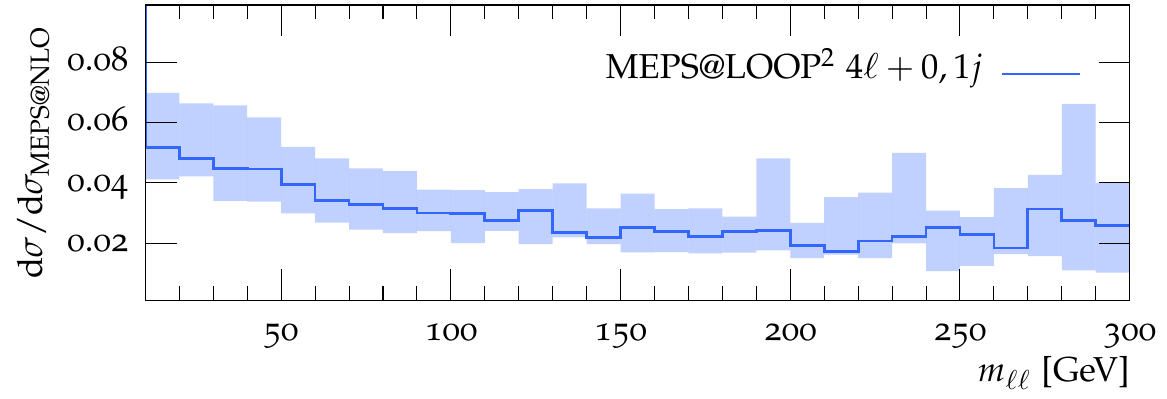}
\includegraphics[width=0.48\textwidth]{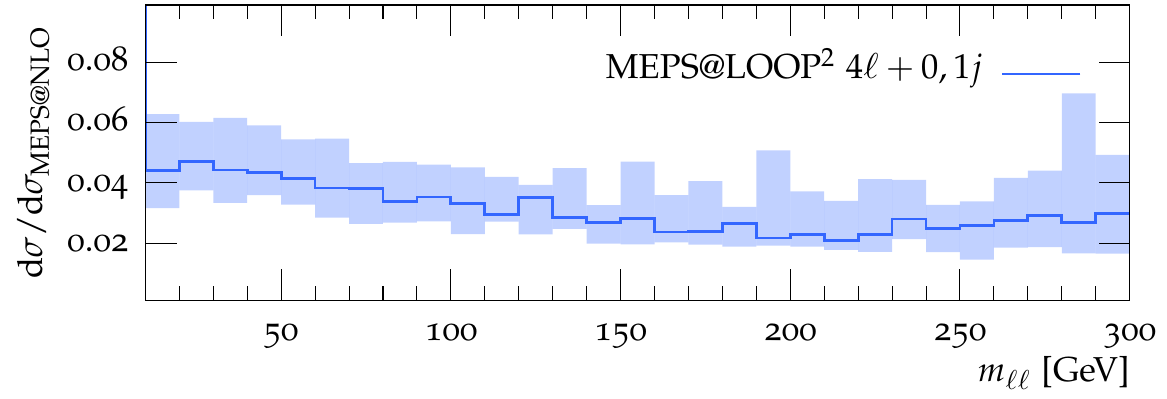}
\end{center}
\vspace*{-0.3cm}
\caption{\ATLAS (left)  and \CMS (right)
analysis at $8\UTeV$ after pre-selection cuts:
dilepton invariant mass distribution in the 0-jet (top) and 1-jet (bottom) bins.
Similar predictions and uncertainty bands as in \reffi{fig:flj_hww_pre_ptjets}.
} 
\label{fig:flj_hww_pre_mll}
\end{figure}

In \reffis{fig:flj_hww_pre_ptjets}--\ref{fig:flj_hww_pre_mll}
we present jet and lepton observables after pre-selection cuts.
The curves for \MEPSatNLO, \MCatNLO, \NLOacc, and \MEPSatLOOPSQ
correspond to the central scale choice \refeq{eq:scale}.
The middle and lower panels show relative \MCatNLO and \NLOacc
deviations from \MEPSatNLO, 
and squared quark-loop contributions normalised to the central
\MEPSatNLO result. 
Scale-variation bands are shown only for \MEPSatLOOPSQ and \MEPSatNLO.
In the latter case, renormalisation- and factorisation-scale variations $\delqcd$ (red band), 
resummation-scale variations
$\delres$ (blue band), and their combination in quadrature $\deltot=(\delqcd^2+\delres^2)^{1/2}$ (yellow band),
are displayed as colour-additive regions.
The various band regions assume different colours corresponding to the various possible overlaps.
The band boundary, corresponding to variations $\delta$ in the range $\delqcd,\delres<\delta<\deltot$, is yellow.
Orange areas appear in kinematic regions dominated by QCD-scale variations ($\delres<\delta<\delqcd$),
while green areas reflect dominant resummation-scale variations ($\delqcd<\delta<\delres$), 
and the central band area ($\delta<\delres,\delqcd$), where all three colours overlap, is brown.
Note that scale-variation bands are somewhat distorted by statistical
fluctuations, which tend to increase in the tails of some distributions.

Before splitting the event sample into exclusive jet bins,
in \reffi{fig:flj_hww_pre_ptjets} we show the transverse momenta of the 
hardest (upper plots) and second-hardest (lower plots) jet. 
Here all \NLOacc curves correspond to $4\ell+1$ jet production.
In the case of the first jet, \MCatNLO predictions are only LO accurate and 
significantly underestimate the tail of the $p_\rT$ distribution.
On the other hand, \NLOacc predictions feature a 20\% excess at high $p_\rT$.
As already observed in \reffi{fig:inclusive_merging}.a,
this behaviour can be explained by the fact that the scale \refeq{eq:scale} 
used in the fixed-order calculation does not adapt to the transverse momentum of the jet.

In the case of the second-jet $p_\rT$, \NLOacc and \MEPSatNLO results are both only LO accurate,
and the shape differences at large $p_\rT$ are more pronounced but qualitatively similar as for
the first jet. The excess of the \NLOacc distribution below $10\UGeV$ reveals the presence of the
infrared singularity at $p_\rT\to 0$. The \MCatNLO prediction for the second jet
is entirely based on the shower approximation.  It remains low over the
entire spectrum, and above $30\UGeV$ the deficit starts to be considerable.

The increase of \MEPSatNLO scale variations from a few percent for the first
jet to 10\% for the second one, is due to the transition from NLO to LO
accuracy.  The abundance of orange and brown areas in the \MEPSatNLO bands
indicates that the uncertainty tends to be dominated by QCD-scale
variations.  Green band areas, which correspond to larger resummation-scale
uncertainties, show up less frequently and only in the leading-jet $p_\perp$
distribution.  Even in the small-$p_\rT$ region, where Sudakov logarithms
have the highest possible impact, QCD- and resummation-scale variations do
not exceed 10\%.  This suggests that subleading-logarithmic corrections
beyond the \MEPSatNLO accuracy should be rather modest.

Squared quark-loop corrections range from 1 to 6 percent and feature a more
pronounced dependence on the jet $p_\rT$ as compared to the inclusive
analysis (\cf \reffi{fig:inclusive_loop2_mepsnlo}). The largest
effects arise around $p_\rT\simeq 20\UGeV$, which corresponds to the 0-jet bin of the
$\hww$ analysis.

Let us now switch to leptonic observables in the exclusive 0- and 1-jet bins
of the $\hww$ analyses.  Distributions in the azimuthal dilepton
separation $\Delta \phi_{\ell\ell'}$ and in the dilepton invariant mass
$m_{\ell\ell'}$ are displayed in \reffis{fig:flj_hww_pre_phill} and \ref{fig:flj_hww_pre_mll}. 
These observables play an important role for the description of the background acceptance 
and for the optimisation of the Higgs-boson sensitivity in the experimental analyses.  
The corresponding \MEPSatNLO distributions are NLO accurate in 
both jet bins. This is very well reflected by the \MEPSatNLO uncertainty bands,
which do not exceed the few-percent level.
Also here, resummation-scale variations tend to be slightly subdominant with
respect to QCD-scale variations.  Comparing \NLOacc, \MCatNLO and \MEPSatNLO
distributions in the 0-jet bin, where none of these approximations loses
NLO accuracy, we find overall agreement at the few-percent level.
In the 1-jet bin, the agreement between \NLOacc and \MEPSatNLO remains, as expected,  quite good.
Due to the lack of NLO accuracy, inclusive \MCatNLO predictions feature the characteristic 
10--15\% deficit in the 1-jet bin, which is accompanied by minor shape distortions.
Given the good agreement with \NLOacc within the small uncertainty band,
the shape of \MEPSatNLO distributions seems to be very well under control.

In the 0-jet bin, \MEPSatLOOPSQ corrections are very sensitive
both to the azimuthal separation and to the invariant mass of the dilepton system.
At small $\Delta \phi_{\ell\ell'}$ and $m_{\ell\ell'}$, which corresponds to the Higgs-signal region,
they reach up to 8\% and 6\%, respectively. A similar but weaker sensitivity is
visible also in the 1-jet bin.

Inspecting the transverse-momentum distributions of the harder and softer charged lepton 
(not shown here) we found that the various NLO corrections behave very similarly as for 
$\Delta \phi_{\ell\ell'}$ and $m_{\ell\ell'}$, while squared quark-loop corrections
are less sensitive to the lepton-$p_\rT$ and vary between $2\%$ and $4\%$ only.

\subsection{Kinematic distributions in control and signal regions}
\label{se:hWW_dist}

\begin{figure}
\begin{center}
\includegraphics[width=0.48\textwidth]{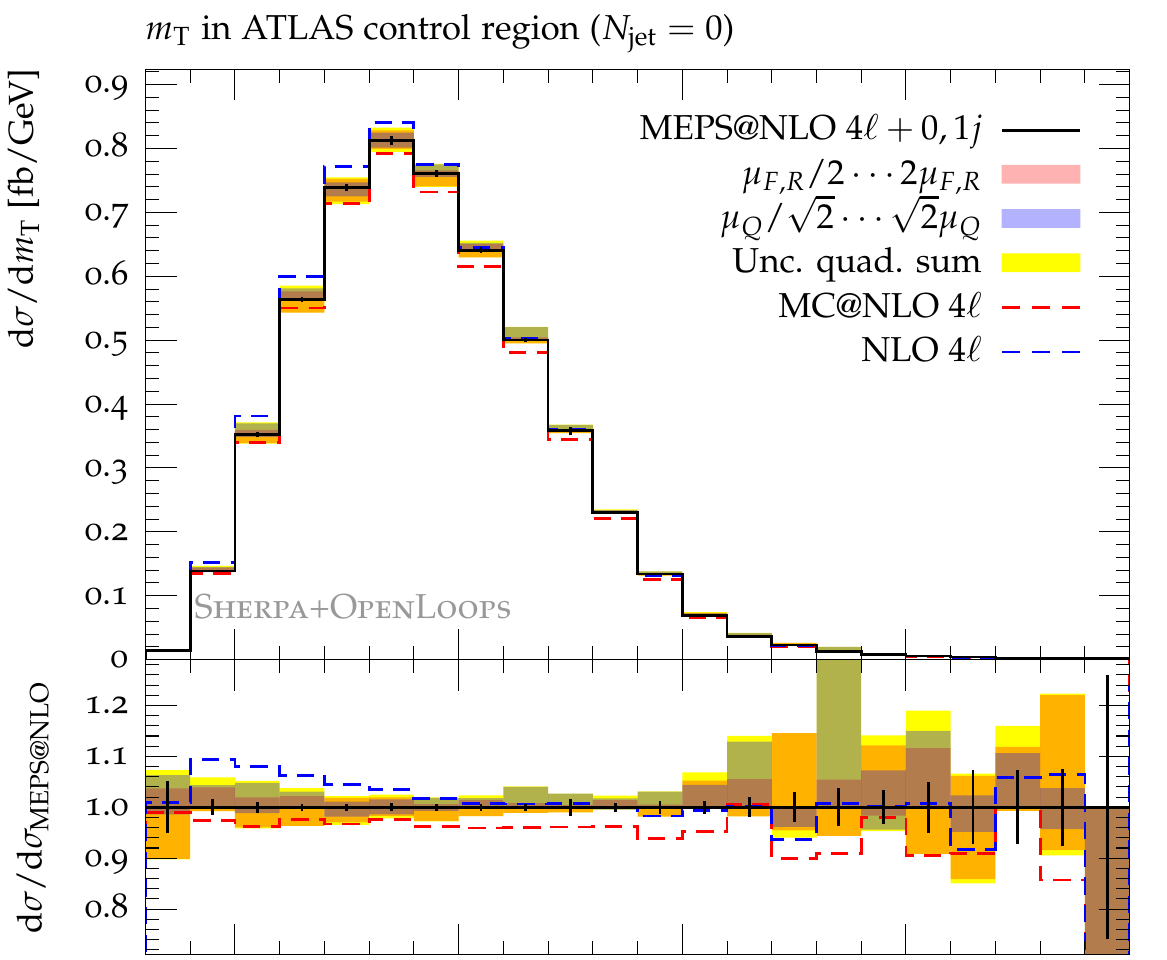}
\includegraphics[width=0.48\textwidth]{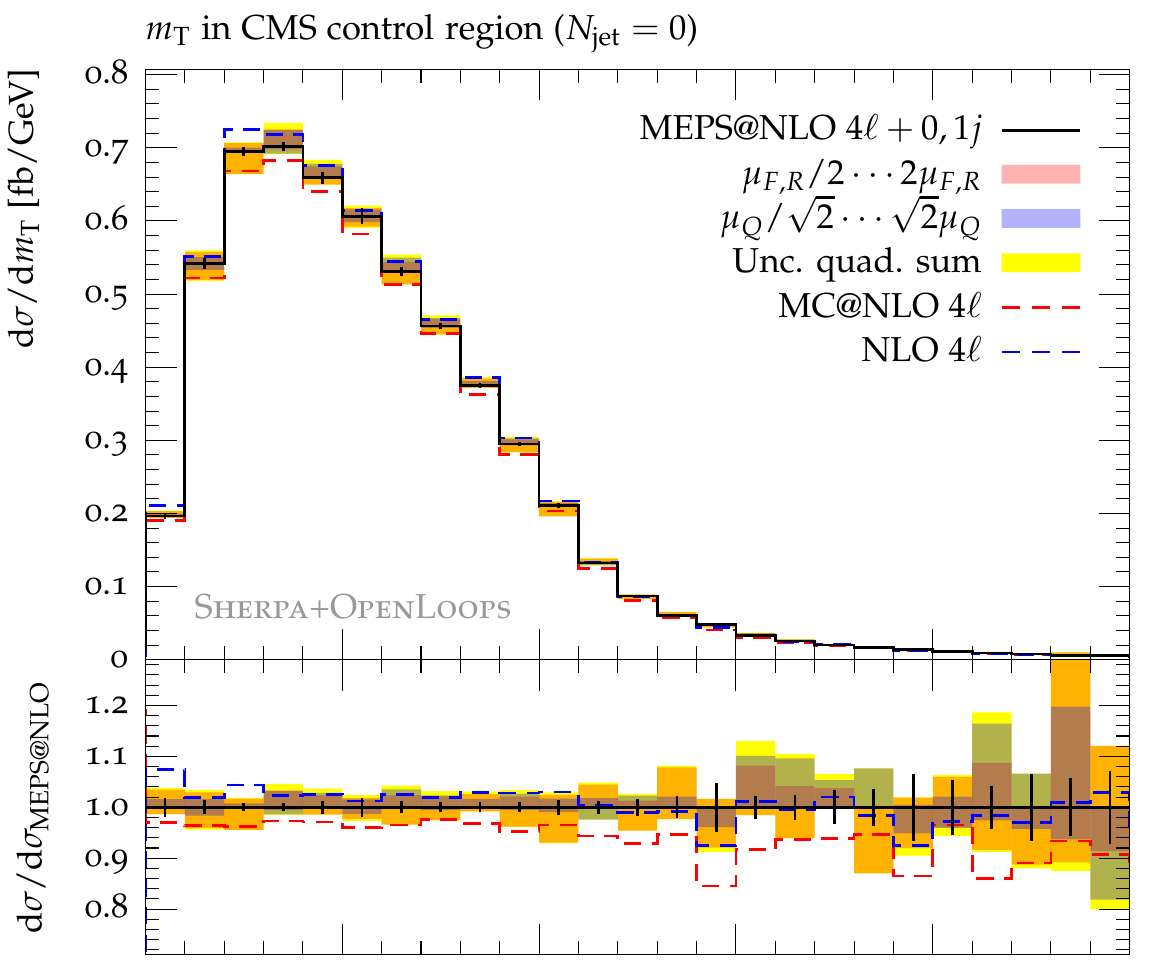}\\\vspace*{-3px}
\includegraphics[width=0.48\textwidth]{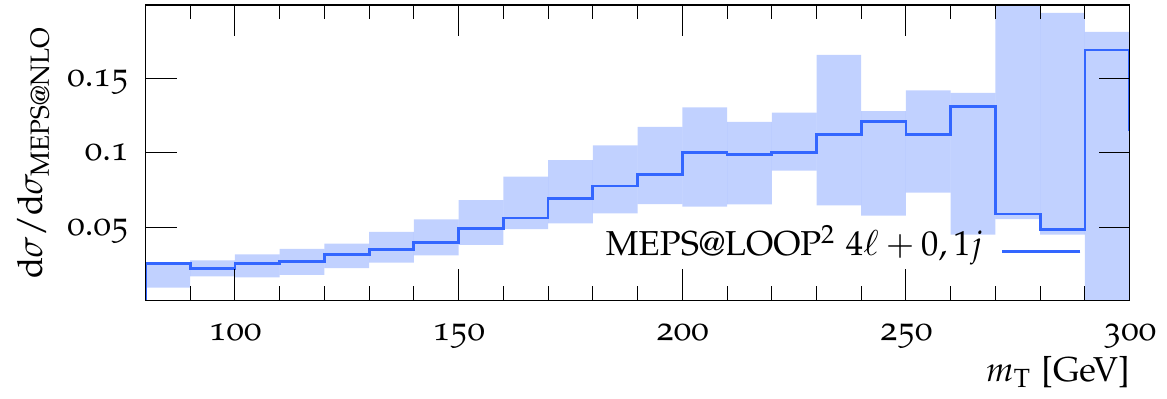}
\includegraphics[width=0.48\textwidth]{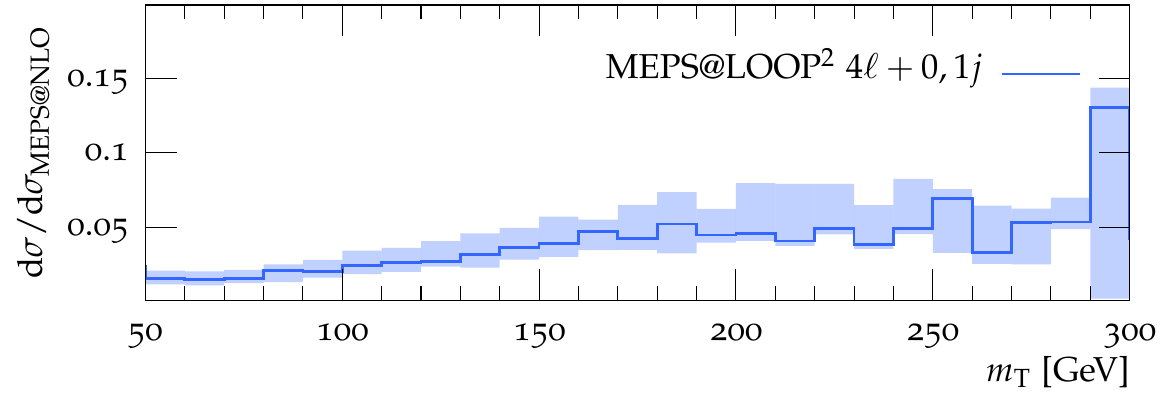}
\includegraphics[width=0.48\textwidth]{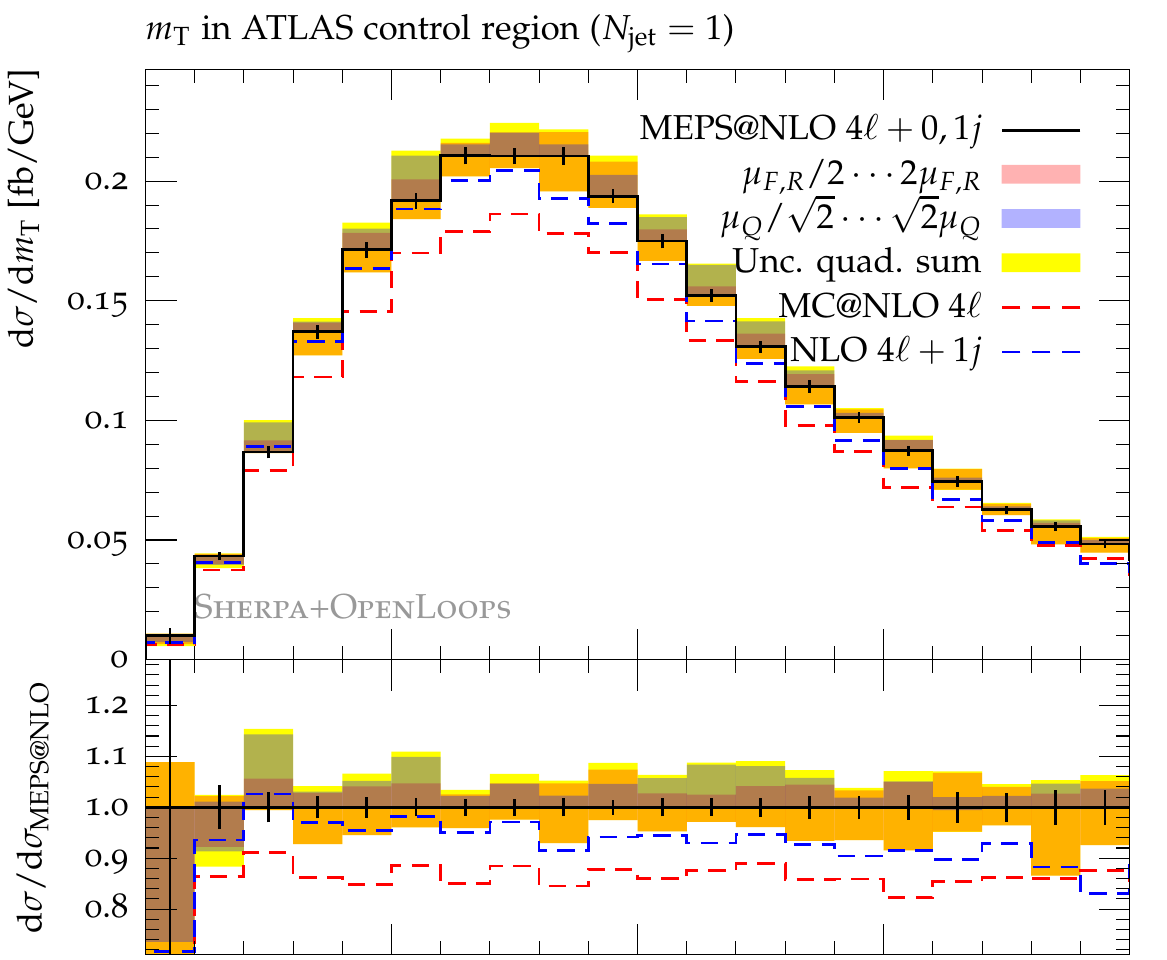}
\includegraphics[width=0.48\textwidth]{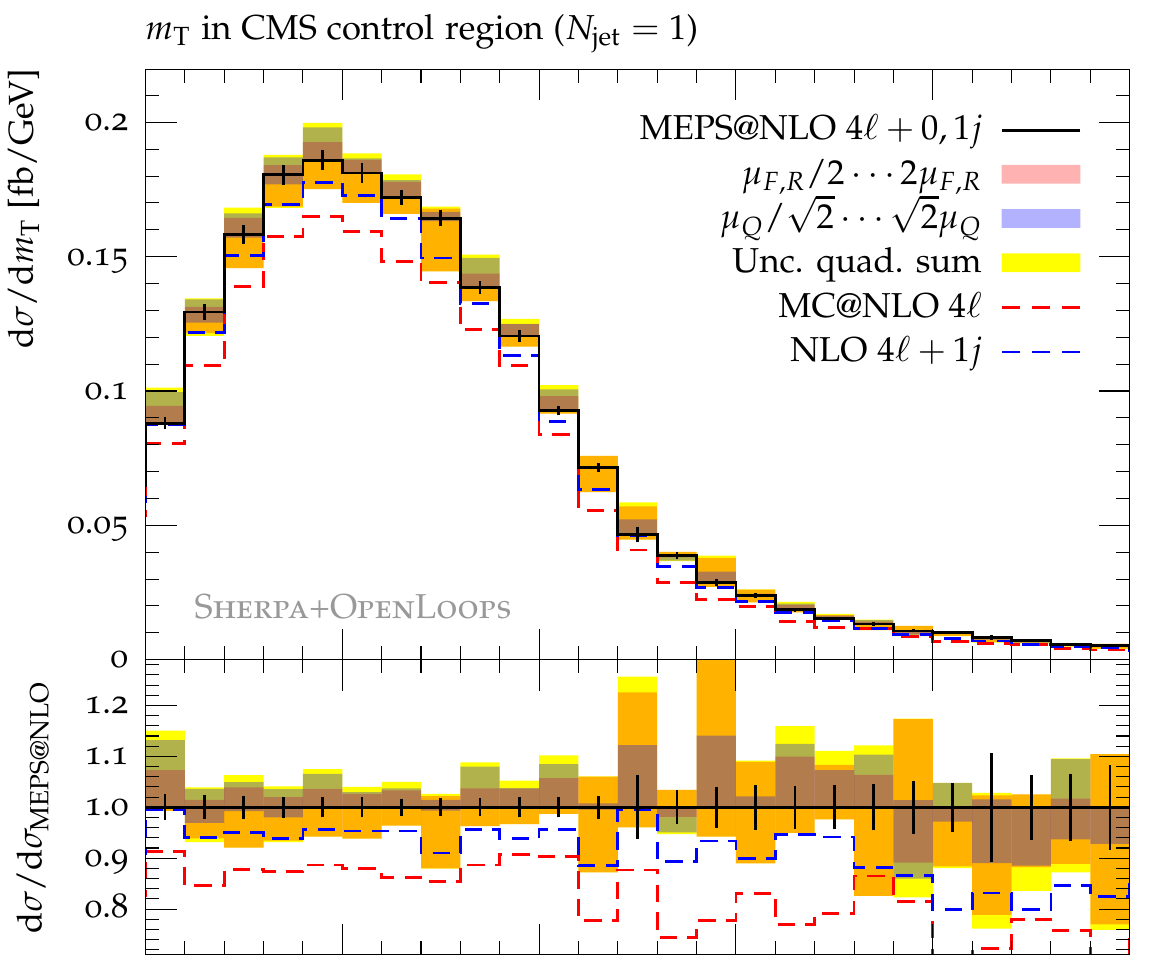}\\\vspace*{-3px}
\includegraphics[width=0.48\textwidth]{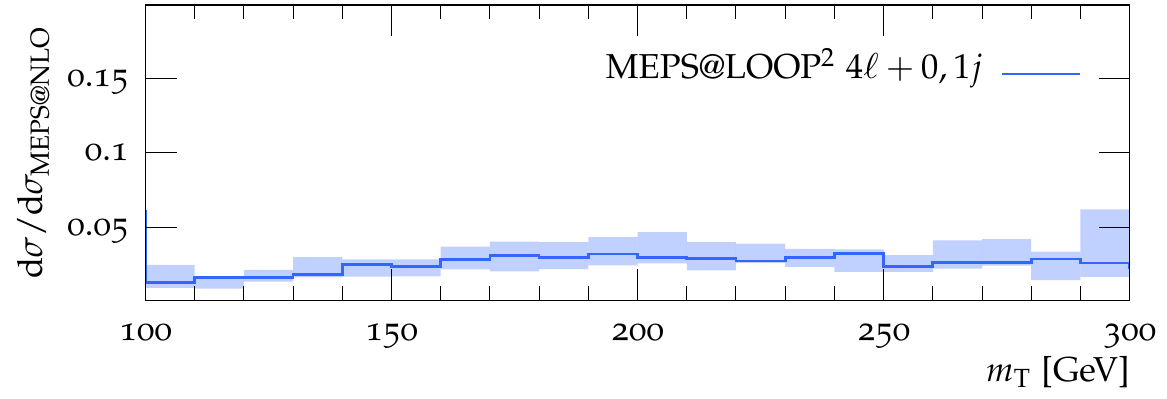}
\includegraphics[width=0.48\textwidth]{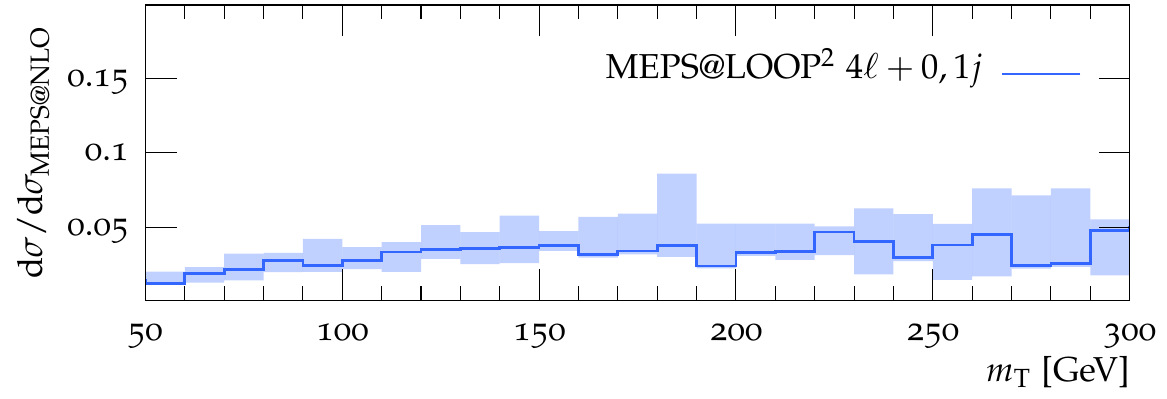}
\end{center}
\vspace*{-0.3cm}
\caption{Control region of the \ATLAS (left)  and \CMS (right)
analysis at $8\UTeV$:
  transverse-mass distribution in the 0-jet (top) and 1-jet (bottom) bins.
Similar predictions and uncertainty bands as in \reffi{fig:flj_hww_pre_ptjets}.
} 
\label{fig:flj_hww_c_mT}
\end{figure}

\begin{figure}
\begin{center}
\includegraphics[width=0.48\textwidth]{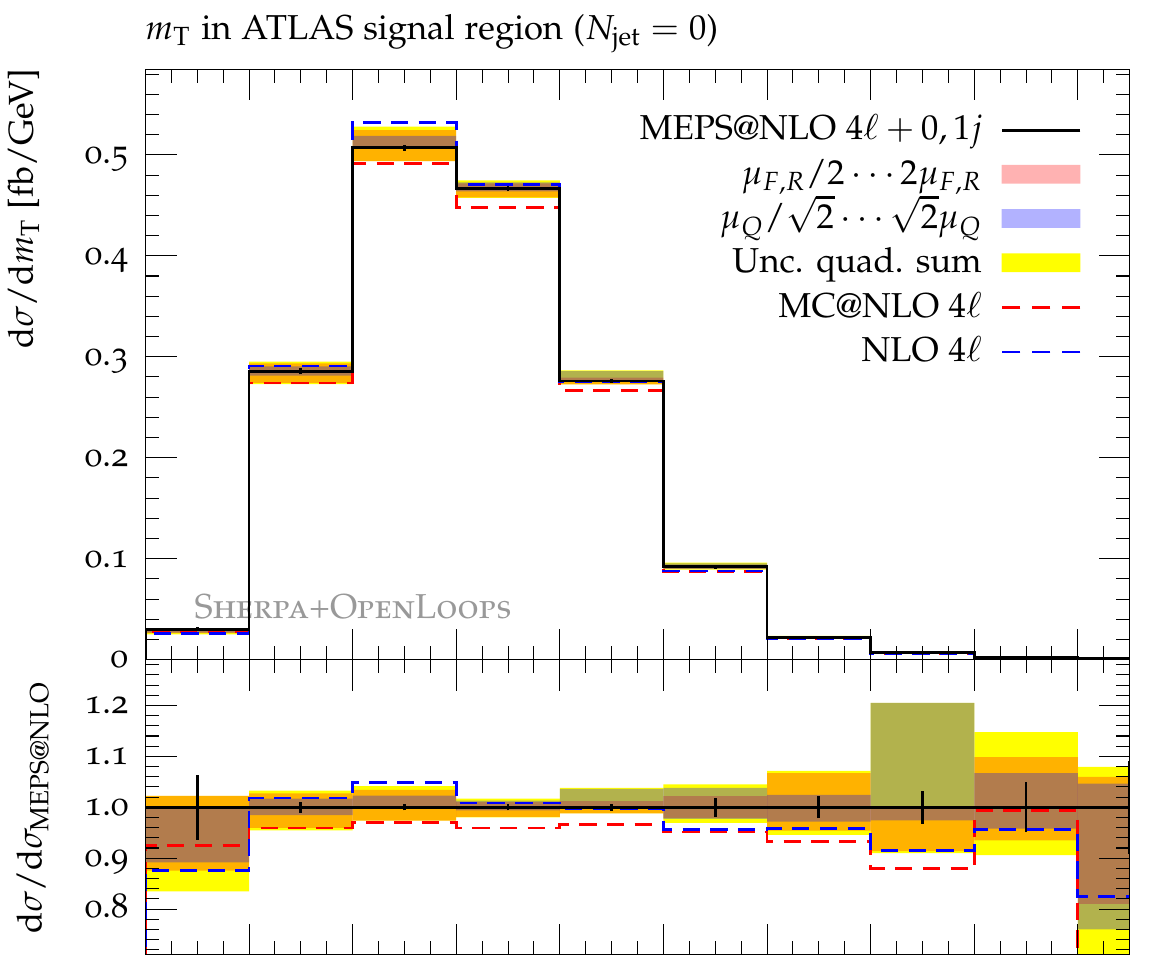}
\includegraphics[width=0.48\textwidth]{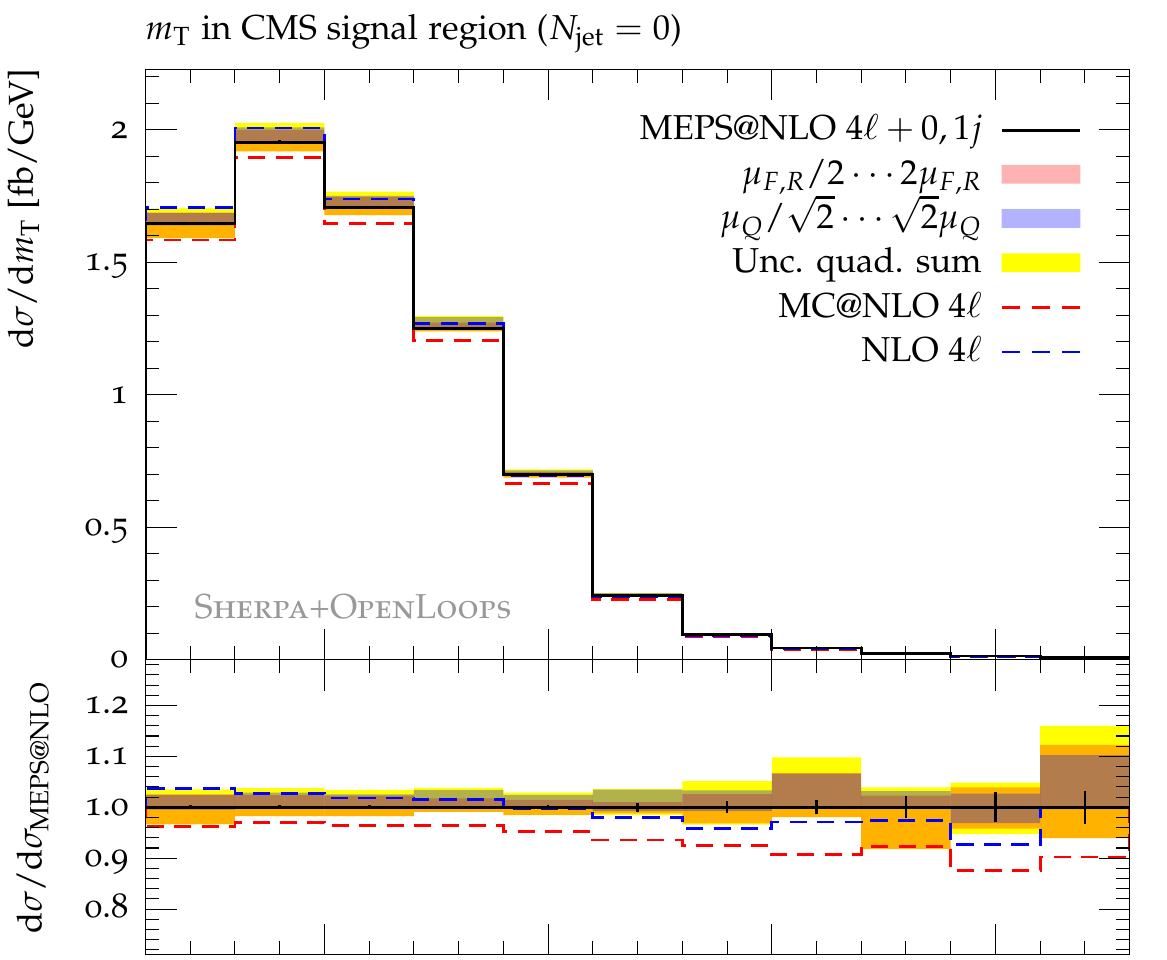}\\\vspace*{-3px}
\includegraphics[width=0.48\textwidth]{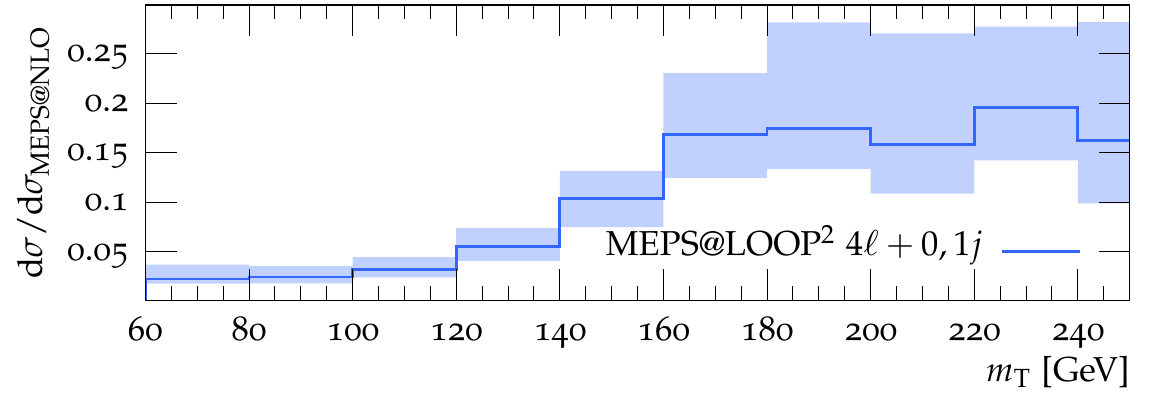}
\includegraphics[width=0.48\textwidth]{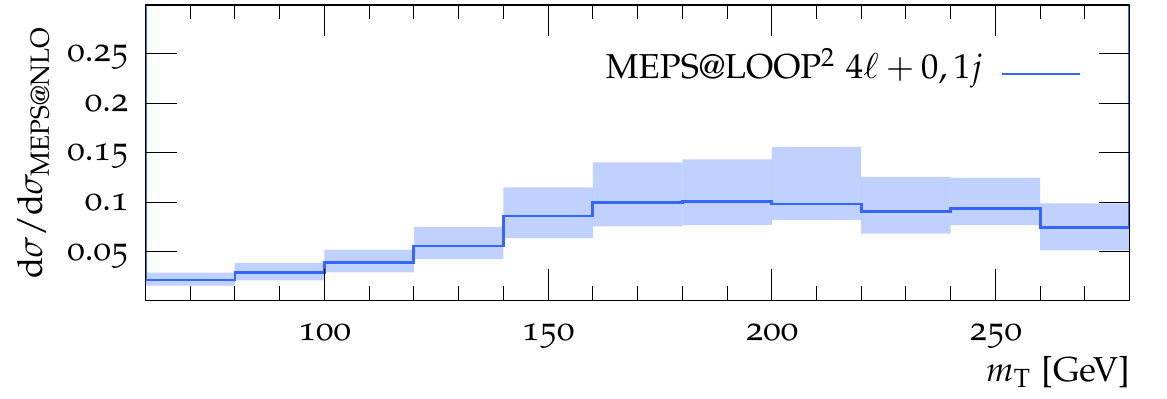}
\includegraphics[width=0.48\textwidth]{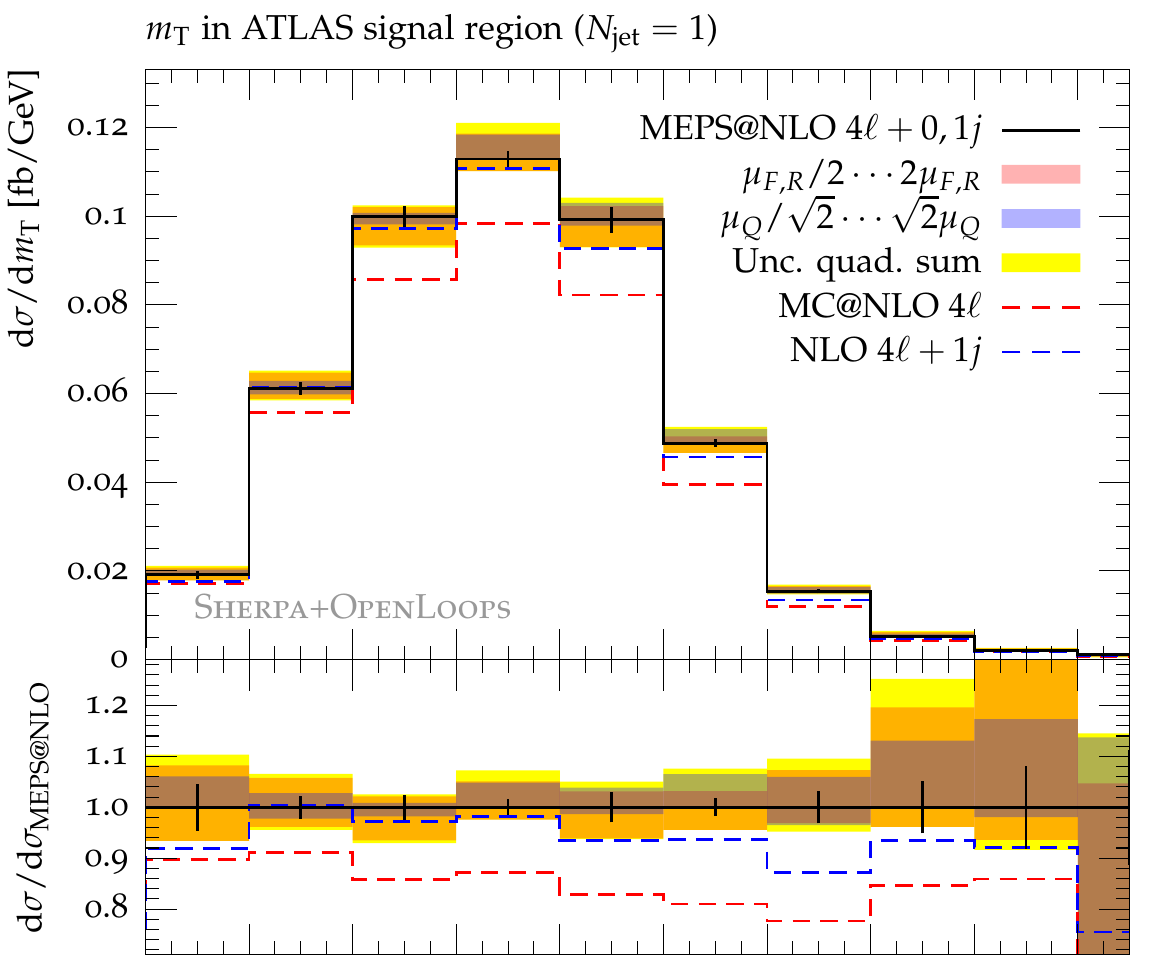}
\includegraphics[width=0.48\textwidth]{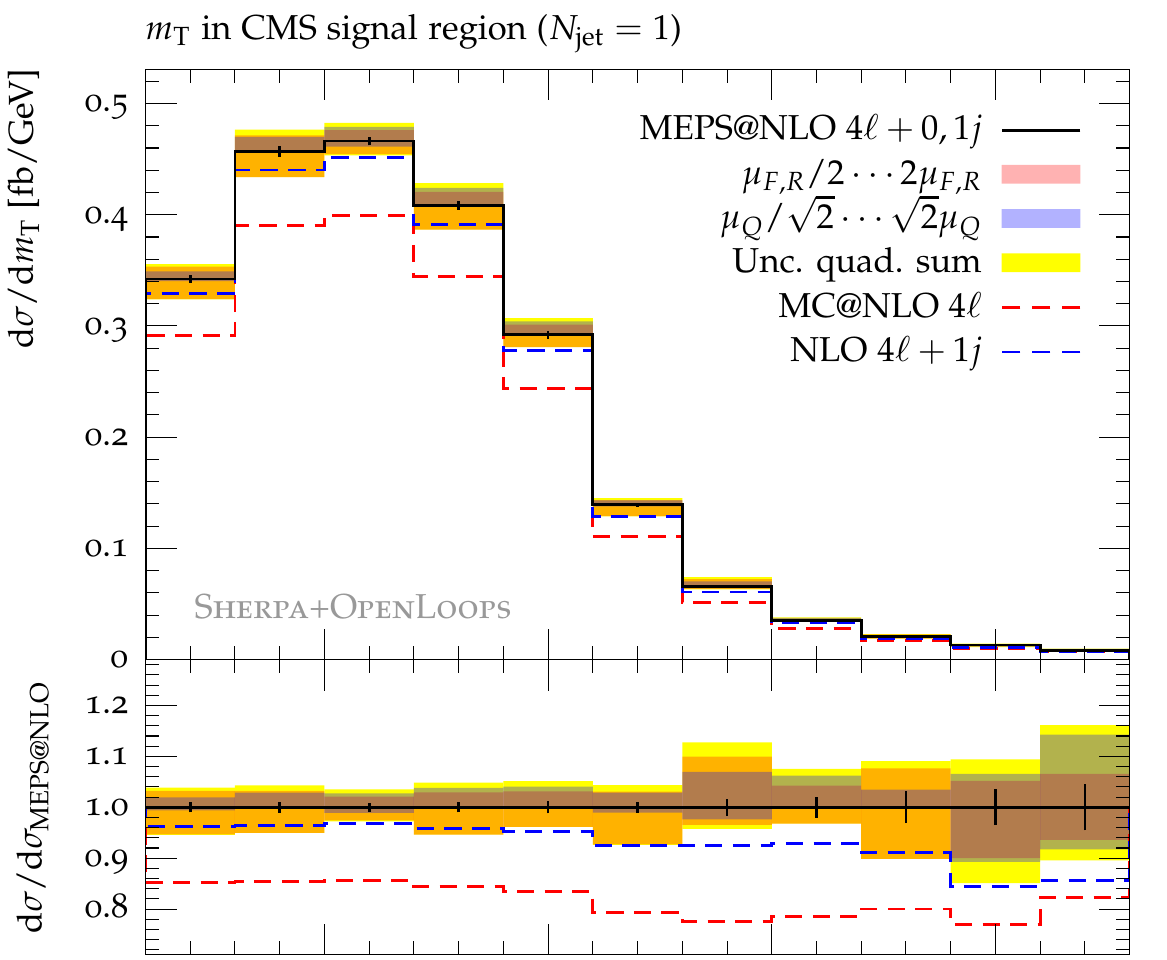}\\\vspace*{-3px}
\includegraphics[width=0.48\textwidth]{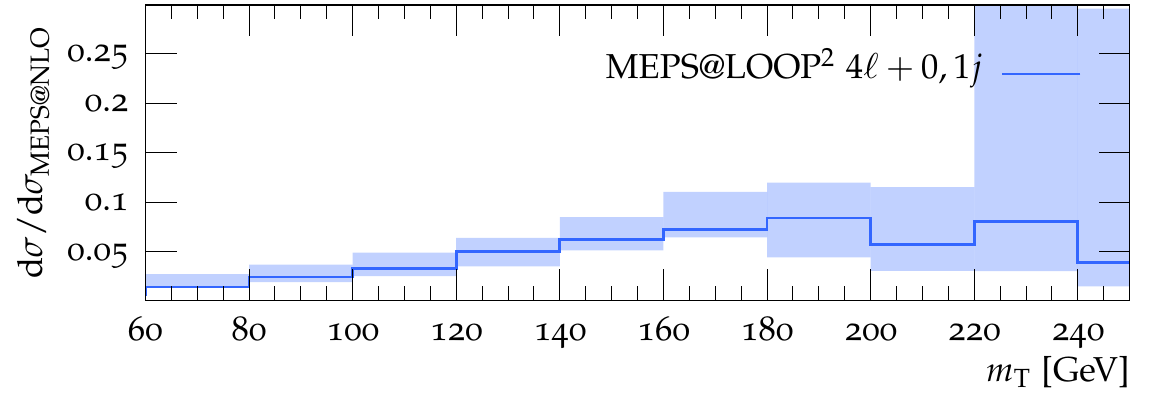}
\includegraphics[width=0.48\textwidth]{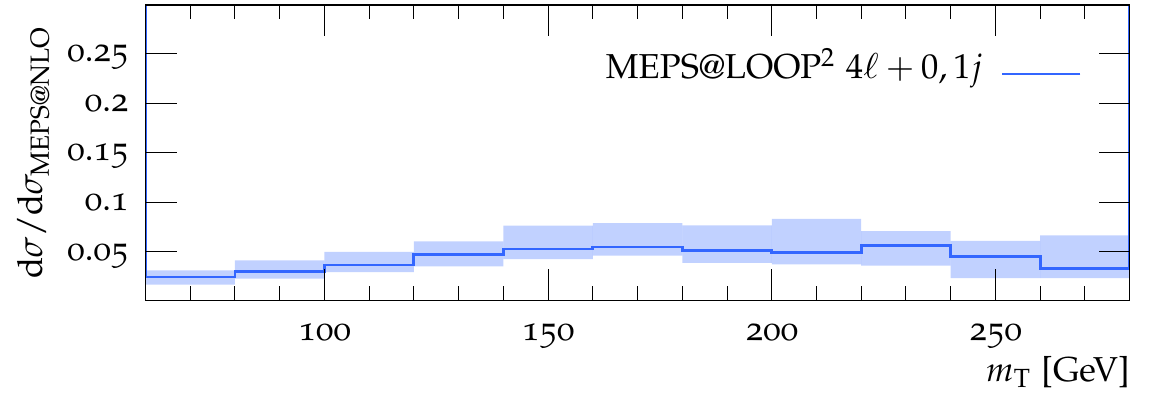}
\end{center}
\vspace*{-0.3cm}
\caption{
Signal region of the \ATLAS (left)  and \CMS (right)
analysis at $8\UTeV$:
  transverse-mass distribution in the 0-jet (top) and 1-jet (bottom) bins.
Similar predictions and uncertainty bands as in \reffi{fig:flj_hww_pre_ptjets}.
}
\label{fig:flj_hww_s_mT}
\end{figure}

\begin{figure}
\begin{center}
\includegraphics[width=0.48\textwidth]{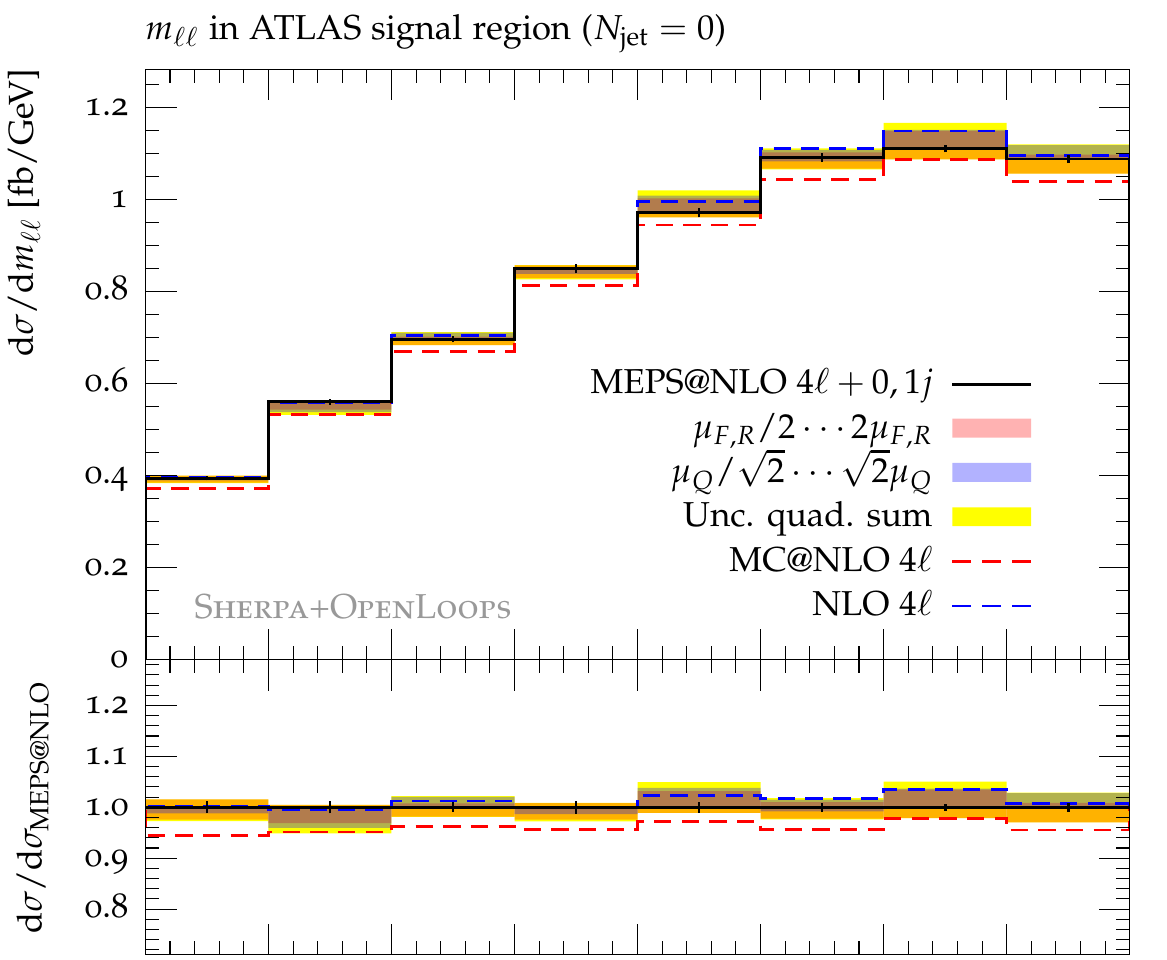}
\includegraphics[width=0.48\textwidth]{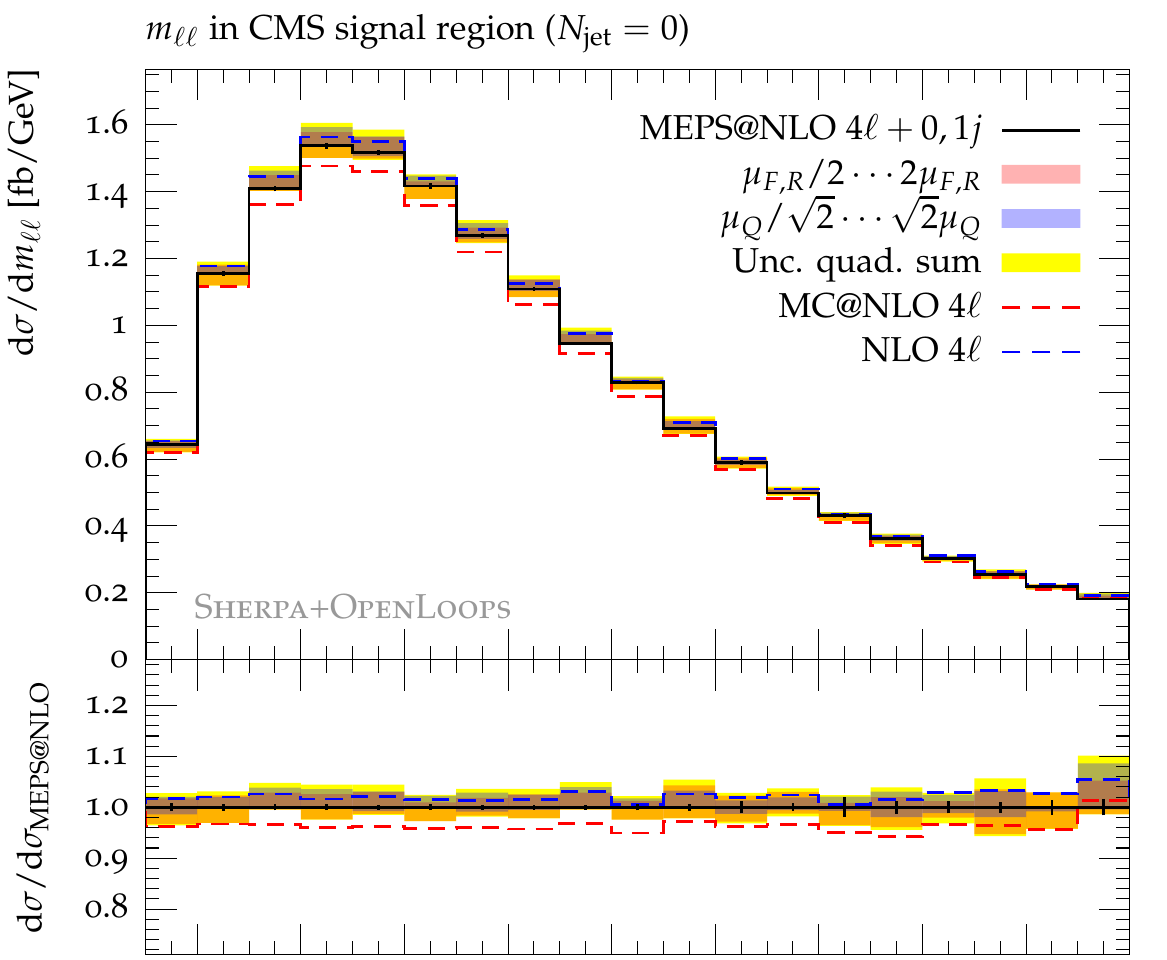}\\\vspace*{-3px}
\includegraphics[width=0.48\textwidth]{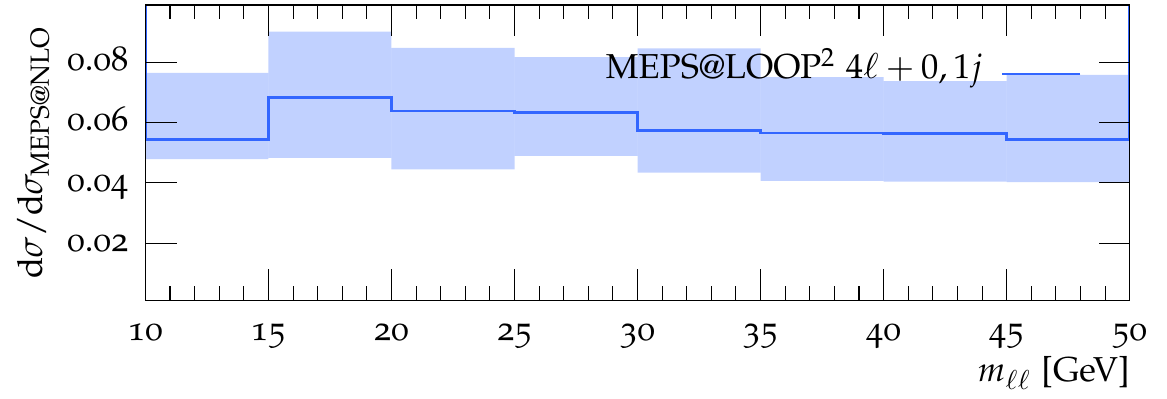}
\includegraphics[width=0.48\textwidth]{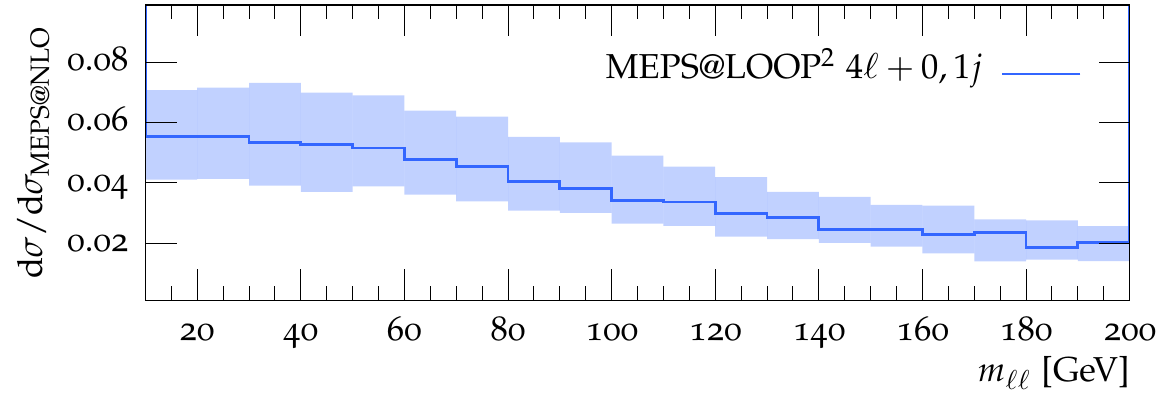}
\includegraphics[width=0.48\textwidth]{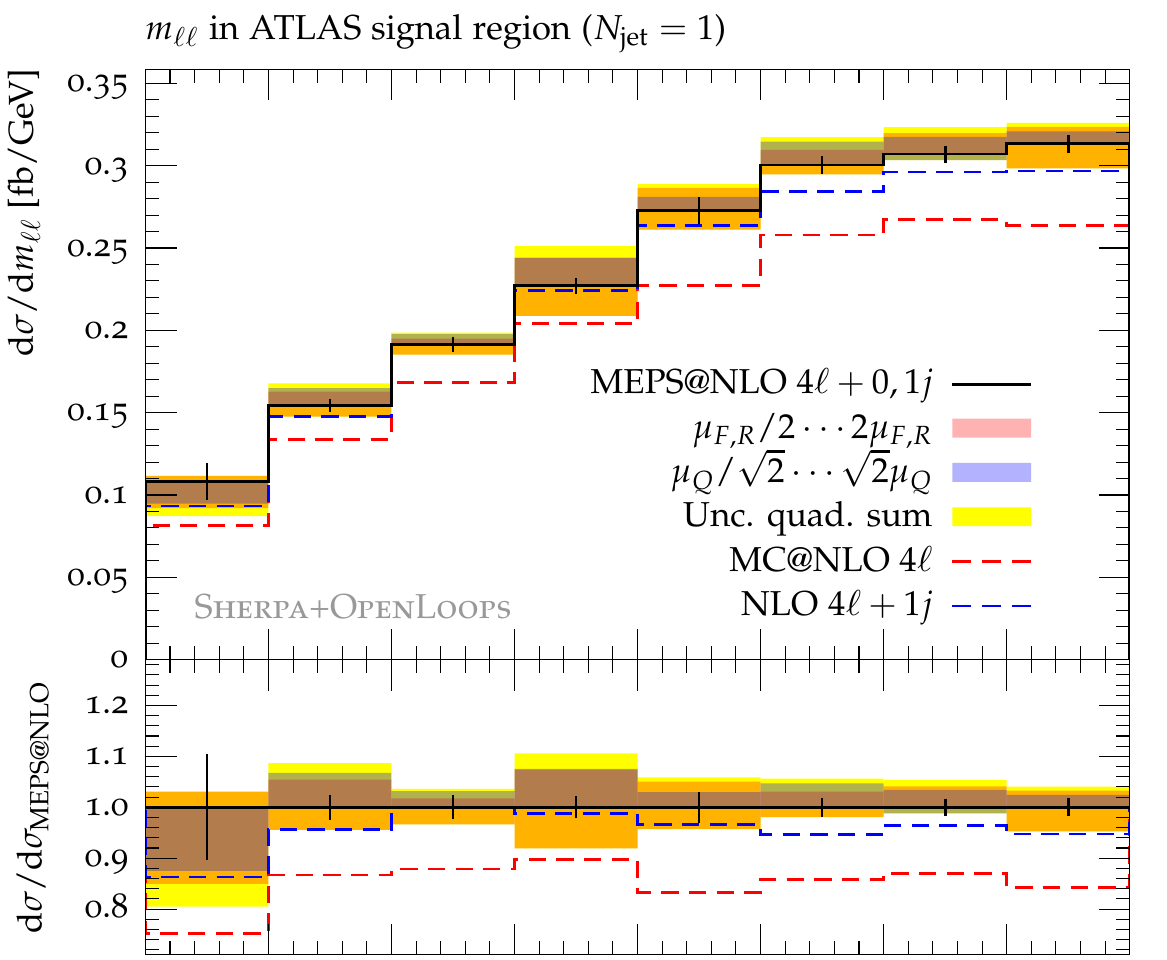}
\includegraphics[width=0.48\textwidth]{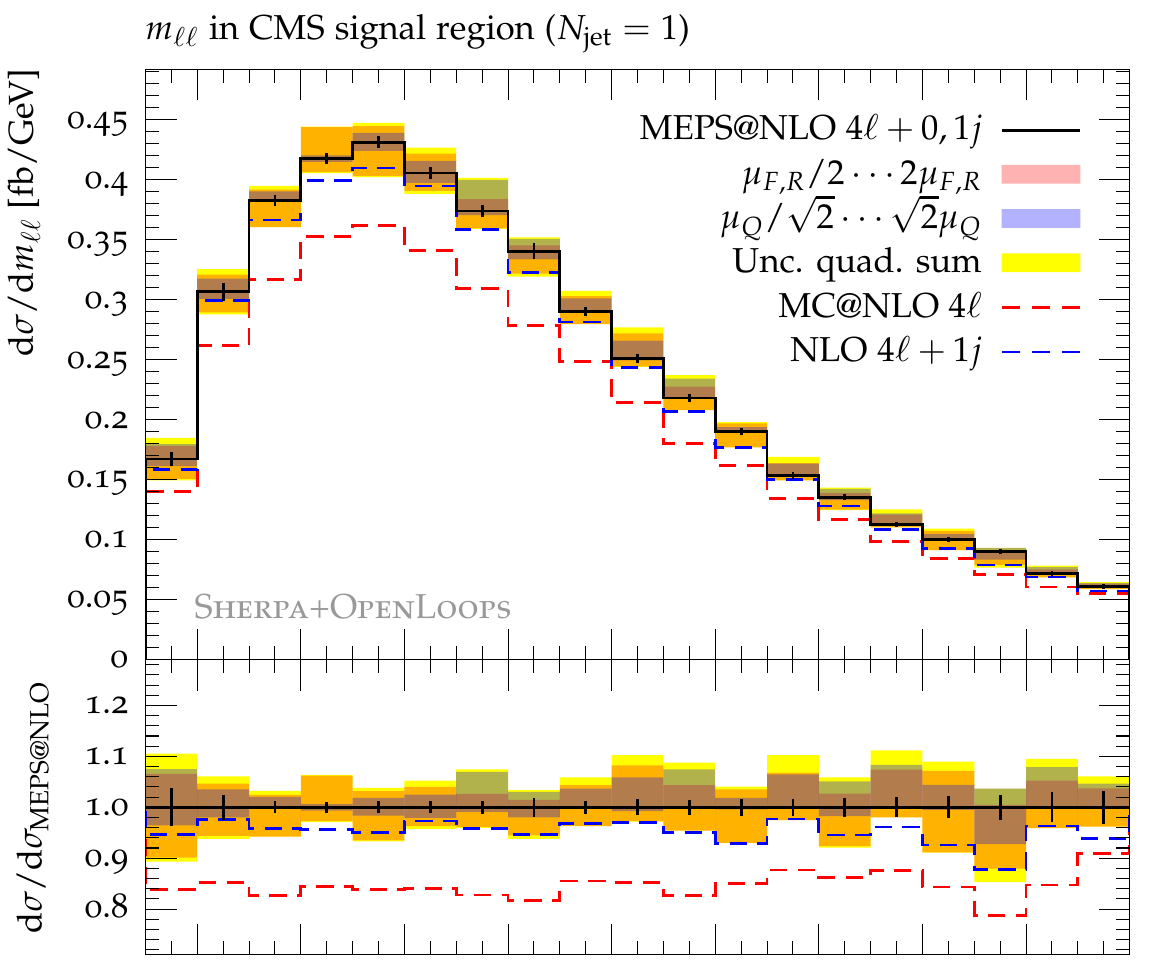}\\\vspace*{-3px}
\includegraphics[width=0.48\textwidth]{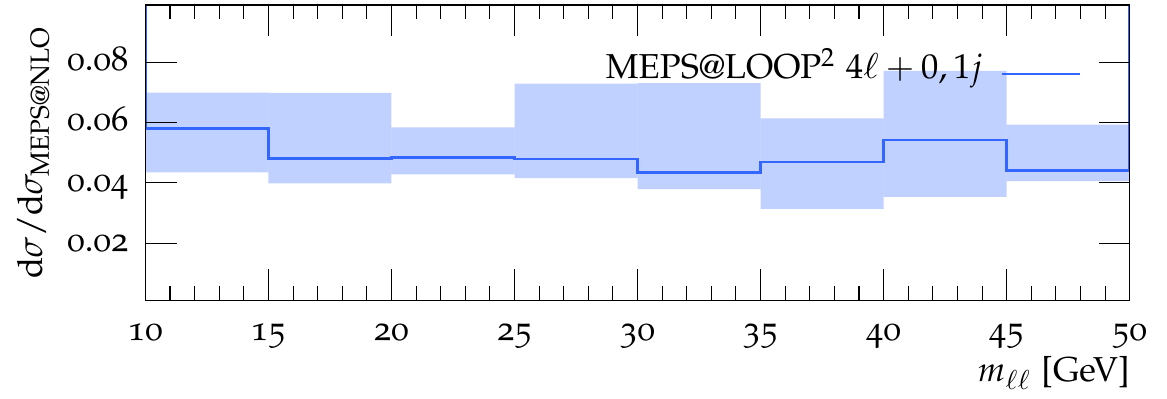}
\includegraphics[width=0.48\textwidth]{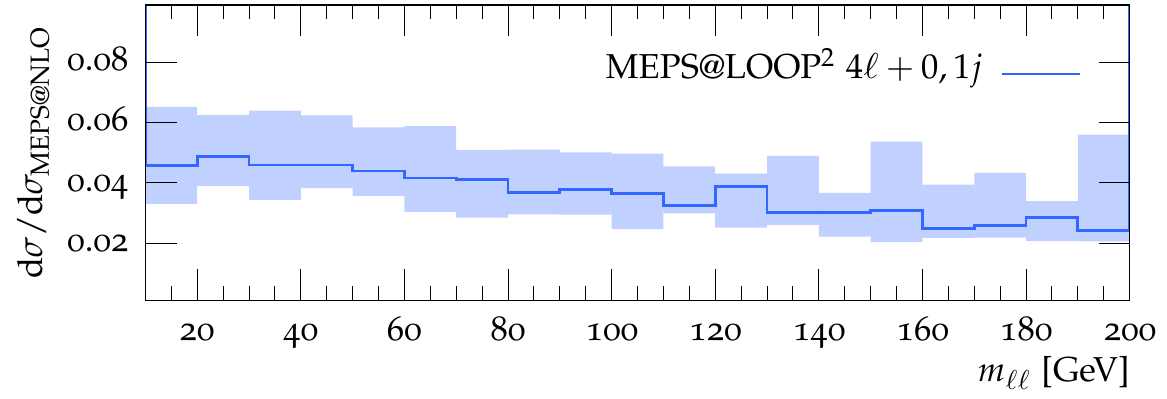}
\end{center}
\vspace*{-0.3cm}
\caption{
Signal region of the \ATLAS (left)  and \CMS (right)
analysis at $8\UTeV$:
dilepton invariant-mass distribution in the 0-jet (top) and 1-jet (bottom) bins.
Similar predictions and uncertainty bands as in \reffi{fig:flj_hww_pre_ptjets}.
} 
\label{fig:flj_hww_s_mll}
\end{figure}

We now turn to the control (C) and signal (S) regions of the experimental
analyses (see \refta{Tab:HWWcuts}) and discuss the
distributions in the $\PW\PW$ transverse mass, $m_\rT$, and in the dilepton
invariant mass, $\mll$.  These observables are sensitive to the Higgs-boson signal,
and their shape permits to increase the signal-to-background discrimination
in the final fit.  Separate distributions for the exclusive 0- and 1-jet
bins and for the two experiments are shown in
\reffis{fig:flj_hww_c_mT}--\ref{fig:flj_hww_s_mll}.

In the signal and control regions, as well as in both jet bins, the size of
the various corrections and the \MEPSatNLO uncertainties behave fairly similar to
what observed at pre-selection level.  The \NLOacc, \MCatNLO and \MEPSatNLO
distributions agree at few-percent level in the 0-jet bin, while in the
1-jet bin discrepancies between \MCatNLO and \MEPSatNLO on the 10--15\% level
and little \MCatNLO shape distortions appear.  The size of the corrections and the
scale uncertainties for the two experimental analyses are qualitatively and
quantitatively similar. Obviously, due to the different cuts,
absolute background predictions for \ATLAS and \CMS  behave differently.
The shapes of \MEPSatNLO distributions are again in excellent agreement with 
\NLOacc, suggesting  moderate Sudakov logarithms beyond 
NLO. This is consistent with the small  scale uncertainty 
of the merged simulation.

Squared quark-loop corrections feature a nontrivial  sensitivity to $m_\rT$
and $\mll$, which varies depending on the experimental analysis, the
selection region, and the jet bin.  The largest squared quark-loop
corrections arise in the 0-jet bin, at small $\mll$ and at large $m_\rT$. 
The corrections to the transverse-mass distribution start growing at
$m_\rT=100$--$150\UGeV$ and for the \ATLAS(\CMS) analysis they reach
10-20\%(5-10\%) in the tail.  The largest effects arise in the signal region
and in the \ATLAS analysis, which implements tighter $\mll$ and $\Delta
\phi_{\ell\ell'}$ cuts.  For what concerns the $\mll$ distribution,
\reffi{fig:flj_hww_s_mll} shows that in the 0-jet bin of the \CMS signal
region squared quark-loop corrections behave similarly as in the inclusive
case (\cf \reffi{fig:inclusive_loop2_mepsnlo}).  The fact that the
characteristic enhancement at small $\mll$ is not visible in the \ATLAS
signal region, is simply due to the cut on $\mll$ at $50\UGeV$.  For what
concerns the 1-jet bin, \MEPSatLOOPSQ corrections are generally slightly
smaller and less dependent on $m_\rT$ and $\mll$.

\subsection{Exclusive 0- and 1-jet bin cross sections in control and signal regions}
\label{se:hWW_XS}

\begin{table}
  \small
  \begin{center}
\begin{tabular}{|@{\;}c@{\;}||@{\;}c@{\;}|@{\;}c@{\;}|@{\;}c@{\;}|@{\;}c@{\;}|}
\hline
  0-jet bin	
& \NLOfljs 
& \MCatNLOfl
& \MEPSatNLOflj
& \MEPSatLOOPSQflj 
\\[1mm]\hline
  $\sigsr$ [fb]	
& $34.28(9)\;^{+2.1\%}_{-1.6\%}$ 
& $32.52(8)\;^{+2.1\%}_{-0.8\%}$$\;^{+1.2\%}_{-0.7\%}$ 
& $33.81(12)\;^{+1.4\%}_{-2.2\%}$$\;^{+2.0\%}_{-0.4\%}$$\;^{+1.6\%}_{-1.7\%}$ 
& $1.98(2) \;^{+23\%}_{-16.5\%}$$\;^{+27\%}_{-20\%}$ 
\\[1mm]\hline
  $\sigcr$ [fb]      
& $55.76(9)\;^{+2.0\%}_{-1.7\%}$ 
& $52.28(9)\;^{+1.4\%}_{-0.7\%}$$\;^{+1.4\%}_{-1.1\%}$ 
& $54.18(15)\;^{+1.4\%}_{-1.9\%}$$\;^{+2.5\%}_{-0.4\%}$$\;^{+1.7\%}_{-2.0\%}$ 
& $2.41(2) \;^{+22\%}_{-17\%}$$\;^{+27\%}_{-18\%}$ 
\\[1mm]\hline\hline
1-jet bin  
& \NLOfljs 
& \MCatNLOfl 
& \MEPSatNLOflj 
& \MEPSatLOOPSQflj 
\\[1mm]\hline
  $\sigsr$ [fb]      
& $8.99(4)\;^{+4.9\%}_{-9.5\%}$ 
& $8.02(4)\;^{+8.5\%}_{-6.4\%}$$\;^{+0\%}_{-3.1\%}$ 
& $9.37(9)\;^{+2.6\%}_{-2.7\%}$$\;^{+2.5\%}_{-0.0\%}$$\;^{-0.1\%}_{-4.1\%}$ 
& $0.46(1)\;^{+40\%}_{-18\%}$$\;^{+2.2\%}_{-6.3\%}$ 
\\[1mm]\hline
  $\sigcr$ [fb]      
& $26.50(8)\;^{+6.4\%}_{-12.5\%}$ 
& $24.58(8)\;^{+6.1\%}_{-6.5\%}$$\;^{+1.2\%}_{-3.0\%}$ 
& $28.32(13)\;^{+3.1\%}_{-4.7\%}$$\;^{+4.1\%}_{-0.0\%}$$\;^{+0.6\%}_{-2.7\%}$ 
& $0.79(1) \;^{+33\%}_{-20\%}$$\;^{+15\%}_{-7\%}$ 
\\[1mm]\hline
\end{tabular}
    \caption{Exclusive 0- and 1-jet bin $\mnen+$jets cross sections in the signal (\ssel) 
and control (\csel) regions of the \ATLAS analysis at
$8\UTeV$.  Fixed-order \NLOacc results (with appropriate jet multiplicity) are compared to
\MCatNLO and \MEPSatNLO predictions.
Squared quark-loop contributions (\MEPSatLOOPSQ) are presented separately.
Scale uncertainties are shown as
$\sigma\pm\delta_\mathrm{QCD}\pm\delta_\mathrm{res}\pm\delta_{\qcut}$,
where $\delta_\mathrm{QCD}$, $\delta_\mathrm{res}$ and $\delta_{\qcut}$ correspond respectively 
to variations of the QCD $(\mu_\mathrm{R}, \mu_{\mathrm{F}})$, resummation ($\mu_Q$) and 
merging ($\qcut$) scales.
Statistical errors are given in parenthesis.}
    \label{Tab:NLOMC_WWATLAS}
  \end{center}
\end{table}

\begin{table}
  \small
  \begin{center}
\begin{tabular}{|@{\,}c@{\,}||@{\,}c@{\,}|@{\,}c@{\,}|@{\,}c@{\,}|@{\,}c@{\,}|}
    \hline
0-jet bin   
& \NLOfljs 
& \MCatNLOfl 
& \MEPSatNLOflj 
& \MEPSatLOOPSQflj 
\\[1mm]\hline
$\sigsr$ [fb]      
& $156.65(18)\;^{+1.7\%}_{-1.6\%}$ 
& $147.8(2)\;^{+1.3\%}_{-0.6\%}$$\;^{+1.2\%}_{-1.0\%}$ 
& $153.6(3)\;^{+2.1\%}_{-1.9\%}$$\;^{+2.8\%}_{-0.0\%}$$\;^{+1.6\%}_{-2.2\%}$ 
& $ 6.65(4)\;^{+22\%}_{-17\%}$$\;^{+26\%}_{-18\%}$ 
\\[1mm]\hline
  $\sigcr$ [fb]      
& $59.26(15)\;^{+1.3\%}_{-1.3\%}$ 
& $55.92(11)\;^{+0.8\%}_{-0.2\%}$$\;^{+0.5\%}_{-0.9\%}$ 
& $58.06(21)\;^{+2.1\%}_{-2.0\%}$$\;^{+2.2\%}_{-0.2\%}$$\;^{+1.5\%}_{-2.1\%}$ 
& $ 1.47(2)\;^{+26\%}_{-17\%}$$\;^{+28\%}_{-16\%}$ 
\\[1mm]\hline\hline
1-jet bin  
& \NLOfljs 
& \MCatNLOfl 
& \MEPSatNLOflj 
& \MEPSatLOOPSQflj 
\\[1mm]\hline  
  $\sigsr$ [fb]      
& $43.01(9)\;^{+3.3\%}_{-7.4\%}$ 
& $37.87(9)\;^{+7.6\%}_{-6.8\%}$$\;^{+0.9\%}_{-3.6\%}$ 
& $44.99(18)\;^{+2.5\%}_{-4.2\%}$$\;^{+2.9\%}_{-0.0\%}$$\;^{+0.5\%}_{-2.5\%}$ 
& $ 1.83(2)\;^{+34\%}_{-20\%}$$\;^{+6\%}_{-7\%}$ 
\\[1mm]\hline
$\sigcr$ [fb]      
& $20.48(6)\;^{+4.8\%}_{-10.3\%}$ 
& $18.90(7)\;^{+7.4\%}_{-7.3\%}$$\;^{+1.8\%}_{-3.3\%}$ 
& $21.70(11)\;^{+3.2\%}_{-4.1\%}$$\;^{+3.4\%}_{-0.0\%}$$\;^{+0.5\%}_{-1.6\%}$ 
& $ 0.62(1)\;^{+39\%}_{-16\%}$$\;^{+16\%}_{-6\%}$ 
\\[1mm]\hline
    \end{tabular}
    \caption{Exclusive 0- and 1-jet bin $\mnen+$jets cross sections in the signal (\ssel) 
and control (\csel) regions of the \CMS analysis at $8\UTeV$. Similar predictions and conventions as in \refta{Tab:NLOMC_WWATLAS}.}
    \label{Tab:NLOMC_WWCMS}
  \end{center}
\end{table}

A precise quantitative assessment of the various correction effects and
residual uncertainties is provided in \reftas{Tab:NLOMC_WWATLAS}
and~\ref{Tab:NLOMC_WWCMS}, where we present exclusive 0- and 1-jet bin cross
sections in the signal and control regions of the two experimental analyses.
The \NLOacc and \MEPSatNLO predictions at the central scale
differ by only 1.5--3\% and 4--6\% in the 0-jet and 1-jet bins, respectively.  
This confirms that the discrepancy of order 5\%
observed in the inclusive 0-jet bin
(\cf \refta{Tab:InclusiveXS1}) is due the NLO corrections to the
first jet emission in \MEPSatNLO.
The differences between \MCatNLO and \MEPSatNLO in the exclusive 0- and 1-jet bins
reach 2--4\% and 13--16\%, respectively, and the discrepancy in the 1-jet bin
is consistent with the deficit of \MCatNLO observed in differential distributions.  
Deviations between the \NLOacc, \MCatNLO and \MEPSatNLO
approximations are fairly similar in the various analyses and kinematic
regions. As compared to corresponding results in~\citere{Heinemeyer:2013tqa},
the various cross sections in \reftas{Tab:NLOMC_WWATLAS} 
and~\ref{Tab:NLOMC_WWCMS} differ by 1-5\% and 1-10\% in the 0- and 1-jet bins, respectively.
These shifts are consistent with scale-variation uncertainties and can be
attributed, as observed in \refse{se:NLOWW}, to the new scale choice
\refeq{eq:scale} used in the present study.

Adding  $(\mu_\rR,\mu_\rF)$, $\mu_Q$ and $\qcut$ variations in quadrature, 
the combined scale uncertainties of \MEPSatNLO cross-section predictions 
do not exceed 4(6)\% in the 0(1)-jet bin.
Renormalisation-, factorisation-, resummation- and merging-scale variations
yield comparable contributions to the total scale uncertainty.
In the 1-jet bin, \MEPSatNLO results feature smaller QCD-scale variations as compared to 
the \NLOacc calculation. This can be
attributed to the variation of extra $\alphaS$ terms originating from the
shower and to the  CKKW scale choice in \MEPSatNLO.

Comparing \NLOacc and \MCatNLO cross sections in the 0-jet bin we observe a
rather constant difference of about 5\% that can be interpreted as the
contribution from resummed Sudakov logarithms beyond NLO.  On the one
hand, this indicates that matching to the parton shower is essential in
order to reach few-percent precision.  On the other hand, the rather mild
impact of Sudakov resummation suggests that subleading Sudakov logarithms
beyond the shower approximation should not have a large impact on the $\hww$
analysis.  This is confirmed by the fact that resummation-scale variations
of \MCatNLO and \MEPSatNLO cross sections do not exceed 2-3\% in the various
jet bins.

The relative impact of squared quark-loop corrections as compared to merged
\NLOacc predictions varies between 2.5 and 6 percent, depending on the experiment, the
kinematic selection region, and the jet bin.  
In both experiments and jet bins,
squared quark-loop effects increase when 
moving from control to signal regions.  
In the case of \CMS they grow from 2-3.5\% to 4\%, while in the
\ATLAS analysis, due to the tighter $\Delta \phi_{\ell\ell'}$ and $m_{\ell\ell'}$
cuts, the effects are more pronounced and increase from 3-4.5\% to
5-6\%.
Squared quark-loop uncertainties amount to 30--40\%,
similarly as for the inclusive analysis of \refse{Sec:WWinc}.

\begin{table}
  \small
  \begin{center}
\begin{tabular}{|@{\:}c@{\:}||@{\:}c@{\:}|@{\:}c@{\:}|@{\:}c@{\:}|@{\:}c@{\:}|@{\:}c@{\:}|}
    \hline
\ATLAS
& \NLOfljs
& \MCatNLOfl
& \MEPSatNLOflj
& \MEPSTOT
& $\dsc$
\\[1mm]\hline
$\sigsr/\sigcr$  
0-jet 
& $0.615\;^{-0.1\%}_{-0.1\%}$ 
& $0.622\;^{-0.7\%}_{+0.1\%}$$\;^{+0.2\%}_{-0.4\%}$ 
& $0.624\;^{+0\%}_{-0.3\%}$$\;^{+0.5\%}_{-0\%}$$\;^{+0.1\%}_{-0.3\%}$ 
& $0.632\;^{-0.3\%}_{+0.5\%}$$\;^{+0.2\%}_{+0.3\%}$ 
& $1.3\%$
\\[1mm]\hline
$\sigsr/\sigcr$  
1-jet 
& $0.339\;^{+1.4\%}_{-3.4\%}$ 
& $0.326\;^{-2.3\%}_{-0.1\%}$$\;^{+1.2\%}_{+0.1\%}$ 
& $0.331\;^{+0.5\%}_{-2.1\%}$$\;^{+1.5\%}_{-0\%}$$\;^{+0.7\%}_{+1.4\%}$ 
& $0.338\;^{-0.4\%}_{-1.8\%}$$\;^{+1.8\%}_{+0.1\%}$ 
& $2.1\%$
\\[1mm]\hline\hline
\CMS
& \NLOfljs
& \MCatNLOfl
& \MEPSatNLOflj
& \MEPSTOT
& $\dsc$
\\[1mm]\hline
$\sigsr/\sigcr$  
0-jet 
& $2.64\;^{-0.4\%}_{+0.3\%}$ 
& $2.64\;^{-0.5\%}_{+0.4\%}$$\;^{-0.7\%}_{+0.1\%}$ 
& $2.65\;^{+0\%}_{-0.1\%}$$\;^{-0.6\%}_{-0.2\%}$$\;^{-0.1\%}_{+0.1\%}$ 
& $2.69\;^{-0.2\%}_{+0.2\%}$$\;^{-0.9\%}_{+0.2\%}$ 
& $1.5\%$
\\[1mm]\hline
$\sigsr/\sigcr$  
1-jet 
& $2.10\;^{+1.4\%}_{-3.2\%}$ 
& $2.00\;^{-0.2\%}_{-0.5\%}$$\;^{+0.9\%}_{+0.3\%}$ 
& $2.07\;^{+0.7\%}_{+0.1\%}$$\;^{+0.5\%}_{-0\%}$$\;^{+0\%}_{+0.9\%}$ 
& $2.10\;^{+0.4\%}_{+0.4\%}$$\;^{+0.7\%}_{+0.1\%}$ 
& $1.4\%$
\\[1mm]\hline
    \end{tabular}
    \caption{ Ratios of signal- to control-region cross sections in the 0-
and 1-jet bins of the two experimental analyses.  Fixed-order
\NLOacc results (with appropriate jet multiplicity) are compared to
\MCatNLO and \MEPSatNLO predictions.
The combination of \NLOacc and squared quark-loop merged results, denoted as
\MEPSTOT, represents the best prediction.
Upper and lower variations are obtained from corresponding QCD-, resummation- and 
merging-scale uncertainties in \reftas{Tab:NLOMC_WWATLAS} and~\ref{Tab:NLOMC_WWCMS} assuming correlated
$\sigsr$ and $\sigcr$  variations.
The last column shows the relative difference between 
\MEPSatNLO and full \MEPSTOT predictions, which corresponds to the shift induced by
squared quark-loop corrections.
}
    \label{Tab:Ratios}
  \end{center}
\end{table}

Detailed results for the ratios of signal- to control-region cross sections,
$\sigsr/\sigcr$, are presented in \refta{Tab:Ratios}.  These ratios and the
related uncertainties play an important role for the extrapolation from
control to signal regions in data-driven WW-background determinations.  In
addition to \NLOacc, \MCatNLO and \MEPSatNLO ratios, we also present results
obtained from the combination of \NLOacc and squared quark-loop merging.
These latter are denoted as \MEPSTOT and represent our best
predictions. 
Upper and lower variations are obtained from corresponding QCD-, 
resummation- and merging-scale variations in \reftas{Tab:NLOMC_WWATLAS} and~\ref{Tab:NLOMC_WWCMS}.
More precisely, the ratios are evaluated at different scales,
\beq
R(\xi_\rR,\xi_\rF,\xi_Q,\qcut) = 
\frac{\sigsr(\xi_\rR\mu_\rR,\xi_\rF\mu_\rF,\xi_Q\mu_Q,\qcut)}
{\sigcr(\xi_\rR\mu_\rR,\xi_\rF\mu_\rF,\xi_Q\mu_Q,\qcut)},
\label{eq:ratiovars}
\eeq
applying correlated variations in signal and control regions.
As shown in \refta{Tab:Ratios}, due to almost complete cancellations between
$\sigsr$ and $\sigcr$ variations this naive approach results in typical
$\sigsr/\sigcr$ shifts at the sub-percent level, which cannot be regarded as
realistic estimates of uncertainties due to unknown higher-order
corrections.
On the other hand, applying uncorrelated scale variations to $\sigsr$ and $\sigcr$ would tend to overestimate 
$\sigsr/\sigcr$ uncertainties. This becomes clear if one considers the ideal limit of identical signal and control regions,
where $\sigsr/\sigcr=1$ and the uncertainty must vanish.
The reason why scale variations
are not adequate to quantify theory uncertainties 
associated to the extrapolation between different kinematic regions,
is that they tend to shift the normalisation of scattering amplitudes 
without altering their kinematic dependence. 
In this 
respect,
 squared quark-loop corrections
provide much more useful insights into kinematic effects 
associated to higher-order corrections. 
As shown in the last column of \refta{Tab:Ratios},
their impact on the $\sigsr/\sigcr$ 
ratios amounts to $\dsc\simeq 1.5\%$,  which largely 
exceeds the typical scale variations of \MEPSatNLO and \MEPSTOT predictions. 
This is due to the fact that squared quark-loop effects 
induce genuine NNLO kinematic distortions. Moreover,
squared quark loops constitute only a subset of the full NNLO corrections, 
and their impact on $\sigsr/\sigcr$ can be assumed to be quantitatively similar
to the still unknown NNLO contributions. With other words,
the $\dsc$ shifts in \refta{Tab:Ratios} can be considered as a realistic 
estimate of the \MEPSTOT uncertainty of the $\sigsr/\sigcr$ ratios.

\section{Conclusions}
\label{Sec:Conclusions}

In this publication we have presented the first results for the simulation
of hadronic four-lepton plus jets production using the novel \MEPSatNLO
multi-jet merging technology at NLO, and including also NNLO contributions
from squared quark loops.  This was also the first phenomenological
application of the fully automated approach provided by the combination of
the \Sherpa Monte Carlo with the \OpenLoops generator of one-loop
amplitudes.  The \OpenLoops algorithm is based on a new numerical approach for
the recursive construction of cut-opened loop diagrams, which allows for a very fast
evaluation of NLO matrix elements within the Standard Model.  For the
calculation of tensor integrals it relies on the \Collier library, which
implements the numerically stable reduction algorithms by Denner--Dittmaier.

Four-lepton plus jets final states are of large topical interest due to
their implications on ongoing Higgs-boson studies, and in this paper we
discussed detailed predictions for the \ATLAS and \CMS $\hww$ analyses at
$8\UTeV$ in the 0- and 1-jet bins.  For a thorough description of
four-lepton production---including off-shell vector-boson effects,
non-resonant topologies, and related interferences---the complex-mass scheme
was applied.  The use of exclusive jet bins, which is mandatory in order to
suppress the background provided by top-quark production and decay,
introduces potentially large theory uncertainties and ultimately requires a
very robust modelling of jet-production properties and related errors.  This
requires an NLO accurate description of jet radiation, with a careful
assessment of the uncertainties stemming from the usual perturbative scale
variations, but also a resummation of Sudakov logarithms arising from jet
vetoes, and an analysis of the related uncertainties.  The \MEPSatNLO
approach as implemented in \Sherpa allows to carry out this program in a
fully automated way.  In particular, the resummation of Sudakov logarithms
is effectively implemented by matching NLO matrix elements to the \Sherpa
parton shower, and uncertainties related to subleading Sudakov logarithms
beyond the shower approximation can be assessed through resummation-scale
variations.

In order to allow precise statements on the impact of jet vetoes and jet
binning on the $\hww$ analyses, we merged matrix elements for four leptons
plus up to one jet at NLO accuracy, thus arriving at a simulation of the WW
background with unprecedented accuracy.  As a result of this calculation the
residual scale uncertainty is reduced to about $5\%$ on observables related
to the hardest jet up to transverse momenta of the order of $200$~GeV.
We note large differences of up to $40\%$ with respect to NLO or
\MCatNLO simulations of the $\Pp\Pp\to 4\ell$ process.  These
differences typically manifest themselves in regions of large jet momentum,
where inclusive NLO or \MCatNLO predictions are bound to undershoot the QCD
activity.  This of course is even more pronounced for observables related to
the subleading jet. 
As compared to NLO predictions for $\Pp\Pp\to4\ell+1j$, apart from a generally 
good agreement, multi-jet merging yields quite significant corrections 
in the tail of the first-jet $p_\rT$ distribution.  
This effect can be attributed to the fact that
the CKKW-merging approach implemented in \MEPSatNLO
consistently adapts the renormalisation scale to the transverse momenta of the emitted jets.

The multi-jet merging thus improves the quality and stability of the
perturbative series, especially for jet observables.  This holds for hard
phase-space regions as well as for low jet momentum, where fixed-order calculations
start to suffer from the missing resummation of potentially large
logarithms.
Studying the case of a jet veto, we found that for veto scales around
30(10)$\UGeV$ resummation effects beyond NLO amount to about 5(20)\% of the
vetoed four-lepton cross section.  Their relatively small magnitude can be
attributed to the limited size of Sudakov logarithms but also to
cancellations between leading- and subleading-logarithmic contributions.

In the case of the inclusive four-lepton cross section, as a result of NLO
corrections to the first QCD emission, \MEPSatNLO results turn out to be 9\%
higher as compared to inclusive NLO and \MCatNLO calculations.  Moreover,
the CKKW scale choice in \MEPSatNLO leads to a milder renormalisation-scale
dependence as compared to fixed-order and \MCatNLO predictions evaluated at
a scale of the order of the W-boson transverse mass.
For leptonic observables in the exclusive jet bins of the $\hww$ analyses,
typically NLO and \MCatNLO provide a good description in the 0-jet bin,
but \MCatNLO exhibits a deficit of about 10--15\% in the 1-jet bin.  It is
notable that, for these observables, we find scale uncertainties of only a
few percent in our best NLO prediction, \ie \MEPSatNLO.
Our analysis indicates that also the uncertainties related to the choice of
resummation scale, and thus due to the parton shower and its resummation
properties, are at the percent level.  This is consistent with the
observation that Sudakov logarithms beyond NLO have a rather moderate impact
on the jet bins of the $\hww$ analysis, and it suggests that subleading
logarithmic corrections beyond the \MEPSatNLO accuracy should not be
important.

In addition to matched and merged NLO simulations, we also studied NNLO
contributions to four-lepton plus jets production that emerge through squared one-loop
amplitudes involving closed quark loops. 
These contributions are dominated by the gluon--gluon channel, which is 
enhanced by the high partonic flux.
Moreover, squared 
quark-loop 
corrections are quite sensitive 
to lepton--\-lepton correlations that play a key role in the $\hww$ analysis.  Their relative
impact as compared to the full NLO contributions amounts to only 3\% in
the inclusive case, but grows to 6\% if Higgs-analysis cuts are applied. 
This corresponds to about $50\%$ of the Higgs-boson signal in the relevant
analysis regions, which calls for a detailed theoretical investigation of
squared quark-loop terms and of their nontrivial kinematic features.
To this end we considered all relevant squared 
quark-loop matrix elements for the production of four leptons plus up to one jet.
In particular, in addition to the well-known gluon--gluon fusion contributions,
for the first time we also studied the $\Pg q\to 4\ell+q$,
$\Pg \bar q\to 4\ell+\bar q$, and $q\bar q\to 4\ell+\Pg$ channels.
In order to merge squared quark-loop corrections with different jet
multiplicity, we extended the tree-level multi-jet merging in \Sherpa to
include also purely loop-induced processes.  In this context, the
inclusion of the quark channels is indispensable for a 
consistent merging.
The net effect of this merging is a visibly harder tail in the jet transverse
momentum distribution with respect to the one obtained from only taking the 
leading $\Pg\Pg\to 4\ell$ contribution supplemented with the parton shower.  
To the best of our knowledge this has not been studied before.

In the $\hww$ analyses, the size of squared quark-loop corrections turns out
to vary from 2\% to 6\%, depending on the jet bin, on the kinematic region
and on the experiment.  The merging approach is especially important in
order to guarantee decent predictions in the 1-jet bin.  Due to their
nontrivial kinematic dependence, squared quark-loop corrections have a quite
significant impact on the extrapolation of the WW-background from control to
signal regions.  The resulting shift in the relevant cross-section ratios is
of order 1.5\%, and we argued that these corrections can be regarded as a realistic
estimate of unknown higher-order effects in the data-driven
determination of the WW-background at the LHC.

At this point it should be stressed that all the studies reported here are at
the parton level only, with one choice of PDFs to facilitate a clear and 
direct comparison between the different approaches.  It is, however, a 
straightforward exercise to allow for different PDFs or to go from the parton 
to the hadron level in a simulation like the one presented here: switching 
on hadronisation and the underlying event modelling allows to assess these 
effects automatically.  
As a further extension, it is possible to extend the current study to cases
including all possible other four-lepton final states or to study in more
detail the two-jet bin of the simulation, which is crucial for the
vector-boson fusion signatures.  For the latter case, the simulation could
be extended to the production of four leptons in association with two jets
at next-to leading order accuracy.  It can be anticipated that a simulation
on the level presented here would certainly lead to a similarly relevant
reduction of QCD uncertainties for this important channel of Higgs physics.

\appendix

\section{Treatment of bottom- and top-quark contributions}
\label{app:b-treatment}
Consistently with the five-flavour evolution of PDFs and
$\alphaS$, for bottom quarks we adopt the massless approximation.  Top
quarks are thus the
only QCD partons that we treat as massive.  They can contribute to $\Pp\Pp\to
\PWp\PWm+$jets through closed quark loops, but also via resonant top
propagators in sub-processes with external b quarks, such as
$\Pg\Pb\to\PWp\PWm\Pb$ and $\Pg\Pg\to\PWp\PWm\Pb\bar\Pb$.  Partonic channels of
this type are dominated by $\PW\Pt$ and $\Pt\bar\Pt$ production, and are more
conveniently handled as separate processes.  Therefore, as operational
definition of $\PWp\PWm+$jets production, we consider only partonic channels
that do not involve b quarks in the initial or final state.
As pointed out in~\citere{Dittmaier:2009un}, when excluding external
b quarks, care must be taken to avoid NLO infrared singularities in
$\Pp\Pp\to\PWp\PWm j$.  This issue is related to the renormalisation of the
external-gluon wave function, which receives a b-quark contribution
\begin{eqnarray}
\dZAb&=&\frac{\alphaS}{6\pi}
\left[\mu^{2\varepsilon}\Delta_\ir - \mu^{2\varepsilon}\Delta_\uv\right] = 0,
\label{eq:dzab}
\end{eqnarray}
where $\mu$ is the scale of dimensional regularisation, and infrared (IR) and
ultraviolet (UV) singularities in $D=4-2\varepsilon$ dimensions yield
\begin{eqnarray}
\Delta_{\ir,\uv}&=& 
\frac{(4\pi)^\varepsilon}{\Gamma(1-\varepsilon)}
\frac{1}{\varepsilon},\qquad
\mu^{2\varepsilon}\Delta_{\ir,\uv}= 
\Delta_{\ir,\uv}+\ln\mu^2+\cal{O}(\varepsilon).
\label{eq:poles}
\end{eqnarray}
The renormalisation constant \refeq{eq:dzab} vanishes due to an exact IR--UV
compensation.  However, while its UV pole $\mu^{2\varepsilon}\Delta_\uv$
cancels in renormalised $q\bar q\to\PWp\PWm \Pg$ amplitudes,\footnote{Here we
discuss only partonic processes with gluons in the final state.  Similar
arguments apply also to the crossing-related $q\Pg\to\PWp\PWm q$ and $\bar
q\Pg\to\PWp\PWm \bar q$ channels.} the compensation of the IR pole
$\mu^{2\varepsilon}\Delta_\ir$ requires a $q\bar q\to \PWp\PWm \Pb\bar\Pb$
real-emission counterpart involving collinear $\Pg\to\Pb\bar \Pb$
splittings.  The inclusion of $\PWp\PWm\Pb\bar\Pb$ final states---at least
in the collinear region---is thus indispensable for an infrared-safe
NLO definition of $\PWp\PWm j$ production in the five-flavour scheme.

In~\citere{Dittmaier:2009un}, the IR cancellation was achieved by including
the contribution of $\Pg\to\Pb\bar\Pb$ splittings to the Catani--Seymour
$\mathcal{I}$-operator~\cite{Catani:1996vz},
\begin{eqnarray} 
\calIb&=&-\frac{\alphaS}{6\pi}\left[ \Delta_\ir +
\frac{1}{2}\left(\ln\frac{\mu^2}{2p_q p_{\Pg}}+\ln\frac{\mu^2}{2p_{\bar{q}}
p_g}\right) +\frac{8}{3} \right], \label{eq:Iop} 
\end{eqnarray}
where $p_{q}, p_{\bar q}$ and $p_{\Pg}$ are the quark, anti-quark and gluon momenta,
respectively. Combining $\dZAb+\calIb$ yields an IR-finite and
$\ln\mu$-independent result. The $\mathcal{I}$-operator
contribution \refeq{eq:Iop} results from dipole-subtraction terms,
which approximate $\Pg\to\Pb\bar\Pb$ splittings in the collinear limit, upon 
integration over the entire $\Pb\bar\Pb$ phase space.  
In principle, it should be combined with a subtracted
real-emission counterpart, which is free from singularities but depends on the
cuts applied to the $\Pb\bar\Pb$ pair.  In~\citere{Dittmaier:2009un},
this finite real-emission part was omitted, arguing that its contribution
should be small if $\Pb\bar\Pb$ pairs are confined in a jet
cone. This kinematic restriction of the $\Pb\bar\Pb$ phase space 
would also suppress $\Pt\bar\Pt$ and $\Pt\PW$ contributions.
However, confining $\Pb\bar\Pb$ pairs in narrow jets would introduce 
potentially
large logarithms of the jet radius.
Moreover, the consistent inclusion of the real-emission part would exactly
cancel the $8/3$ term in \refeq{eq:Iop}, which results from the unphysical
dipoles, and replace it by an unknown cut-dependent contribution.  The
inclusion of $\mathcal{I}$-operator terms \refeq{eq:Iop} without
corresponding real-emission parts should thus be regarded as a
regularisation {\it prescription}, which guarantees the correct cancellation
of poles and large logarithms corresponding to inclusive $\Pb\bar \Pb$
emission, but involves ad-hoc constant parts.  This ambiguity can be removed
only upon inclusion of the dipole-subtracted $\PWp\PWm\Pb\bar\Pb$ remnant.

Based on these considerations, we adopt a splitting approach similar
to~\citere{Dittmaier:2009un}, but we prefer to subtract only the singular
and logarithmically-enhanced terms arising from inclusive $\Pg\to\Pb\bar\Pb$
emissions.  More precisely, instead of the subtraction term \refeq{eq:Iop}
we use\footnote{Technically, we circumvent the explicit implementation of
the subtraction term \refeq{eq:Imod} by assigning the values $\Delta_\ir\to
0$ and $\mu\to\mu_\rR$ to the dimensional-regularisation parameters.}
\begin{eqnarray} 
\calIb_{\irct}&=&-\frac{\alphaS}{6\pi}\left( \Delta_\ir +
\ln\frac{\mu^2}{\mu_\rR^2}\right).  
\label{eq:Imod}
\end{eqnarray}
Since the renormalisation scale $\mu_\rR$ is typically of the same order of
the kinematic invariants in \refeq{eq:Iop}, the main difference between
\refeq{eq:Iop} and \refeq{eq:Imod} amounts to 
\begin{equation}
\calIb-\calIb_{\irct}
=
-\frac{\alphaS}{6\pi}\left[ 
\frac{1}{2}\left(\ln\frac{\mu_\rR^2}{2p_q p_{\Pg}}+\ln\frac{\mu_\rR^2}{2p_{\bar{q}}
p_g}\right) +\frac{8}{3} \right]
\simeq -\frac{4\alphaS}{9\pi}\simeq - 1.7\%, 
\label{eq:subunc}
\end{equation}
and can be regarded as the typical ambiguity inherent in the 
separation
of
the $\PWp\PWm j$ and $\PWp\PWm\Pb\bar\Pb$ cross sections.  Note that, in
order to reflect this kind of uncertainty in standard scale-variation
studies, we intentionally introduce a fake $\ln\mu_\rR$ dependence in the 
IR-subtraction term \refeq{eq:Imod}.

This small ambiguity is due to the absence of the dipole-subtracted
$\PWp\PWm\Pb\bar\Pb$ emission, which is supposed to be included in a
separate calculation of 
$\PWp\PWm\Pb\bar\Pb$ production, \ie of
$\Pt\bar\Pt$ and $\PW\Pt$ off-shell production. 
It can be removed by combining
the $\PWp\PWm+$jets and $\PWp\PWm\Pb\bar\Pb$ calculations 
in a single simulation.
For a consistent matching of the two processes, the $\mathcal{I}$-operator
term \refeq{eq:Iop} in 
the 
$\Pp\Pp\to\PWp\PWm\Pb\bar\Pb$ 
calculation
should be replaced by the
finite shift\footnote{Here we assume that $\Pp\Pp\to\PWp\PWm\Pb\bar\Pb$ is
computed using dipole subtraction, but the matching procedure can be
obviously adapted to any other subtraction method.} 
\refeq{eq:subunc}.  

In summary, due to collinear $\Pg\to\Pb\bar\Pb$ singularities,
the splitting of $\Pp\Pp\to\PWp\PWm j$ and $\Pp\Pp\to\PWp\PWm\Pb\bar\Pb$ 
is not unique, and the subtraction term \refeq{eq:Imod}
corresponds to a natural matching prescription, which is free from 
large logarithms and ad-hoc constants.

\section{Cuts of the \ATLAS and \CMS \texorpdfstring{H$\to$WW$^*$}{H->WW*} analyses in 0- and 1-jet bins}
\label{se:hww_setup}

The cuts of the \ATLAS~\cite{ATLAS-CONF-2013-030} and 
\CMS~\cite{CMS-PAS-HIG-13-003} $\hww\to\mnen$ 
analyses at 8 TeV in the exclusive 0- and 1-jet bins
are listed  in \refta{Tab:HWWcuts}. 
To be close to
the experimental definitions of both \ATLAS and \CMS,
lepton isolation is implemented at the particle level. The scalar sum of
the transverse momenta of all visible particles within a $R=0.3$ cone
around the lepton candidate is not allowed to exceed 15\% of the
lepton $p_\perp$.
Partons are recombined 
into jets using the anti-$k_\perp$ algorithm~\cite{Cacciari:2008gp}.
The different $\PW\PW$ transverse-mass definition employed in \ATLAS and \CMS
is consistently taken into account,
\beq
    m_\rT^2 = \left\{\begin{array}{lcl}
    \left(\sqrt{p_{\perp,\ell\ell'}^2+m_{\ell\ell'}^2}+\met\right)^2-
    \left|p_{\perp,\ell\ell'}+\met
    \vphantom{\sqrt{p_{\perp,\ell\ell'}^2}}\right|^2 &
    \mathrm{for} & \mathrm{ATLAS}\\[2mm]
    2 |p_{\perp,\ell\ell'}|\,|\met|\,(1-\cos\Delta\phi_{\ell\ell',\,\met}) &
    \mathrm{for} & \mathrm{CMS}
    \end{array}\right.,
\label{eq:mt}
\eeq
where $p_{\perp,\ell\ell'}$ and $m_{\ell\ell'}$ are the transverse momentum and the 
mass of the di-lepton system, respectively, $\met$ is the missing transverse momentum, and 
$\Delta\phi_{\ell\ell',\met}$ is the difference in azimuth between $\met$ 
and $p_{\perp,\ell\ell'}$.
After a pre-selection ({\bf \psel}), additional cuts are applied that define 
a signal ({\bf \ssel}) and a control ({\bf \csel}) region. The latter is
exploited to normalise background simulations to data in the experimental 
analyses in each jet bin.  In the \ATLAS analysis, different cuts are 
applied in the 0- and 1-jet bins. 
All cuts have been implemented in 
form of a \Rivet~\cite{Buckley:2010ar} 
analysis.
\begin{table}
  \begin{center}
    \begin{tabular}{|ll||l|l|}
      \hline
      {{ anti-$k_{\mathrm{T}}$ jets}} && \multicolumn{1}{p{5cm}|}{\ATLAS} & 
      \multicolumn{1}{p{5cm}|}{\CMS} \\\hline 
      $R$ &  $=$    &  0.4          & 0.5\\
      $p_{\perp, j}(|\eta_{j}|)$ &$ >$  & $25\UGeV$ \hfill($|\eta_j|<2.4$) 
                                     & $30\UGeV$ \hfill($|\eta_j|<4.7$)\\ 
                                    && $30\UGeV$ \hfill($2.4<|\eta_j|<4.5$) &
      \\[2mm]\hline\hline
      {{\bf \psel} selection} && \multicolumn{1}{p{5cm}|}{\ATLAS} & 
      \multicolumn{1}{p{5cm}|}{\CMS} \\\hline 
      $p_{\perp,\{\ell_1,\,\ell_2\}}$ &$>$  & $25, 15\UGeV$   & $20, 10 \UGeV$\\
      $|\eta_{\{\Pe,\,\mu\}}|$ &$ <$       & 2.47, 2.5    & 2.5, 2.4\\
      $|\eta_{\Pe}|$           & $\notin$    & $[1.37,1.57]$   & \\
      $p_{\perp,\ell\ell'}$ &$ >$         & see {{\bf \ssel}, {\bf \csel}} 
                                                     &  $30 \UGeV$\\
      $m_{\ell\ell'}$ &$ >$              & $10\UGeV$       & $12\UGeV$\\
      $E\!\!\!\!/_T^{\mathrm(proj)}$ &$ >$ & $25\UGeV$      & $20\UGeV$
      \\[2mm]\hline\hline
      {\bf \ssel} region &  & \multicolumn{1}{p{5cm}|}{\ATLAS} & 
      \multicolumn{1}{p{5cm}|}{\CMS} \\\hline
      $\Delta\phi_{\ell\ell',\,E\!\!\!/_T}$&$>$ & $\pi/2$ \hfill(0 jets only) & \\ 
      $p_{\perp,\ell\ell'}$ &$ >$         & $30\UGeV$ \hfill(0 jets only) & \\
      $\Delta\phi_{\ell\ell'}$ &$ <$    &  1.8 rad       &  \\
      $m_{\ell\ell'}$ &$ <$             &  $50\UGeV$        & $200\UGeV$\\
      $m_T$ &$ \in$                   &                & $[60\UGeV,\,
                                                           280\UGeV]$\\[2mm]\hline\hline
      {{\bf \csel} region} &  & \multicolumn{1}{p{5cm}|}{\ATLAS} & 
      \multicolumn{1}{p{5cm}|}{\CMS} \\\hline 
      $\Delta\phi_{\ell\ell',\,E\!\!\!/_T}$&$>$ & $\pi/2$ \hfill(0 jets only) & \\ 
      $p_{\perp,\ell\ell'}$ &$ >$         & $30\UGeV$ \hfill(0 jets only) & \\
      $m_{\ell\ell'}$ &            & $\in [50,100]\UGeV$ \hfill(0 jets only)     
      & $>100\UGeV$\\
      &             & $>80\UGeV$ \hfill(1 jet only)   & \\
      \hline
    \end{tabular}
    \caption{Jet definitions and selection cuts in the 
      \ATLAS and \CMS analyses of  
      $\hww\to \mnen$ at $8\UTeV$. 
The cuts refer to various levels and 
      regions, namely event pre-selection ({\bf \psel} cuts), the signal region 
      ({\bf \psel} and {\bf \ssel} cuts) and the control region 
      ({\bf \psel} and {\bf \csel} cuts).  
The projected missing transverse energy $E\!\!\!/_T^{\mathrm(proj)}$ 
is defined as 
$E\!\!\!/_\rT^{\mathrm(proj)} = E\!\!\!/_\rT\cdot\sin\left(\min\{\Delta\phi_{\mathrm{near}},\,\pi/2\}\right)$,
where $\Delta\phi_{\mathrm{near}}$ denotes the angle between the missing
transverse momentum $E\!\!\!/_\rT$ and the nearest lepton in the transverse plane.
}
    \label{Tab:HWWcuts}
  \end{center}
\end{table}

\label{App:Cuts}

\acknowledgments

We are grateful to A.~Denner, S.~Dittmaier and L.~Hofer for providing us with the
one-loop tensor-integral library \Collier.
We thank T.~Gehrmann for discussions and S.~Kallweit for cross-checking four-lepton plus 0- and 1-jet
NLO
cross sections.  The research of F.~Cascioli, P.~Maierh\"ofer and S.~Pozzorini
is supported by the SNSF.  Stefan H\"oche was supported by the
U.S.\ Department of Energy under Contract No.\ DE--AC02--76SF00515. Frank
Siegert's work was supported by the German Research Foundation (DFG) via grant
DI 784/2-1. We gratefully thank the bwGRiD project for computational resources.
This research used resources of the National Energy Research 
Scientific Computing Center, which is supported by the Office of Science 
of the U.S. Department of Energy under Contract No. DE-AC02-05CH11231.

\bibliographystyle{JHEP}  
\bibliography{FourLeptons}

\providecommand{\href}[2]{#2}\begingroup\raggedright\begin{thebibliography}{10}

\bibitem{Aad:2012tfa}
{\bf ATLAS Collaboration} Collaboration, G.~Aad et~al., {\it {Observation of a
  new particle in the search for the Standard Model Higgs boson with the ATLAS
  detector at the LHC}},  {\em Phys.Lett.} {\bf B716} (2012) 1--29,
  [\href{http://xxx.lanl.gov/abs/1207.7214}{{\tt arXiv:1207.7214}}].

\bibitem{Chatrchyan:2012ufa}
{\bf CMS Collaboration} Collaboration, S.~Chatrchyan et~al., {\it {Observation
  of a new boson at a mass of 125 GeV with the CMS experiment at the LHC}},
  {\em Phys.Lett.} {\bf B716} (2012) 30--61,
  [\href{http://xxx.lanl.gov/abs/1207.7235}{{\tt arXiv:1207.7235}}].

\bibitem{ATLAS-CONF-2013-030}
{\bf ATLAS} Collaboration, {\em {Measurements of the properties of the
  Higgs-like boson in the $WW^{*}\to\ell\ell\nu\nu$ decay channel with the
  ATLAS detector using 25 fb$^{-1}$ of proton-proton collision data}}.
\newblock ATLAS-CONF-2013-030.

\bibitem{CMS-PAS-HIG-13-003}
{\bf CMS} Collaboration, {\em {Evidence for a particle decaying to $W^+W^-$ in
  the fully leptonic final state in a standard model Higgs boson search in pp
  collisions at the LHC}}.
\newblock CMS-PAS-HIG-13-003.

\bibitem{Ohnemus:1991kk}
J.~Ohnemus, {\it {An Order $\alpha_s$ calculation of hadronic $W^{-} W^{+}$
  production}},  {\em Phys.Rev.} {\bf D44} (1991) 1403--1414.

\bibitem{Frixione:1993yp}
S.~Frixione, {\it {A Next-to-leading order calculation of the cross-section for
  the production of W+ W- pairs in hadronic collisions}},  {\em Nucl.Phys.}
  {\bf B410} (1993) 280--324.

\bibitem{Ohnemus:1994ff}
J.~Ohnemus, {\it {Hadronic $Z Z$, $W^{-} W^{+}$, and $W^\pm Z$ production with
  QCD corrections and leptonic decays}},  {\em Phys.Rev.} {\bf D50} (1994)
  1931--1945, [\href{http://xxx.lanl.gov/abs/hep-ph/9403331}{{\tt
  hep-ph/9403331}}].

\bibitem{Dixon:1999di}
L.~J. Dixon, Z.~Kunszt, and A.~Signer, {\it {Vector boson pair production in
  hadronic collisions at order $\alpha_s$: Lepton correlations and anomalous
  couplings}},  {\em Phys.Rev.} {\bf D60} (1999) 114037,
  [\href{http://xxx.lanl.gov/abs/hep-ph/9907305}{{\tt hep-ph/9907305}}].

\bibitem{Campbell:1999ah}
J.~M. Campbell and R.~K. Ellis, {\it {An Update on vector boson pair production
  at hadron colliders}},  {\em Phys.Rev.} {\bf D60} (1999) 113006,
  [\href{http://xxx.lanl.gov/abs/hep-ph/9905386}{{\tt hep-ph/9905386}}].

\bibitem{Campbell:2011bn}
J.~M. Campbell, R.~K. Ellis, and C.~Williams, {\it {Vector boson pair
  production at the LHC}},  {\em JHEP} {\bf 1107} (2011) 018,
  [\href{http://xxx.lanl.gov/abs/1105.0020}{{\tt arXiv:1105.0020}}].

\bibitem{Frixione:2002ik}
S.~Frixione and B.~R. Webber, {\it {Matching NLO QCD computations and parton
  shower simulations}},  {\em JHEP} {\bf 0206} (2002) 029,
  [\href{http://xxx.lanl.gov/abs/hep-ph/0204244}{{\tt hep-ph/0204244}}].

\bibitem{Frixione:2007vw}
S.~Frixione, P.~Nason, and C.~Oleari, {\it {Matching NLO QCD computations with
  Parton Shower simulations: the POWHEG method}},  {\em JHEP} {\bf 0711} (2007)
  070, [\href{http://xxx.lanl.gov/abs/0709.2092}{{\tt arXiv:0709.2092}}].

\bibitem{Melia:2011tj}
T.~Melia, P.~Nason, R.~Rontsch, and G.~Zanderighi, {\it {W+W-, WZ and ZZ
  production in the POWHEG BOX}},  {\em JHEP} {\bf 1111} (2011) 078,
  [\href{http://xxx.lanl.gov/abs/1107.5051}{{\tt arXiv:1107.5051}}].

\bibitem{Frederix:2011ss}
R.~Frederix, S.~Frixione, V.~Hirschi, F.~Maltoni, R.~Pittau, et~al., {\it
  {Four-lepton production at hadron colliders: aMC@NLO predictions with
  theoretical uncertainties}},  {\em JHEP} {\bf 1202} (2012) 099,
  [\href{http://xxx.lanl.gov/abs/1110.4738}{{\tt arXiv:1110.4738}}].

\bibitem{Campbell:2007ev}
J.~M. Campbell, R.~K. Ellis, and G.~Zanderighi, {\it {Next-to-leading order
  predictions for $WW+1$ jet distributions at the LHC}},  {\em JHEP} {\bf 0712}
  (2007) 056, [\href{http://xxx.lanl.gov/abs/0710.1832}{{\tt
  arXiv:0710.1832}}].

\bibitem{Dittmaier:2007th}
S.~Dittmaier, S.~Kallweit, and P.~Uwer, {\it {NLO QCD corrections to WW+jet
  production at hadron colliders}},  {\em Phys.Rev.Lett.} {\bf 100} (2008)
  062003, [\href{http://xxx.lanl.gov/abs/0710.1577}{{\tt arXiv:0710.1577}}].

\bibitem{Dittmaier:2009un}
S.~Dittmaier, S.~Kallweit, and P.~Uwer, {\it {NLO QCD corrections to
  $pp/p\bar{p} \to WW+jet+X$ including leptonic W-boson decays}},  {\em
  Nucl.Phys.} {\bf B826} (2010) 18--70,
  [\href{http://xxx.lanl.gov/abs/0908.4124}{{\tt arXiv:0908.4124}}].

\bibitem{Jager:2006zc}
B.~Jaeger, C.~Oleari, and D.~Zeppenfeld, {\it {Next-to-leading order QCD
  corrections to W+W- production via vector-boson fusion}},  {\em JHEP} {\bf
  0607} (2006) 015, [\href{http://xxx.lanl.gov/abs/hep-ph/0603177}{{\tt
  hep-ph/0603177}}].

\bibitem{Jager:2013mu}
B.~Jaeger and G.~Zanderighi, {\it {Electroweak W+W-jj prodution at NLO in QCD
  matched with parton shower in the POWHEG-BOX}},  {\em JHEP} {\bf 1304} (2013)
  024, [\href{http://xxx.lanl.gov/abs/1301.1695}{{\tt arXiv:1301.1695}}].

\bibitem{Melia:2011dw}
T.~Melia, K.~Melnikov, R.~Rontsch, and G.~Zanderighi, {\it {NLO QCD corrections
  for $W^+W^-$ pair production in association with two jets at hadron
  colliders}},  {\em Phys.Rev.} {\bf D83} (2011) 114043,
  [\href{http://xxx.lanl.gov/abs/1104.2327}{{\tt arXiv:1104.2327}}].

\bibitem{Greiner:2012im}
N.~Greiner, G.~Heinrich, P.~Mastrolia, G.~Ossola, T.~Reiter, et~al., {\it {NLO
  QCD corrections to the production of W+ W- plus two jets at the LHC}},  {\em
  Phys.Lett.} {\bf B713} (2012) 277--283,
  [\href{http://xxx.lanl.gov/abs/1202.6004}{{\tt arXiv:1202.6004}}].

\bibitem{Melia:2010bm}
T.~Melia, K.~Melnikov, R.~Rontsch, and G.~Zanderighi, {\it {Next-to-leading
  order QCD predictions for $W^+W^+jj$ production at the LHC}},  {\em JHEP}
  {\bf 1012} (2010) 053, [\href{http://xxx.lanl.gov/abs/1007.5313}{{\tt
  arXiv:1007.5313}}].

\bibitem{Denner:2012dz}
A.~Denner, L.~Hosekova, and S.~Kallweit, {\it {NLO QCD corrections to W+ W+ jj
  production in vector-boson fusion at the LHC}},  {\em Phys.Rev.} {\bf D86}
  (2012) 114014, [\href{http://xxx.lanl.gov/abs/1209.2389}{{\tt
  arXiv:1209.2389}}].

\bibitem{Jager:2011ms}
B.~Jaeger and G.~Zanderighi, {\it {NLO corrections to electroweak and QCD
  production of W+W+ plus two jets in the POWHEGBOX}},  {\em JHEP} {\bf 1111}
  (2011) 055, [\href{http://xxx.lanl.gov/abs/1108.0864}{{\tt
  arXiv:1108.0864}}].

\bibitem{Campanario:2013qba}
F.~Campanario, M.~Kerner, L.~D. Ninh, and D.~Zeppenfeld, {\it {WZ production in
  association with two jets at NLO in QCD}},  {\em Phys.Rev.Lett.} {\bf 111}
  (2013) 052003, [\href{http://xxx.lanl.gov/abs/1305.1623}{{\tt
  arXiv:1305.1623}}].

\bibitem{Binoth:2005ua}
T.~Binoth, M.~Ciccolini, N.~Kauer, and M.~Kramer, {\it {Gluon-induced WW
  background to Higgs boson searches at the LHC}},  {\em JHEP} {\bf 0503}
  (2005) 065, [\href{http://xxx.lanl.gov/abs/hep-ph/0503094}{{\tt
  hep-ph/0503094}}].

\bibitem{Binoth:2006mf}
T.~Binoth, M.~Ciccolini, N.~Kauer, and M.~Kramer, {\it {Gluon-induced W-boson
  pair production at the LHC}},  {\em JHEP} {\bf 0612} (2006) 046,
  [\href{http://xxx.lanl.gov/abs/hep-ph/0611170}{{\tt hep-ph/0611170}}].

\bibitem{Campbell:2011cu}
J.~M. Campbell, R.~K. Ellis, and C.~Williams, {\it {Gluon-Gluon Contributions
  to W+ W- Production and Higgs Interference Effects}},  {\em JHEP} {\bf 1110}
  (2011) 005, [\href{http://xxx.lanl.gov/abs/1107.5569}{{\tt
  arXiv:1107.5569}}].

\bibitem{Melia:2012zg}
T.~Melia, K.~Melnikov, R.~Rontsch, M.~Schulze, and G.~Zanderighi, {\it {Gluon
  fusion contribution to W+W- + jet production}},  {\em JHEP} {\bf 1208} (2012)
  115, [\href{http://xxx.lanl.gov/abs/1205.6987}{{\tt arXiv:1205.6987}}].

\bibitem{Kauer:2012hd}
N.~Kauer and G.~Passarino, {\it {Inadequacy of zero-width approximation for a
  light Higgs boson signal}},  {\em JHEP} {\bf 1208} (2012) 116,
  [\href{http://xxx.lanl.gov/abs/1206.4803}{{\tt arXiv:1206.4803}}].

\bibitem{Heinemeyer:2013tqa}
{\bf The LHC Higgs Cross Section Working Group} Collaboration, S.~Heinemeyer
  et~al., {\it {Handbook of LHC Higgs Cross Sections: 3. Higgs Properties}},
  \href{http://xxx.lanl.gov/abs/1307.1347}{{\tt arXiv:1307.1347}}.

\bibitem{Gleisberg:2008ta}
T.~Gleisberg, S.~Hoeche, F.~Krauss, M.~Schonherr, S.~Schumann, et~al., {\it
  {Event generation with SHERPA 1.1}},  {\em JHEP} {\bf 0902} (2009) 007,
  [\href{http://xxx.lanl.gov/abs/0811.4622}{{\tt arXiv:0811.4622}}].

\bibitem{Cascioli:2011va}
F.~Cascioli, P.~Maierhoefer, and S.~Pozzorini, {\it {Scattering Amplitudes with
  Open Loops}},  {\em Phys.Rev.Lett.} {\bf 108} (2012) 111601,
  [\href{http://xxx.lanl.gov/abs/1111.5206}{{\tt arXiv:1111.5206}}].

\bibitem{collier}
A.~Denner, S.~Dittmaier, and L.~Hofer.
\newblock In preparation.

\bibitem{Denner:2002ii}
A.~Denner and S.~Dittmaier, {\it {Reduction of one loop tensor five point
  integrals}},  {\em Nucl.Phys.} {\bf B658} (2003) 175--202,
  [\href{http://xxx.lanl.gov/abs/hep-ph/0212259}{{\tt hep-ph/0212259}}].

\bibitem{Denner:2005nn}
A.~Denner and S.~Dittmaier, {\it {Reduction schemes for one-loop tensor
  integrals}},  {\em Nucl.Phys.} {\bf B734} (2006) 62--115,
  [\href{http://xxx.lanl.gov/abs/hep-ph/0509141}{{\tt hep-ph/0509141}}].

\bibitem{Denner:2010tr}
A.~Denner and S.~Dittmaier, {\it {Scalar one-loop 4-point integrals}},  {\em
  Nucl.Phys.} {\bf B844} (2011) 199--242,
  [\href{http://xxx.lanl.gov/abs/1005.2076}{{\tt arXiv:1005.2076}}].

\bibitem{Hoeche:2011fd}
S.~Hoeche, F.~Krauss, M.~Schonherr, and F.~Siegert, {\it {A critical appraisal
  of NLO+PS matching methods}},  {\em JHEP} {\bf 1209} (2012) 049,
  [\href{http://xxx.lanl.gov/abs/1111.1220}{{\tt arXiv:1111.1220}}].

\bibitem{Hoeche:2012ft}
S.~Hoeche, F.~Krauss, M.~Schonherr, and F.~Siegert, {\it {W+n-jet predictions
  at the Large Hadron Collider at next-to-leading order matched with a parton
  shower}},  {\em Phys.Rev.Lett.} {\bf 110} (2013) 052001,
  [\href{http://xxx.lanl.gov/abs/1201.5882}{{\tt arXiv:1201.5882}}].

\bibitem{Gehrmann:2012yg}
T.~Gehrmann, S.~Hoeche, F.~Krauss, M.~Schonherr, and F.~Siegert, {\it {NLO QCD
  matrix elements + parton showers in $e^+e^-\to$ hadrons}},  {\em JHEP} {\bf
  1301} (2013) 144, [\href{http://xxx.lanl.gov/abs/1207.5031}{{\tt
  arXiv:1207.5031}}].

\bibitem{Hoeche:2012yf}
S.~Hoeche, F.~Krauss, M.~Schonherr, and F.~Siegert, {\it {QCD matrix elements +
  parton showers: The NLO case}},  {\em JHEP} {\bf 1304} (2013) 027,
  [\href{http://xxx.lanl.gov/abs/1207.5030}{{\tt arXiv:1207.5030}}].

\bibitem{Denner:2005fg}
A.~Denner, S.~Dittmaier, M.~Roth, and L.~Wieders, {\it {Electroweak corrections
  to charged-current $e^+ e^-\to$ 4 fermion processes: Technical details and
  further results}},  {\em Nucl.Phys.} {\bf B724} (2005) 247--294,
  [\href{http://xxx.lanl.gov/abs/hep-ph/0505042}{{\tt hep-ph/0505042}}].

\bibitem{Bredenstein:2009aj}
A.~Bredenstein, A.~Denner, S.~Dittmaier, and S.~Pozzorini, {\it {NLO QCD
  corrections to $pp \to t \bar{t} b \bar{b} + X$ at the LHC}},  {\em
  Phys.Rev.Lett.} {\bf 103} (2009) 012002,
  [\href{http://xxx.lanl.gov/abs/0905.0110}{{\tt arXiv:0905.0110}}].

\bibitem{Bredenstein:2010rs}
A.~Bredenstein, A.~Denner, S.~Dittmaier, and S.~Pozzorini, {\it {NLO QCD
  Corrections to Top Anti-Top Bottom Anti-Bottom Production at the LHC: 2. full
  hadronic results}},  {\em JHEP} {\bf 1003} (2010) 021,
  [\href{http://xxx.lanl.gov/abs/1001.4006}{{\tt arXiv:1001.4006}}].

\bibitem{Denner:2010jp}
A.~Denner, S.~Dittmaier, S.~Kallweit, and S.~Pozzorini, {\it {NLO QCD
  corrections to WWbb production at hadron colliders}},  {\em Phys.Rev.Lett.}
  {\bf 106} (2011) 052001, [\href{http://xxx.lanl.gov/abs/1012.3975}{{\tt
  arXiv:1012.3975}}].

\bibitem{Denner:2012yc}
A.~Denner, S.~Dittmaier, S.~Kallweit, and S.~Pozzorini, {\it {NLO QCD
  corrections to off-shell top-antitop production with leptonic decays at
  hadron colliders}},  {\em JHEP} {\bf 1210} (2012) 110,
  [\href{http://xxx.lanl.gov/abs/1207.5018}{{\tt arXiv:1207.5018}}].

\bibitem{vanHameren:2009vq}
A.~van Hameren, {\it {Multi-gluon one-loop amplitudes using tensor integrals}},
   {\em JHEP} {\bf 0907} (2009) 088,
  [\href{http://xxx.lanl.gov/abs/0905.1005}{{\tt arXiv:0905.1005}}].

\bibitem{Draggiotis:2009yb}
P.~Draggiotis, M.~Garzelli, C.~Papadopoulos, and R.~Pittau, {\it {Feynman Rules
  for the Rational Part of the QCD 1-loop amplitudes}},  {\em JHEP} {\bf 0904}
  (2009) 072, [\href{http://xxx.lanl.gov/abs/0903.0356}{{\tt
  arXiv:0903.0356}}].

\bibitem{Ossola:2006us}
G.~Ossola, C.~G. Papadopoulos, and R.~Pittau, {\it {Reducing full one-loop
  amplitudes to scalar integrals at the integrand level}},  {\em Nucl.Phys.}
  {\bf B763} (2007) 147--169,
  [\href{http://xxx.lanl.gov/abs/hep-ph/0609007}{{\tt hep-ph/0609007}}].

\bibitem{Catani:1991hj}
S.~Catani, Y.~L. Dokshitzer, M.~Olsson, G.~Turnock, and B.~R. Webber, {\it {New
  clustering algorithm for multijet cross sections in $e^+e^-$ annihilation}},
  {\em Phys. Lett.} {\bf B269} (1991) 432--438.

\bibitem{Catani:1990rr}
S.~Catani, B.~R. Webber, and G.~Marchesini, {\it {QCD coherent branching and
  semiinclusive processes at large $x$}},  {\em Nucl. Phys.} {\bf B349} (1991)
  635--654.

\bibitem{Stewart:2011cf}
I.~W. Stewart and F.~J. Tackmann, {\it {Theory Uncertainties for Higgs and
  Other Searches Using Jet Bins}},  {\em Phys.Rev.} {\bf D85} (2012) 034011,
  [\href{http://xxx.lanl.gov/abs/1107.2117}{{\tt arXiv:1107.2117}}].

\bibitem{Catani:1996vz}
S.~Catani and M.~Seymour, {\it {A General algorithm for calculating jet
  cross-sections in NLO QCD}},  {\em Nucl.Phys.} {\bf B485} (1997) 291--419,
  [\href{http://xxx.lanl.gov/abs/hep-ph/9605323}{{\tt hep-ph/9605323}}].

\bibitem{Schumann:2007mg}
S.~Schumann and F.~Krauss, {\it {A Parton shower algorithm based on
  Catani-Seymour dipole factorisation}},  {\em JHEP} {\bf 0803} (2008) 038,
  [\href{http://xxx.lanl.gov/abs/0709.1027}{{\tt arXiv:0709.1027}}].

\bibitem{Hoeche:2009xc}
S.~Hoeche, S.~Schumann, and F.~Siegert, {\it {Hard photon production and
  matrix-element parton-shower merging}},  {\em Phys.Rev.} {\bf D81} (2010)
  034026, [\href{http://xxx.lanl.gov/abs/0912.3501}{{\tt arXiv:0912.3501}}].

\bibitem{Hoeche:2009rj}
S.~Hoeche, F.~Krauss, S.~Schumann, and F.~Siegert, {\it {QCD matrix elements
  and truncated showers}},  {\em JHEP} {\bf 0905} (2009) 053,
  [\href{http://xxx.lanl.gov/abs/0903.1219}{{\tt arXiv:0903.1219}}].

\bibitem{Catani:2001cc}
S.~Catani, F.~Krauss, R.~Kuhn, and B.~Webber, {\it {QCD matrix elements +
  parton showers}},  {\em JHEP} {\bf 0111} (2001) 063,
  [\href{http://xxx.lanl.gov/abs/hep-ph/0109231}{{\tt hep-ph/0109231}}].

\bibitem{Agrawal:2012df}
P.~Agrawal and A.~Shivaji, {\it {Di-Vector Boson + Jet Production via Gluon
  Fusion at Hadron Colliders}},  {\em Phys.Rev.} {\bf D86} (2012) 073013,
  [\href{http://xxx.lanl.gov/abs/1207.2927}{{\tt arXiv:1207.2927}}].

\bibitem{Lai:2010vv}
H.-L. Lai, M.~Guzzi, J.~Huston, Z.~Li, P.~M. Nadolsky, et~al., {\it {New parton
  distributions for collider physics}},  {\em Phys.Rev.} {\bf D82} (2010)
  074024, [\href{http://xxx.lanl.gov/abs/1007.2241}{{\tt arXiv:1007.2241}}].

\bibitem{Campbell:2010ff}
J.~M. Campbell and R.~Ellis, {\it {MCFM for the Tevatron and the LHC}},  {\em
  Nucl.Phys.Proc.Suppl.} {\bf 205-206} (2010) 10--15,
  [\href{http://xxx.lanl.gov/abs/1007.3492}{{\tt arXiv:1007.3492}}].

\bibitem{Krauss:2001iv}
F.~Krauss, R.~Kuhn, and G.~Soff, {\it {AMEGIC++ 1.0: A Matrix element generator
  in C++}},  {\em JHEP} {\bf 0202} (2002) 044,
  [\href{http://xxx.lanl.gov/abs/hep-ph/0109036}{{\tt hep-ph/0109036}}].

\bibitem{Gleisberg:2008fv}
T.~Gleisberg and S.~Hoeche, {\it {Comix, a new matrix element generator}},
  {\em JHEP} {\bf 0812} (2008) 039,
  [\href{http://xxx.lanl.gov/abs/0808.3674}{{\tt arXiv:0808.3674}}].

\bibitem{Gleisberg:2007md}
T.~Gleisberg and F.~Krauss, {\it {Automating dipole subtraction for QCD NLO
  calculations}},  {\em Eur.Phys.J.} {\bf C53} (2008) 501--523,
  [\href{http://xxx.lanl.gov/abs/0709.2881}{{\tt arXiv:0709.2881}}].

\bibitem{Lonnblad:2012ng}
L.~Lonnblad and S.~Prestel, {\it {Unitarising Matrix Element + Parton Shower
  merging}},  {\em JHEP} {\bf 1302} (2013) 094,
  [\href{http://xxx.lanl.gov/abs/1211.4827}{{\tt arXiv:1211.4827}}].

\bibitem{Lonnblad:2012ix}
L.~Lonnblad and S.~Prestel, {\it {Merging Multi-leg NLO Matrix Elements with
  Parton Showers}},  {\em JHEP} {\bf 1303} (2013) 166,
  [\href{http://xxx.lanl.gov/abs/1211.7278}{{\tt arXiv:1211.7278}}].

\bibitem{Buckley:2010ar}
A.~Buckley, J.~Butterworth, L.~Lonnblad, H.~Hoeth, J.~Monk, et~al., {\it {Rivet
  user manual}},  \href{http://xxx.lanl.gov/abs/1003.0694}{{\tt
  arXiv:1003.0694}}.

\bibitem{Cacciari:2008gp}
M.~Cacciari, G.~P. Salam, and G.~Soyez, {\it {The Anti-k(t) jet clustering
  algorithm}},  {\em JHEP} {\bf 0804} (2008) 063,
  [\href{http://xxx.lanl.gov/abs/0802.1189}{{\tt arXiv:0802.1189}}].

\end{thebibliography}\endgroup

\end{document}